\documentclass[11pt]{article}
\usepackage{amssymb,latexsym, amsmath}
\newtheorem{corollary}{Corollary}
\newtheorem{lemma}{Lemma}
\newtheorem{thm}{Theorem}[section]

\makeatletter
\@addtoreset{equation}{section}

\makeatother

\newcommand{\comment}[1]{\vspace{5 mm}\par 
\marginpar{\large\underline{}}
\framebox{\begin{minipage}[c]{0.95 \textwidth}
\rm #1 \end{minipage}}\vspace{5 mm}\par}
\def\boxeq#1{\\ \nopagebreak
\framebox{\rule[-#1\baselineskip]{\textwidth}{0pt}
\rule[-#1\baselineskip]{0pt}{#1.5\baselineskip}}
\vspace{-#1\baselineskip}\vspace{-1\baselineskip}
}
\def\mathbb#1{\mathbf{ #1}}
\def\mathfrak#1{\mathbf{ #1}}

\begin{document}

\title{\vbox to 0pt 
{\vskip -1cm \rlap{\hbox to .5\textwidth 
{\rm{
{\small SUBMITTED TO }
{\it Physica D}
} } 
} }
\vspace{-6mm}
Averaged Lagrangians and the mean dynamical effects 
of fluctuations in continuum mechanics
}
\author{
Darryl D. Holm
\\Theoretical Division and Center for Nonlinear Studies
\\Los Alamos National
Laboratory, MS B284
\\ Los Alamos, NM 87545
\\ {\footnotesize email: dholm@lanl.gov}
}

\date{February 19, 2001\vspace{-6mm}}

\maketitle

\large


\begin{abstract}

We begin by placing the Generalized Lagrangian Mean (GLM)
equations for a compressible adiabatic fluid into the
Euler-Poincar\'e (EP) variational framework of fluid dynamics, for an
averaged Lagrangian. We then state the EP Averaging Lemma -- that
GLM averaged equations arise from GLM averaged Lagrangians in
the EP framework. Next, we derive a set of approximate small amplitude
GLM equations ($g\ell{m}$ equations) at second order in the
fluctuating displacement of a Lagrangian trajectory from its mean
position. These equations express the linear and nonlinear
back-reaction effects on the Eulerian mean fluid quantities by the
fluctuating displacements of the Lagrangian trajectories in terms of
their Eulerian second moments. 

The derivation of the $g\ell{m}$ equations uses the linearized
relations between Eulerian and Lagrangian fluctuctions, in the
tradition of Lagrangian stability analysis for fluids. The
$g\ell{m}$ derivation also uses the method of averaged
Lagrangians, in the tradition of wave, mean flow interaction (WMFI).
The new $g\ell{m}$ EP motion equations for compressible and
incompressible ideal fluids are compared with the Euler-alpha
turbulence closure equations. An alpha model is a GLM (or
$g\ell{m}$) fluid theory with a Taylor hypothesis closure (THC). Such
closures are based on the linearized fluctuation relations that
determine the dynamics of the Lagrangian statistical quantities in
the Euler-alpha equations. Thus, by using the EP Averaging Lemma, we
bridge between the GLM equations and the Euler-alpha closure
equations, upon making the small-amplitude approximation resulting in
the new $g\ell{m}$ equations in the EP variational framework. The
$g\ell{m}$ equations also lead to generalizations of the Euler-alpha
models to include compressibility and magnetic fields.

\end{abstract}

\clearpage
\tableofcontents

\vfill

\comment{
{\bf \Large Topical outline:}

\begin{description}
\item 
Defining relations for Generalized Lagrangian mean
\item
Transport properties of the GLM equations
\begin{description}
\item
Scalar advection of specific entropy for adiabatic fluids
\item
Mass conservation
\item
Magnetic flux advection
\item
GLM circulation theorem
\end{description}
\item
EP formulation of the GLM equations using an averaged variational
principle
\item
EP Averaging Lemma
\item
Discussion of properties of EP fluid theories inherited by the GLM
equations -- Kelvin circulation theorem and conservation laws
\item
Linear fluctuation relations obtained from Taylor series
approximations of the fluid variables
\\
-- the $\chi^{\,\prime}$, $D^{\,\prime}$ and $\mathbf{u}^{\,\prime}$
equations and their relation to Lagrangian fluctuating displacements
and gradients of mean fluid quantities.
\item
Truncation at second order in these fluctuations and averaging in the
GLM variational principle
\item
EP derivation of the new $g\ell{m}$ equations
\item
Particular solutions of the $\mathbf{u}^{\,\prime}$ equation
and their use in making {\bf Taylor hypothesis closures (THCs)}
\item
EP Variants of the alpha models arising from substituting Taylor
hypotheses based on the linear fluctuation relations into the averaged
variational principle at second order
\item
Adding the effects of compressibility and magnetic fields to the
Euler-alpha models
\end{description}

}

\vfill\eject


\section{Introduction}

\subsection{Decomposition of multiscale problems \& scale-up}

In turbulence, in climate modeling and in other multiscale fluids
problems, a major challenge is ``scale-up.'' This is the challenge of
deriving models that correctly capture the mean, or large scale flow
-- including the influence on it of the rapid, or small scale
dynamics. 

In classical mechanics this sort of problem has been approached by
choosing a proper ``slow + fast'' decomposition and deriving
evolution equations for the slow mean quantities by using, say, the 
method of averages. See e.g., Sanders \& Verhulst [1985], and
Lochak \& Meunier [1988] for descriptions of this method.  For
nondissipative systems in classical mechanics that arise from
Hamilton's variational principle, the method of averages may extend
to the averaged Lagrangian method, under certain conditions. See,
e.g., Whitham [1965, 1970] for discussions of these conditions. In
applying this method, the averaged Lagrangian acquires additional
symmetries that are induced by the averaging process. These
symmetries imply additional conservation laws for ``adiabatic
invariants'' obtained via Noether's theorem. Thus, perhaps not
unexpectedly, the system that results after averaging has fewer
actively coupled degrees of freedom than in the original system,
because averaging creates ignorable coordinates. This is an example
of Lagrangian reduction by symmetries -- the counterpart for
Lagrangian systems of Marsden-Weinstein reduction for Hamiltonian
systems, introduced in Marsden \& Scheurle [1993]. 

\paragraph{Lagrangian reduction for mechanics on Lie groups.}

Some additional features arise in Lagrangian reduction when the
system's state space is a Lie group, $G$, as in the case of fluid
dynamics. In particular, suppose that a Lagrangian defined on the
tangent space of a Lie group is also invariant under the group's
action. Then a reduced Lagrangian may be defined on the group's Lie
algebra (i.e., the tangent space at its identity). Such a reduction
can be accomplished, for example, by transforming the Lagrangian to
group invariant variables. The Euler-Lagrange equations on the
group's cotangent space then reduce to {\bf Euler-Poincar\'e (EP)
equations} describing motion on the dual of its Lie algebra. This EP
framework is now well developed. See Marsden \& Ratiu [1999] for the
an excellent introduction to the classical theory. For continuum
mechanics, see Holm, Marsden \& Ratiu [1998a], who provide proofs of
the basic EP theorems and present many applications of them for ideal
fluids and plasmas with advected quantities. 

In fluid dynamics, $G$ is the group of diffeomorphisms -- the smooth
invertible maps that take the reference configuration of the fluid
into its current configuration. Correspondingly, the Euler-Lagrange
dynamics occurs in the Lagrangian picture of fluid motion, and the
Euler-Poincar\'e dynamics occurs in the Eulerian picture. The
invariance of the Lagrangian under the action of the group $G$ arises
in fluids whenever the variational principle depends only on Eulerian
variables. These Eulerian variables are invariant under redefinitions
of the Lagrangian coordinates by the action of the group (particle
relabeling symmetry). 

The EP theorems in the forms we shall use them for
modeling in fluid dynamics are stated in Appendix \#1 (section
\ref{EP-appendix}). There are two main EP theorems for continuum
theories. The first EP theorem establishes that the two motion
equations and also the two variational principles for the Lagrangian
and Eulerian pictures of continua are {\it all} equivalent. The
second EP theorem establishes the Kelvin-Noether circulation law as a
result of Lagrangian reduction by particle relabeling symmetry. These
two main EP theorems describe how the transport structure of Eulerian
fluid dynamics arises variationally, after factoring the Lagrangian
by its particle relabeling symmetry. For the continuum theory, see
Holm, Marsden \& Ratiu [1998a], and for its basis in classical
mechanics, see Marsden \& Ratiu [1999]. 

\paragraph{Eulerian vs Lagrangian means.}

In meteorology and oceanography, the averaging approach has a
venerable history and many facets. Often this averaging is applied in
the geosciences in combination with additional approximations
involving force balances (for example, geostrophic and hydrostatic
balances). It is also sometimes discussed as an initialization
procedure that seeks a nearby invariant ``slow manifold'' Leith
[1980].  Moreover, in meteorology and oceanography, the averaging may
be performed in either the Eulerian, or the Lagrangian description.
The relation between averaged quantities in the Eulerian and
Lagrangian descriptions is one of the classical problems of fluid
dynamics.

\paragraph{Wave, mean-flow interaction (WMFI).}

An example of a slow + fast decomposition appearing in meteorology and
oceanography is the wave, mean-flow interaction (WMFI) problem. For
reports of recent progress in the WMFI problem see, e.g., Grimshaw
[1984] and Gjaja \&  Holm [1996]. In the WMFI problem, Lagrangian
trajectories are represented as a mean-flow trajectory plus a WKB wave
packet. The wave packet is taken as a small-amplitude rapid
fluctuation with a slowly varying envelope. That is, the wave packet
has slowly varying amplitude, frequency and wavenumber, but it has
rapidly varying phase. The rapid phase sets the fast time scale over
which one averages to obtain the equations for the wave, mean-flow
interaction. Averaging in the Lagrangian over the fast phase of the
wave in the WMF decomposition introduces a phase symmetry that leads
via Noether's theorem to conservation of the adiabatic invariant
called wave action density. The WMF approach for waves in fluids has
its counterpart in the eikonal approximation for the physics of waves
propagating in slowly varying media. As in the eikonal approximation,
averaging over rapid phases in the WMFI problem leads to modulation
equations. These WMFI modulation equations describe the fluid
interacting with wave packets in the WKB geometrical ray optics
approximation at leading order, with small corrections for
dispersion. 

The standard WMFI equations are a special case of the GLM equations
of  Andrews \& McIntyre [1978a,b] in which the fluctuating
displacement of the Lagrangian trajectory is given as a WKB wave
packet. The GLM equations give an exact theory of nonlinear waves on a
Lagrangian-mean flow for an arbitrary, but prescribed, representation
of the fluctuations. See Gjaja \&  Holm [1996] for a detailed
description of the various levels of approximation in the WMFI theory
in terms of their variational principles. This reference also
discusses the relation of the WMFI theory to the GLM equations,
expressed in terms of the Lagrangian-mean of Hamilton's principle for
incompressible rotating stratified fluids. For convenience, these
levels of approximation in WMFI theory are also listed briefly in
Appendix \#3 (section
\ref{WMFI-appendix}).

\paragraph{Generalized Lagrangian mean (GLM).}

The GLM equations of Andrews \& McIntyre [1978a] systematize the
approach to Lagrangian fluid modeling by introducing a slow + fast
decomposition of the Lagrangian particle trajectory in general form,
then relating the Lagrangian mean of a fluid quantity at the mean
particle position to its Eulerian mean, evaluated at the displaced
fluctuating position. The GLM equations are expressed directly in the
Eulerian representation. The Lagrangian mean has the advantage of 
preserving the fundamental transport structure of fluid dynamics.
For example, the Lagrangian mean commutes with the scalar advection
operator and it preserves the Kelvin circulation property of the
fluid motion equation.

\paragraph{Compatibility of averaging and reduction of
Lagrangians.}

In making slow + fast decompositions and constructing averaged
Lagrangians for fluid dynamics, care must generally be taken to see
that the averaging and reduction procedures do not interfere with
each other. (Reduction in this context refers to symmetry reduction
of the action principle by the subgroup of the diffeomorphisms that
takes the Lagrangian representation to the Eulerian representation of
the flow field.)  

This compatibility requirement is handled  automatically in the GLM
approach. The Lagrangian mean of the action principle for fluids does
not interfere with its reduction to the Eulerian representation,
since the averaging process is performed at {\it fixed Lagrangian
coordinate}. After the Lagrangian mean is taken, the remaining
particle relabeling symmetry in the action principle may be
modded-out by reducing with respect to {\it another} diffeomorphism
subgroup; namely, the subgroup that leaves invariant the quantities
in the Eulerian specification of the {\it averaged} fluid flow. Thus,
the process of taking the Lagrangian mean is compatible with
reduction by the particle-relabeling group. 

We shall perform this reduction of the action principle and
thereby place the GLM equations into the EP framework. In doing this,
we shall demonstrate the {\bf variational reduction property of the
Lagrangian mean}. This is encapsulated in the 
\begin{quote}
{\bf EP Averaging Lemma.} {\it GLM averaging preserves the
Euler-Poincar\'e (EP) variational framework that implies the GLM
fluid equations.}
\end{quote}
Thus, the Lagrangian mean's preservation of the
fundamental transport structure of fluid dynamics also extends to
preserving the EP variational structure of these equations. Of
course, this extension is not possible with the Eulerian mean,
because the Eulerian mean does {\it not} preserve the transport
structure of Eulerian fluid mechanics.

\paragraph{Approach and main results.}

This paper begins by placing the exact nonlinear GLM equations
for a compressible adiabatic fluid into the
Euler-Poincar\'e (EP) variational framework of fluid dynamics for the
corresponding GLM averaged Lagrangian. This result demonstrates the
EP Averaging Lemma, that GLM averaged equations follow from GLM
averaged Lagrangians in the EP framework. Thus, the EP Averaging Lemma  puts the GLM averaged-Lagrangian approach and the method of  
GLM-averaged equations onto equal footing.%
\footnote{ This is quite a bonus for {\it both} approaches to
modeling fluids. The averaged-Lagrangian theory produces dynamics
that can be verified directly by averaging the original equations,
and the GLM-averaged equations inherit the conservation laws that are
available from the symmetries of the Lagrangian.}
We also compare the GLM result with WMFI equations in
the EP framework. 

Staying in the EP framework, we shall derive a set of approximate
small amplitude GLM equations ($g\ell{m}$ equations) by expanding the
GLM Hamilton's principle to second order in the fluctuating
displacement of a Lagrangian trajectory from its mean position. The
resulting $g\ell{m}$ equations possess linear and nonlinear terms that
express back-reaction effects on the mean motion due to the
fluctuating displacements of the Lagrangian trajectories. The
coefficients in these back-reaction terms involve {\it Eulerian mean}
second moments of the fluctuating Lagrangian displacements. These
back-reaction terms appear both in the mean stress tensor and in the
total mean momentum. Thus, these terms have both linear and nonlinear
effects.

The EP derivation of the $g\ell{m}$ equations uses the linearized
relations between Eulerian and Lagrangian fluctuctions, in the
tradition of Lagrangian stability analysis for fluids. The
$g\ell{m}$ derivation also uses the method of averaged
Lagrangians, in the tradition of wave, mean flow interaction (WMFI).
We compare the new $g\ell{m}$ EP motion equations for
compressible and incompressible ideal fluids with the Euler-alpha
turbulence closure equations. We discuss closure hypotheses based on
the linearized fluctuation relations that relate the $g\ell{m}$
equations to the Euler-alpha equations. 

Thus, the new $g\ell{m}$ equations result as a bridge between the
GLM equations and the Euler-alpha closure equations. This bridge is
constructed by making a small-amplitude approximation of the GLM
equations in the EP variational framework. In this framework, we also
derive another variant of the Euler-alpha model that includes the
effects of {\bf compressibility}, or {\bf magnetic fields}. 

\paragraph{Compressible $\overline{g\ell{m}}$ closure equations.}
The equations we propose for the compressible $\overline{g\ell{m}}$
closure model for the Eulerian mean fluid velocity $\bar\mathbf{u}$
consist of the EP motion equation and two auxiliary equations. The EP
motion equation is given by\\
\boxeq3{
\begin{eqnarray}\label{glm-close-mot-eqn-intro}
(\partial_t+\bar\mathbf{u}\cdot\nabla)
\big(
\bar\mathbf{u}
-
\bar\mathbf{p}
\big)
+
\big(
\bar{u}_k
-
\bar{p}_k
\big)
\nabla
\bar{u}^k
\!\!&-&\!\!
\nabla\,\bar\Pi^{g\ell{m}}
\nonumber\\
+\
\tfrac{1}{\bar{D}}
\Gamma_{k\,l}\nabla\overline{\xi^k\xi^l}
\!\!&=&\!\!
\tfrac{1}{\bar{D}}
\big(\,\nabla\!\cdot\nu\,(\bar{D}-\hat{\cal O})
\nabla\big)\bar\mathbf{u}
\,.
\nonumber
\end{eqnarray}
}
Here the pseudomomentum density $\bar{D}\bar\mathbf{p}$ and the
symmetric operator $\hat{\cal O}$ are defined by
\begin{equation}\label{glm-pseudomom-intro}
\bar{D}\bar\mathbf{p}
=
\hat{\cal O}\bar\mathbf{u}
=
\tfrac{1}{2}\, 
\big[
\hat\Delta(\bar{D}\bar\mathbf{u})
+
\bar{D}\hat\Delta\bar\mathbf{u}
+
\bar\mathbf{u}\,\hat\Delta\bar{D}
\big]
\,,
\nonumber
\end{equation}
where 
$\hat\Delta = \partial_{\,l}\,\overline{\xi^k\xi^l}\,\partial_k$
is a generalized Laplacian operator.

The Eulerian mean density $\bar{D}$ and statistical moments
$\overline{\xi^k\xi^l}$ of the Lagrangian fluctuating particle 
displacements satisfy two auxiliary equations,
\begin{equation}\label{cont+xixi-eqn-intro}
\partial_t\,\bar{D}
+
{\rm div}\bar{D}\bar\mathbf{u}
=
0
\quad\hbox{and}\quad
(\partial_t+\bar\mathbf{u}\cdot\nabla)
\,\overline{\xi^k\xi^l}
=
0
\,.
\nonumber
\end{equation}
The boundary conditions in the absence of kinematic shear viscosity
$\nu$ are
\begin{equation}\label{bc-intro}
\hat\mathbf{n}\cdot\bar\mathbf{u}
=
0
\quad\hbox{and}\quad
\hat\mathbf{n}\cdot\overline{\xi\xi}
=
0
\quad\hbox{on the boundary.}
\nonumber
\end{equation}
Here $\xi(\mathbf{x},t)$ is the fluctuating displacement of a
Lagrangian trajectory away from its mean position, $\mathbf{x}$.
These boundary conditions require the mean velocity and the
Lagrangian fluctuating displacement both to be tangential at the
boundary. (An additional boundary condition is needed when kinematic
shear viscosity is present.) 

\paragraph{Lagrangian averaged Navier-Stokes$-\alpha$ (LANS$-\alpha$)
models.}  The barotropic {\it compressible} LANS$-\alpha$
model results from these $\overline{g\ell{m}}$ equations when the
preserved initial condition 
$\overline{\xi^k\xi^l}=\alpha^2\delta^{kl}$ is chosen and the
{\bf Lagrangian correlation length scale} $\alpha$ is taken to be a
constant, by choosing it to be so initially. For a constant $\alpha$,
the generalized Laplacian $\hat\Delta$ in the operator
$\hat{\cal O}$ reduces to $\hat\Delta\to\alpha^2\Delta$,
where $\Delta$ is the ordinary Laplacian. 

The {\it incompressible} LANS$-\alpha$ model results when,
in addition, the relation $\bar{D}=1$ is also enforced in these
equations through the pressure constraint. In this case, the operator
$\bar{D}-\hat{\cal O}$ reduces to the Helmholtz operator,
$1-\alpha^2\Delta$. The incompressible LANS$-\alpha$ model
using this operator has recently been proposed and tested as a
{\bf turbulence closure model} in Chen et al. [1998], [1999a,b,c].
See Foias, Holm \& Titi [2001a,b] for reviews of the mathematical
and physical properties of the incompressible LANS$-\alpha$
model. See also Marsden and Shkoller [2001] and references therein 
for additional analysis and discussions, as well as alternative
derivations and interpretations of this model.

\paragraph{Outline of the paper.} 

Section \ref{GLM-rev} introduces the
information from the Generalized Lagrangian Mean (GLM) theory that we
will need in the remainder of the paper. Section \ref{EPform-of-GLM}
begins by placing the GLM equations for a rotating adiabatic
compressible fluid into the Euler-Poincar\'e (EP) variational
framework of fluid dynamics in the Eulerian description. This is
first done explicitly by re-deriving the GLM equations using
Hamilton's principle with a GLM averaged Lagrangian. We then explain
that every ideal GLM continuum equation follows from a GLM averaged
variational principle, via the EP Averaging Lemma.

After re-framing GLM theory as an EP variational problem, we identify
the parallels and similarities between the GLM and WMFI theories in
the EP framework. We pause to correct an omission in the Andrews \&
McIntyre [1978a] GLM theory of stratified Boussinesq fluids that
restores the Kelvin circulation theorem for these equations and its
implication of local conservation for potential vorticity. We then
move on to develop and discuss the $g\ell{m}$ small-amplitude
approximation of GLM that also possesses this fundamental property. 

In section \ref{Lin-fluct-rel}, we review the standard linearized
Eulerian/Lagrangian fluctuation relations. This is done in
preparation for section \ref{glm-der} in which we construct a
small-amplitude approximation of the GLM equations for a compressible
adiabatic fluid at second order in the fluctuating displacement of a
Lagrangian trajectory from its mean position. Substituting the {\it
linear} fluctuation relations into the GLM action principle in the EP
framework turns out to have both linear and nonlinear effects on
the resulting EP equations. Another characteristic feature of these
small amplitude GLM equations ($g\ell{m}$ equations) is that they
involve second-gradients of Eulerian mean flow quantities, in
combination with quadratic moments of the  Lagrangian displacement
statistics. The latter must be modeled in closing the system.

In section \ref{alpha} we introduce several additional modeling
decisions to arrive at second-order closures for the $g\ell{m}$
equations. These modeling decisions are formulated as variants of
the Taylor-hypothesis for frozen-in turbulence. The resulting
Taylor-hypothesis closure (THC) models recover the recently
discovered Euler-alpha equations for incompressible ideal
fluids. The THC models also provide a systematic basis for extending
the Euler-alpha equations to the {\bf compressible case}. Thus, the
Euler-alpha equations reappear in a more general  context than in
their original derivation in Holm, Marsden \& Ratiu [1998a,b]. Section
\ref{conclusions} summarizes these Taylor-hypothesis closure results,
provides additional discussion and makes suggestions for diagnosing
direct numerical simulations for the purpose of determining the
accuracy of these various approximate models. 

Four appendices collect additional related material. Appendix
\#1  (section \ref{EP-appendix}) states the two main EP
theorems of Holm, Marsden \& Ratiu [1998a] for fluids with advected
properties. Appendix \#2 (section \ref{GEOM-appendix}) collects the
linearized relations between the Eulerian and Lagrangian fluctuating
fluid quantities and expresses them in a convenient geometrical
(Lie-algebraic) form. Appendix \#3 (section \ref{WMFI-appendix}) lists
the Lagrangian-mean Hamilton's principles for several levels of
approximation in WMFI theory. (The GLM theory appears on this
list of WMFI approximations at a certain level, because the
{\it self-consistent} WMFI theory is more general than the GLM theory
with {\it prescribed} fluctuation properties.) Appendix \#4 (section
\ref{GLM-MHD-appendix}) highlights some of the results in the rest of
the paper by setting up the corresponding results for the GLM,
$g\ell{m}$, and $\alpha-$models of ideal MHD (MagnetoHydroDynamics).


\subsection{Eulerian and Lagrangian means}

In fluids, averages over the fast variables in a slow + fast
decomposition may be taken in either an Eulerian sense, or in a
Lagrangian sense. The traditional Reynolds turbulence decomposition,
for example, is taken in an Eulerian sense. In the Reynolds
decomposition, the fluid velocity $\mathbf{u}$ at a given position,
$\mathbf{x}$, is decomposed as 
$\mathbf{u}=\bar\mathbf{u}+\mathbf{u}^{\,\prime}$ 
where $\overline{\mathbf{u}^{\,\prime}}=0$ and
overbar (as in $\bar\mathbf{u}$) denotes the Eulerian
mean. The Eulerian mean commutes with space and time
partial derivatives, but it has the {\bf disadvantage} of not
commuting with the advection operator
$D/Dt=\partial_t+\mathbf{u}\cdot\nabla$, that appears thoughout the
fluid equations. This disadvantage leads to the well-known closure
problem -- that the Eulerian mean fluid equations in general do not
form a closed system. For example, the Eulerian mean equation for
advection of a scalar fluid quantity
$\chi$ gives
\begin{eqnarray}
\overline{\Big(\frac{D\chi}{Dt}\Big)}
=
\frac{D^E\bar{\chi}}{Dt}
\
+\
\underbrace{\,
\overline{\mathbf{u}^{\,\prime\,}\cdot\nabla\chi^{\,\prime}}
\,}_{\hbox{Extra}}
\,,
\quad\hbox{with}\quad
\frac{D^E}{Dt}
=
\frac{\partial}{\partial t}
+
\bar\mathbf{u}\cdot\nabla
\,,
\nonumber
\end{eqnarray}
in which the extra term must be modeled to obtain closure.

The Lagrangian mean -- denoted as $\bar{\chi}^L(\mathbf{x},t)$
for a fluid quantity $\chi$ -- is taken at constant Lagrangian
coordinate, $\mathbf{x}_0$. Consequently, by its definition the
Lagrangian mean commutes with the advective derivative, written in
the Lagrangian picture as
\[
\overline{\big(\partial_t\big|_{\mathbf{x}_0}\chi\big)}^L
=
\partial_t\big|_{\mathbf{x}_0}\bar{\chi}^L
\,.\]
Written in the Eulerian picture, this is
\[
\overline{\Big(\frac{D\chi}{Dt}\Big)}^L
=
\frac{\partial\bar{\chi}^L}{\partial t}
+
\bar\mathbf{u}^L\cdot\nabla\bar{\chi}^L
\equiv
\frac{D^L\bar{\chi}^L}{Dt}
\,,\]
where $\bar\mathbf{u}^L$ is the Lagrangian mean velocity, i.e., 
$\bar\mathbf{u}^L={D^L \mathbf{x}}/{Dt}$ is the tangent vector
along the mean Lagrangian trajectory. 

\paragraph{Defining relation for Lagrangian mean.}
The Lagrangian mean for a fluid quantity $\chi$ is defined in terms of
the Eulerian mean operation as 
\begin{equation}\label{LM-def-rel}
\bar\chi^L(\mathbf{x},t)
=
\overline{\chi \big(\mathbf{x} + \xi(\mathbf{x},t),t\big)}
\,,\nonumber
\end{equation}
where $\overline{(\,\cdot\,)}$ denotes the Eulerian mean and 
$\mathbf{x} + \xi(\mathbf{x},t)$ is the current
position of a fluid parcel whose mean position is
$\mathbf{x}$. See Andrews \& McIntyre [1978a,b] for discussions of
the many implications of this defining relation. For example, one sees
immediately from this defining relation that, relative to the Eulerian
mean, the Lagrangian mean has {\bf two disadvantages}: it is history
dependent and it does not commute with the spatial gradient.

Thus, the Eulerian mean commutes with gradients, but it interferes
with the advection operator and fails to produce a closed system of
equations. In contrast, the Lagrangian mean commutes with the
advection operator and produces the GLM equations. The GLM equations,
however, are also not closed. For closure, they require the
statistical properties of the Lagrangian disturbance $\xi$ to be
prescribed.

\paragraph{Levels of approximation.}
In principle, the GLM theory is more accurate than an Eulerian-mean
theory, because it exactly preserves scalar advection relations and,
thus, yields an exact nonlinear theory. As we shall see in section
\ref{EPform-of-GLM}, the GLM theory also preserves the
EP mathematical structure of the original unapproximated equations --
including the Kelvin circulation theorem and the balance laws for
energy and momentum. However, the results of any Lagrangian mean
theory may be difficult to interpret accurately in an Eulerian
setting. 

For its completion, the GLM theory still requires {\bf closure
assumptions} about the Lagrangian-mean statistics of its
wave properties -- the Lagrangian fluctuating displacements, $\xi$.
For sufficiently small disturbances, plausible assumptions
based on WMFI theory may assist in this closure. However, at finite 
disturbance amplitude, the nonlinear effects of the mean motion on
the statistics of the Lagrangian fluctuating displacements must
become part of a {\it dynamically self-consistent} solution, and
these statistics can no longer be taken as prescribed quantities.
See Gjaja and Holm [1996] for initial investigations of a
dynamically self-consistent WMFI theory at the level of the WKB
envelope approximation.

Therefore, we see the need for developing an approximate small
amplitude counterpart to the GLM theory, in preparation for
developing a set of dynamically self-consistent closure assumptions.
For this, we shall begin in section \ref{EPform-of-GLM} by placing
the the GLM equations themselves into the larger framework of EP
variational principles.

In section \ref{Lin-fluct-rel} we shall recall the linearized
Eulerian/Lagrangian fluctuation relations that are familiar
from traditional Lagrangian stability analysis for fluids.
In section \ref{glm-der}, we shall apply these linearized fluctuation
relations within the EP framework to develop a
{\bf small-amplitude approximation} that minimizes the disadvantages
of interpreting the GLM description in an Eulerian setting. Namely, we
shall assume that the fluctuating displacement of the fluid
trajectory around its mean, $\xi(\mathbf{x},t)$, is sufficiently small
and also that the mean flow is sufficiently smooth, that we may
linearize the relations for  Eulerian fluctuating quantities in terms
of the Lagrangian displacement fluctuation, $\xi$. This is standard
practice in traditional Lagrangian stability theory. These standard
linearized relations will  allow us to identify which moments of the
Lagrangian statistics must be modeled in making an approximate {\bf
second-order Eulerian-mean closure} of the GLM equations of motion at
order $O(|\xi|^2)$. These second-order small-amplitude GLM
equations (denoted $g\ell{m}$ equations) are derived by applying the
EP framework to the Eulerian mean of the second-variation Lagrangian
for GLM. 

In section \ref{alpha}, we shall seek {\bf variational closures} of
the $g\ell{m}$ equations. The closed $\overline{g\ell{m}}$
equations result, upon introducing an additional {\bf Taylor
hypothesis} into the derivation of $g\ell{m}$ theory. The $g\ell{m}$
theory is derived in section \ref{glm-der} by making approximations,
averaging and taking variations of the GLM Lagrangian. The family of
closed $\overline{g\ell{m}}$ equations will be derived in the same EP
framework. However, we shall first introduce particular (or
partial) solutions of the linearized fluctuation relations into the
derivation of $g\ell{m}$ theory, before averaging and taking
variations in the EP variational principle. The closed family of 
$\overline{g\ell{m}}$ equations includes the {\bf Euler-alpha model}
for incompressible ideal fluids discovered in Holm, Marsden
\& Ratiu [1998a,b], as well as recent extensions of it derived in
Marsden \& Shkoller [2001]. 

Finally, the $\overline{g\ell{m}}$ theory given here also extends
these Euler-alpha equations to include the effects of 
{\bf compressibility} and {\bf magnetic fields}.


\section{Brief review of GLM theory for compressible fluids}
\label{GLM-rev}

An exceptional accomplishment in formulating averaged motion equations
for fluid dynamics is the Generalized Lagrangian Mean (GLM) theory of
nonlinear waves on a Lagrangian-mean flow, as explained in two
consecutive papers of Andrews \& McIntyre [1978a,b]. This
section introduces what we shall need later from the rather complete
description given in these papers. (Even now, these fundamental
papers still make worthwhile reading and are taught in many
atmospheric science departments.) In section \ref{EPform-of-GLM} we
shall place the GLM equations into the Euler-Poincar\'e
variational framework. Then in section \ref{glm-der}, we shall derive
the second-order approximate $g\ell{m}$ equations in this variational
framework by making a small-amplitude approximation.


\subsection{Relevant information from the GLM theory}


\subsubsection*{Defining relations for Lagrangian mean
\& Stokes correction in terms of Eulerian mean}

The GLM equations are based on defining fluid quantities at a
displaced fluctuating position. In the GLM description, $\bar{\chi}$
denotes the Eulerian mean of a fluid quantity
$\chi=\bar\chi+\chi^{\,\prime}$ while $\bar{\chi}^L$ denotes the
Lagrangian mean of the same quantity, defined by
\[
\bar{\chi}^L(\mathbf{x})
\equiv
\overline{\chi^\xi(\mathbf{x})}
\,,\quad\hbox{with}\quad
\chi^\xi(\mathbf{x})
\equiv
\chi(\mathbf{x}+\xi(\mathbf{x},t))
\,.
\]
Here $\mathbf{x}^\xi\equiv\mathbf{x}+\xi(\mathbf{x},t)$ is the current
position of a Lagrangian fluid trajectory whose mean position is 
$\mathbf{x}$. Thus, $\xi(\mathbf{x},t)$ with vanishing Eulerian 
mean $\bar\xi=0$ denotes the fluctuating displacement of a Lagrangian
particle trajectory about its mean position $\mathbf{x}$. 

\begin{quote}
{\bf Remark.}
Fortunately, this notation is also {\it standard} in the
stability analysis of fluid equilibria in the Lagrangian picture.
See, e.g., the classic works of Bernstein et al. [1958], Frieman \&
Rotenberg [1960] and Newcomb [1962]. See Jeffrey \& Taniuti [1966]
for a collection of reprints showing applications of this approach
in controlled thermonuclear fusion research. For insightful reviews,
see Bernstein [1983], Chandrasekhar [1987] and, more recently, Hameiri
[1998]. Rather than causing confusion, this confluence of notation
encourages the transfer of ideas between traditional Lagrangian
stability analysis for fluids and GLM theory, as we shall see in 
section
\ref{glm-der}.
\end{quote}

In GLM theory, the difference  $\chi^\xi-\bar{\chi}^L=\chi^\ell$ is
called the {\bf Lagrangian disturbance} of the quantity $\chi$. One
finds $\overline{\chi^\ell}=0$, since the Eulerian mean possesses the
{\bf projection property} $\bar{\bar{\chi}}=\bar{\chi}$ for any
quantity $\chi$ (and, in particular, it possesses that property
for $\chi^\xi$).%
\footnote{Note that spatial filtering in general does {\it not}
possess the projection property.}
Andrews \& McIntyre [1978a] show that, provided the
smooth map $\mathbf{x}\to\mathbf{x}+\xi(\mathbf{x},t)$ is invertible
(that is, provided the vector field $\xi(\mathbf{x},t)$ generates a
diffeomorphism), then the Lagrangian disturbance velocity
$\mathbf{u}^\ell$ may be expressed in terms of $\xi$ by
\begin{equation}\label{u-ell-def}
\mathbf{u}^\ell
=
\mathbf{u}^\xi
-
\bar\mathbf{u}^L
=
\frac{D^L\xi}{Dt}
\,,\quad\hbox{where}\quad
\frac{D^L\xi}{Dt}
\equiv
\frac{\partial\xi}{\partial t}
+
\bar\mathbf{u}^L\cdot\nabla\xi
\,.\nonumber
\end{equation}
Consequently, the Lagrangian disturbance velocity $\mathbf{u}^\ell$ is
a genuine fluctuation quantity satisfying
$\overline{\mathbf{u}^\ell}=0$, since
$\overline{\mathbf{u}^\xi} - \overline{\bar\mathbf{u}^L} 
= \overline{\mathbf{u}^\xi} - \overline{\overline{\mathbf{u}^\xi}}
=0$, by the projection property. (Alternatively,
$\overline{\mathbf{u}^\ell}=\overline{D^L\xi/Dt}=0$ also follows,
since the Eulerian mean commutes with ${D^L}/{Dt}$ and $\xi$ has
mean zero.)

The difference between the Eulerian and Lagrangian means is called 
the {\bf Stokes correction}, e.g.,
\[
\bar{\chi}^S(\mathbf{x})
=
\bar{\chi}^L(\mathbf{x})
-
\bar{\chi}(\mathbf{x})
\,.
\]
In a Taylor series approximation, one finds
\[
\bar{\chi}^S 
=
\overline{\xi\cdot\nabla\chi^{\,\prime}}
+
\tfrac{1}{2}\overline{\xi\xi}:\nabla\nabla\bar\chi
+
O(|\xi|^3)
\,.
\]
The order $O(\overline{\xi\xi})$ terms in $\bar\chi^S$ may be
neglected, provided the second gradients of the mean
$\nabla\nabla\bar\chi$ are sufficiently small, as we shall assume
henceforth.


\subsection{GLM preserves advective transport relations, modulo their
tensor transformation factors}\label{GLM-advection-sec}


\subsection*{GLM scalar advection relations}

At position $\mathbf{x}$ the velocity $\mathbf{u}^\xi =
\bar\mathbf{u}^L + \mathbf{u}^\ell$ is the sum of the
Lagrangian mean velocity $\bar\mathbf{u}^L$ and the Lagrangian
disturbance velocity $\mathbf{u}^\ell$. Thus, 
$\mathbf{u}^\xi=D^L\mathbf{x}^\xi/Dt$ and for any scalar field
$\chi(\mathbf{x},t)$ one has, 
\begin{equation}
\Big(\frac{D\chi}{Dt}\Big)^\xi
=
\frac{D^L}{Dt}(\chi^\xi)
\,.\nonumber
\end{equation}
Because $\bar\mathbf{u}^L$ appearing in the advection operator 
${D^L}/{Dt}=\partial_t+\bar\mathbf{u}^L\cdot\nabla$ is a mean
quantity, one then finds, as expected, that the Lagrangian mean
$\overline{(\,\cdot\,)}^L$ commutes with the material derivative
$D/Dt$. That is,
\begin{equation}
\overline{\Big(\frac{D\chi}{Dt}\Big)}^L
=
\frac{D^L}{Dt}(\bar{\chi}^L)
\,,\quad\hbox{and}\quad
\Big(\frac{D\chi}{Dt}\Big)^\ell
=
\frac{D^L}{Dt}\chi^\ell
\,,\nonumber
\end{equation}
where $\chi^\ell=\chi^\xi-\bar{\chi}^L$ is the Lagrangian
disturbance of $\chi$ satisfying $\overline{\chi^\ell}=0$. For
example, in an adiabatic compressible flow, the specific entropy $s$
is advected as a scalar. That is, it satisfies $Ds/Dt=0$ and,
consequently, $D^L\bar{s}^L/Dt=0$, as well. Hence,
$s^\xi=\bar{s}^L$ follows, by integration of
$D^L(\bar{s}^L-s^\xi)/Dt=0$ along mean trajectories and invertibility
of the map $\mathbf{x}\to\mathbf{x}+\xi(\mathbf{x},t)$. 

\begin{quote}
{\bf Remark.}
Of course, this identification is also obvious
physically, since the Lagrangian mean $\bar{s}^L$ and the current
value $s^\xi$ refer to the {\it same} Lagrangian fluid label. 
That is, we initialize with $\xi(\mathbf{x}_0,0)=0$, for a Lagrangian
coordinate $\mathbf{x}_0=\mathbf{x}(\mathbf{x}_0,0)$.
(This initialization still allows for random initial velocities.)
\end{quote}


\subsection*{Mass conservation: the GLM continuity equation}

Remarkably, $\bar{D}^L$ is not the density advected in the GLM theory.
That is,
\begin{equation}
\partial_t\,\bar{D}^L
+
{\rm div}\bar{D}^L\bar\mathbf{u}^L
\ne
0
\,.
\nonumber
\end{equation}
Instead, GLM satisfies another density advection relation --
the {\bf GLM continuity equation},
\begin{equation}\label{GLM-cont-eqn}
\partial_t\,\tilde{D}
+
{\rm div}\tilde{D}\bar\mathbf{u}^L
=
0
\,,
\end{equation}
for a density $\tilde{D}$, which is also a mean quantity.
That is, $\bar{\tilde{D}}=\tilde{D}$, where we invoke again the
projection property of the Eulerian mean. The GLM conserved density
$\tilde{D}$ is given by
\boxeq2{
\begin{equation}
\tilde{D}=
D^\xi{\cal J}
\,,\quad\hbox{where}\quad
{\cal J}
=
\det\big(\nabla_\mathbf{x}(\mathbf{x}+\xi)\big)
\,.
\end{equation}
}

\noindent
The GLM continuity equation for the density $\tilde{D}$ may be shown
by transforming the instantaneous mass conservation relation
$D^\xi\,d^3x^\xi = D(x_0)d^3x_0$
into 
\[
D^\xi{\cal J}
\equiv
D^\xi(x)\det(\nabla_\mathbf{x}(\mathbf{x}+\xi))
= 
D(x_0)d^3x_0/d^3x 
\equiv \tilde{D}
\] 
and then using the defining relation (\ref{LM-def-rel}) for the
Lagrangian mean in terms of the Eulerian mean. In taking the Eulerian
mean of this relation, we keep in mind that $\mathbf{x}$ is an
average quantity, so the right hand side is {\it already} an average
quantity. Thus, $\tilde{D}=D^\xi{\cal J}$ satisfies
$\bar{\tilde{D}}=\tilde{D}$, as claimed, and we note that
$\tilde{D}\ne\bar{D}^L$, in general. The mean mass conservation
relation $\tilde{D}d^3x=D(x_0)d^3x_0$ then implies the continuity
equation for $\tilde{D}$,
\begin{equation}
\partial_t\,\tilde{D}
+
{\rm div}\tilde{D}\bar\mathbf{u}^L
=
0
\,,
\nonumber
\end{equation}
upon recalling that $\bar\mathbf{u}^L$ is the
velocity tangent to the mean Lagrangian position $\mathbf{x}$.

\paragraph{Remarks.}
\begin{description}
\item$\quad\bullet$
The advective transport result for the density $\tilde{D}$ is
especially clear when expressed in Lie derivative form as,
\begin{eqnarray}\label{GLM-Lie-der-density}
0
&=&
\frac{d}{d t}
\Big[
\Big(D(x_0)d^3x_0\Big)\cdot g^{-1}(t)
\Big]
\nonumber\\
&=&
\Big(\frac{\partial}{\partial t}+\pounds_{\bar\mathbf{u}^L}\Big)
(\tilde{D}d^3x)
=
\Big(\partial_t\,\tilde{D}
+
{\rm div}\tilde{D}\bar\mathbf{u}^L\Big)
d^3x
\,,\nonumber
\end{eqnarray}
where $\cdot g^{-1}(t)$ denotes the right action of the group $G$ and
$\pounds_{\bar\mathbf{u}^L}$ denotes the Lie derivative with respect
to $\bar\mathbf{u}^L$ (in the notation explained in Appendices \#1
and \# 2, sections \ref{EP-appendix} and \ref{GEOM-appendix}).
\item$\quad\bullet$
For a fluid with {\bf constant unit density}, $D^\xi=1$,
the GLM theory gives
\begin{equation}\label{GLM-density}
\tilde{D}
=
\overline{ D^\xi{\cal J} }
=
\overline{ \det\big(\nabla_\mathbf{x}(\mathbf{x}+\xi)\big) }
=
1 - \tfrac{1}{2}
\big(\,\overline{ \xi^k\xi^\ell}\,\big)_{\!\!,\,k\ell}
+
O(|\xi|^3)
\,.\nonumber
\end{equation}
Hence, for constant instaneous density, the Lagrangian mean velocity
$\bar\mathbf{u}^L$ has an order $O(|\xi|^2)$ divergence, 
\[
{\rm div}\,\bar\mathbf{u}^L
=
-\,\frac{1}{\tilde{D}}
\frac{D^L\tilde{D}}{Dt}
=
\frac{1}{2}\frac{D^L}{Dt}
\big(\,\overline{ \xi^k\xi^\ell}\,\big)_{\!\!,\,k\ell}
+
O(|\xi|^3)
\,,\]
as shown in Andrews \& McIntyre [1978a].

\end{description}


\subsection*{Magnetic flux advection: the GLM frozen-in magnetic
field}

A preserved mean magnetic flux equation is provided as a basic mean
advection relation in developing a GLM theory of ideal MHD.

The magnetic flux advection relation for a magnetic field in ideal
MHD (MagnetoHydroDynamics) is 
\begin{equation}\label{mag-advect}
\partial_t\, \mathbf{B}^\xi
=
 \big({\rm curl}\,(\mathbf{u}\times\mathbf{B})\big)^\xi
\quad\hbox{with}\quad
({\rm div}\,\mathbf{B})^\xi = 0
\,.\nonumber
\end{equation}
The corresponding GLM advection law may be obtained by averaging the
Cauchy solution for this relation
\begin{equation}\label{mag-Cauchy}
\frac{\mathbf{B}^\xi}{D^\xi}
\cdot
\frac{\partial{x}^i}{\partial\mathbf{x}^\xi}
=
\frac{\mathbf{B}(\mathbf{x}_0)}{D(\mathbf{x}_0)}
\cdot
\frac{\partial{x}^i}{\partial\mathbf{x}_0}
\,,\nonumber
\end{equation}
in which the initial condition $\mathbf{B}(\mathbf{x}_0)$ has zero
divergence. Upon taking the mean of this relation and using
$\tilde{D} = D^\xi{\cal J}$ for the mean GLM density, we find that
the divergenceless vector
$\tilde\mathbf{B}$ whose components are
\begin{equation}\label{mag-XXX}
\tilde{B}^i = K^i_j \, {B}^{\xi\,j}
\quad\hbox{with}\quad
K^i_j
=
{\cal J}\,\frac{\partial{x}^i}{\partial{x}^{\xi\,j}}
\,,\quad
i,j=1,2,3
\,,
\nonumber
\end{equation}
is also a mean quantity. That is, 
$\bar{\tilde\mathbf{B}}=\tilde\mathbf{B}$
and we note that
$\tilde\mathbf{B}\ne\bar\mathbf{B}^L$. 
Here $K^i_j$ satisfying
$
K^i_j\,(\partial{x}^{\xi\,j}/\partial{x}^k)
=
{\cal J}\delta_k^i
$
is the matrix $\mathsf{K}$ of cofactors of the Jacobian
$\mathsf{J}\equiv\partial\mathbf{x}^\xi/\partial\mathbf{x}$. The mean
vector quantity
$\tilde\mathbf{B}=\mathsf{K}\!\cdot\!\mathbf{B}^\xi$ then
satisfies the magnetic flux advection relation,
\begin{equation}\label{GLM-mag-advect}
\partial_t\, \tilde\mathbf{B}
=
{\rm curl}\,
(\bar\mathbf{u}^L\times\tilde\mathbf{B})
\quad\hbox{with}\quad
{\rm div}\,\tilde\mathbf{B} = 0
\,.\nonumber
\end{equation}
That is, the flux of the divergenceless mean quantity
$\tilde\mathbf{B}$ is ``frozen'' into the Lagrangian mean motion of
the fluid.

A fully three dimensional nonlinear GLM theory of MHD is
still under development. See, however, Grappin, E. Cavillier \& M.
Velli [1997] for a recent account, based on the WKB approximation, of
the propagation of acoustic waves incident on the base of a stellar
wind and their back-reaction on the mean flow, in the spherically
symmetric, isothermal case.


\subsection*{Advection of a covariant symmetric tensor}

When nonlinear elasticity is also a factor in the fluid
evolution, there is an additional advection relation,
\begin{equation}
\frac{ \partial}{\partial t}\,S_{ab}^\xi=
-\left(u^k S_{ab,k}+S_{kb}u^k_{,a}+S_{ka}u^k_{,b}\right)^\xi
.
\nonumber
\end{equation}
This is the advection relation for the Cauchy-Green strain tensor
$S_{ab}(\mathbf{x}^\xi,t)$, which measures nonlinear strain in {\it
Eulerian} coordinates. The corresponding Cauchy solution of this
advection relation is expressed in a Cartesian tensor basis as
\begin{equation}
(S_{kl}\,dx^k\otimes dx^l)^\xi
=
S_{ab}(\mathbf{x}_0)\,dx_0^a\otimes dx_0^b
\,.
\nonumber
\end{equation}
The GLM advection law for such a covariant symmetric tensor may be
obtained by rearranging this Cauchy solution. Hence, the
symmetric tensor,
\begin{equation}
\tilde\mathsf{S}_{ij}
\equiv
(\mathsf{J}^T\!\cdot\!\mathsf{S}^\xi\!\cdot\!\mathsf{J})_{ij}
=
S^\xi_{kl}\,\frac{ \partial x^{\xi\,k}}{\partial x^i}
\frac{ \partial x^{\xi\,l}}{\partial x^j}
=
S_{ab}(\mathbf{x}_0)
\,
\frac{ \partial x_0^a}{\partial x^i}
\frac{ \partial x_0^b}{\partial x^j}
\,,
\nonumber
\end{equation}
is a mean quantity, where the Jacobian matrix $\mathsf{J}$ has
elements $\mathsf{J}^k_i\equiv\partial x^{\xi\,k}/\partial x^i$. The
corresponding GLM advection law for the mean Cauchy-Green strain
tensor $\tilde\mathsf{S}$ is given by
\begin{equation}
\frac{ \partial}{\partial t}\,\tilde{S}_{ab}
=
-\,
\bar{u}^{L\,k}\, \tilde{S}_{ab,k}
-
\tilde{S}_{kb}\, \bar{u}^{L\,k}_{,a}
-
\tilde{S}_{ka}\, \bar{u}^{L\,k}_{,b}
\,.
\end{equation}
Note that $\tilde\mathsf{S}=\bar{\tilde\mathsf{S}}$ is a
mean fluid quantity, but $\tilde\mathsf{S}\ne\bar\mathsf{S}^L$ because
of the additional factors involving the Jacobian $\mathsf{J}$.


\noindent
{\bf Remark.}
\begin{quote}
These examples demonstrate that the GLM theory preserves the advective
transport structure of fluid dynamics, modulo the means of their
tensor transformation factors. On a Riemannian manifold, additional
metric terms are also present in these tensor transformation factors.
See, e.g., Marsden and Shkoller [2001] and references therein.
\end{quote}


\subsection{GLM motion equations for adiabatic compressible fluids}

The GLM motion equation for adiabatic compressible fluids in a
frame rotating with constant frequency $\Omega$ are given in Andrews
\& McIntyre [1978a] in Cartesian coordinates as\\
\boxeq2{
\begin{eqnarray}\label{GLM-eqns}
\frac{D^L}{Dt}\big(
\bar\mathbf{u}^L
-
\bar\mathbf{p}
\big)
+
\big(
\bar{u}^L_k
-
\bar{p}_k
\big)
\nabla
\bar{u}^L_k
+
2\Omega\times\bar\mathbf{u}^L
+
\nabla\,\Pi
-
\bar{T}^L\nabla\bar{s}^L
=
0
\,.
\end{eqnarray}
}
Here $\bar\mathbf{p}$ is the {\bf pseudomomentum vector}, a mean
quantity defined by,
\begin{equation}\label{pseudomom-def}
\bar\mathbf{p}
\equiv
-\,\overline{[u^\ell_k+(\Omega\times\xi)_k]\nabla\xi^k}
\,.
\end{equation}
The mean potential $\Pi$ has the form,
\begin{equation}\label{Pi-def}
\Pi
=
\overline{h(p^\xi,s^\xi)}
+
\bar{\Phi}^L(\mathbf{x})
-
\frac{1}{2}
\overline{
\mathbf{u}^\xi
\cdot
[\mathbf{u}^\xi
+
2\Omega\times\xi]
}
\,,
\end{equation}
in which 
\begin{equation}
\overline{h(p^\xi,s^\xi)}
\equiv
\overline{e(D^\xi,s^\xi)}
+
\overline{(p^\xi/D^\xi)}
\nonumber
\end{equation}
is the mean specific enthalpy and $\bar{\Phi}^L$ is the
Lagrangian mean of an external potential $\Phi$. We note that
$s^\xi=\bar{s}^L$ since the specific entropy is a Lagrangian
variable in the adiabatic case. The partial derivative
$\bar{T}^L=\partial\overline{e(D^\xi,\bar{s}^L)}
/\partial\bar{s}^L$ is the Lagrangian mean temperature.

For an adiabatic compressible fluid, the thermodynamic First Law
following a fluid parcel is
\begin{equation}
de(D^\xi,\bar{s}^L)
=
-\,p^\xi \,  d\Big(\frac{1}{D^\xi}\Big)
+
T^\xi  d\bar{s}^L
\,.
\nonumber
\end{equation}
Hence, its Eulerian mean becomes, upon using 
$\tilde{D}=D^\xi{\cal J}$ from mass conservation,
\boxeq2{
\begin{equation}\label{1st-law-thermo}
d\,\overline{e(D^\xi,\bar{s}^L)}
=
-\,\frac{1}{\tilde{D}}
\,\overline{(p^\xi \, d{\cal J})}
+
\frac{1}{\tilde{D}}
\,\overline{(p^\xi/D^\xi)}
\,d\tilde{D}
+
\bar{T}^L  d\bar{s}^L
\,.
\end{equation}
}
$\!\!$

\noindent
{\bf Remarks.}
\begin{description}
\item$\quad\bullet$
The determinant ${\cal J} =
\det\big(\nabla_\mathbf{x}(\mathbf{x}+\xi)\big)$ is a 
fluctuating quantity, not a mean fluid quantity. Therefore, 
${\cal J}$ will not contribute to variations with respect to mean
fluid quantities. However, $\delta{\cal J}
= K^j_k(\partial\,\delta\xi^k/\partial x^j)$
with cofactor 
$K^j_k = {\cal J}\partial\,x^j/\partial (x^k+\xi^k)$ does contribute
to variations with respect to $\xi$ in the self-consistent WMFI
theory of Gjaja \& Holm [1996]. Such variations also arise, for
example, in the Lagrangian stability analysis of the equilibrium
solutions of the GLM  equations. See, e.g., Andrews \& McIntyre
[1978b] for a discussion of Hamilton's principle for the Lagrangian
disturbance
$\xi$ and its relation to the wave action density of the WMFI theory.

\item$\quad\bullet$
For an elastic medium, the specific internal energy $e(D,b,S_{ab})$
also depends on the Cauchy-Green strain tensor
$S_{ab}$. In this case, the stress tensor per unit mass
$\sigma^{ab}$ is determined from the equation of state by the
Doyle-Erickson formula
$\sigma^{ab}\equiv \partial e/\partial S_{ab}$. The Eulerian mean
First Law relation for the specific internal energy of such an
elastic medium then contains an additional stress term,
\begin{equation}
d\overline{e^\xi} 
=
 -\, \overline{p^\xi d(1/D^\xi)} 
+
\bar{T}^L d\bar{s}^L
+ 
\overline{(\sigma^\xi)^{ab} dS^\xi_{ab}}
\,.
\nonumber
\end{equation}
The corresponding mean stresses will be defined from the average
quantity 
\begin{equation}
\overline{(\sigma^\xi)^{ab} 
d\,
(\mathsf{J}^{-T}\!\cdot\!
\tilde\mathsf{S}
\!\cdot\!\mathsf{J}^{-1})_{ab}}
=
\overline{(\sigma^\xi)^{ab} 
(\mathsf{J}^{-T}\!\cdot\!
d\,\tilde\mathsf{S}
\!\cdot\!\mathsf{J}^{-1})_{ab}}
+\
\hbox{terms in }\xi
\,.
\nonumber
\end{equation}

\item$\quad\bullet$
Thus, the preservation of advective transport relations in the GLM
theory enables proper definitions of the {\bf thermodynamic
derivatives of mean constitutive relations} with respect to GLM
average fluid variables.

\end{description}


\subsection{Pseudomomentum \& the
transport structure of the GLM motion
equation}\label{Kel-transport-structure-sec}

The significance of the pseudomomentum $\mathbf{p}$ to the
transport structure of the GLM equations can be understood from
the Lagrangian mean of the contour integral appearing in Kelvin's
circulation theorem for fluid motion in a rotating frame. 
The rotation frequency $\Omega$ is allowed to depend on position and
is given by
$2\Omega(\mathbf{x}^\xi) = ({\rm curl}\,\mathbf{R})^\xi$. The
rotation potential $\mathbf{R}(\mathbf{x}^\xi)$ is decomposed in
standard GLM fashion as  
$\mathbf{R}^\xi=\bar\mathbf{R}^L + \mathbf{R}^\ell$. 
(A constant rotation frequency is recovered from specializing to
$\mathbf{R}^\xi=\Omega\times\mathbf{x}^\xi$.)

The GLM average of Kelvin's circulation integral is defined as,
\begin{eqnarray}
\overline{I(t) }
&=&
\overline{ \oint_{\gamma^\xi(t)} 
\big(\mathbf{u}^\xi+\mathbf{R}(\mathbf{x}^\xi)\big)
\cdot d \mathbf{x}^\xi 
}
\nonumber\\
&=& 
\overline{\oint_{\gamma^\xi(t)} 
\big(\bar\mathbf{u}^L + \bar\mathbf{R}^L
+
\mathbf{u}^\ell +\mathbf{R}^\ell\big) 
\cdot
(d\mathbf{x}+d\xi) }
\nonumber\\
&=& 
\oint_{ \bar\gamma^L(t) }
\big(\bar\mathbf{u}^L + \bar\mathbf{R}^L
+ \overline{[u^\ell_k+R^\ell_k]\nabla\xi^k}\ \big) 
\cdot  d\mathbf{x} 
\nonumber\\
&=& 
\oint_{\bar\gamma^L(t)} 
(\bar\mathbf{u}^L + \bar\mathbf{R}^L - \mathbf{p}) 
\cdot  d\mathbf{x} 
\,,\nonumber
\label{kelnt-psm}
\end{eqnarray}
where the contour $\bar\gamma^L(t)$ moves with velocity 
$\bar\mathbf{u}^L$, since it follows the fluid parcels as the average
is taken. Thus, the Lagrangian mean leaves invariant the {\it form}
of the Kelvin integral, while averaging the {\it velocity} of its
contour. In addition, the pseudomomentum vector $\mathbf{p}$ defined
in (\ref{pseudomom-def}) appears in the GLM averaged Kelvin integral
as the Lagrangian mean contribution of the fluctuations to the GLM
averaged {\it integrand}.

The time derivative of the GLM averaged Kelvin circulation
integral is,
\begin{equation}
\frac{d}{dt}\overline{I(t) }
=
\oint_{\bar\gamma^L(t)} \!\!
\Big[(\partial_t+\bar\mathbf{u}^L\cdot\nabla)
(\bar\mathbf{u}^L - \mathbf{p}) 
+
(\bar{u}^L_k - p_k) \nabla \bar{u}^{L\,k}
+
2\Omega\!\times\!\bar\mathbf{u}^L
\Big]
\!\!\cdot\!d\mathbf{x} 
\,.\nonumber
\label{kel-dot}
\end{equation}
The combination of terms in the integrand defines the {\bf transport
structure} of the GLM theory. From the GLM motion equation
(\ref{GLM-eqns}) one now finds the GLM Kelvin circulation theorem for 
adiabatic compressible flow,
\begin{equation}\label{GLM-comp-Kel-thm}
\frac{d}{dt}\overline{I(t) }
=
\frac{d}{dt}\oint_{c(\bar\mathbf{u}^L)}
\Big( \bar\mathbf{u}^L + \bar\mathbf{R}^L - \mathbf{p} 
\Big)
\cdot
d\mathbf{x}
=
\oint_{c(\bar\mathbf{u}^L)}
\bar{T}^L\,d\bar{s}^L
\,.
\nonumber
\end{equation}
Thus, the Lagrangian mean {\it averages the velocity} of the fluid
parcels on the Kelvin circulation loop, while it {\it adds the mean
contribution} of the fluctuations to the Kelvin circulation integrand.
In particular, upon taking the Lagrangian mean, the velocity of fluid
parcels on the circulation loop and the velocity appearing in the
circulation integrand are {\it different}. 

In the isentropic case (or, if the loop $c(\bar\mathbf{u}^L)$ moving
with the Lagrangian mean flow lies entirely on a level surface of
$\bar{s}^L$) then the right hand side vanishes, and one finds the
``generalized Charney-Drazin theorem'' for transient waves discussed
in Andrews \& McIntyre [1978a].


\section{EP formulation of the GLM equations using an averaged
variational principle}\label{EPform-of-GLM}

\subsection{EP Averaging Lemma for GLM equations}

Most of the important properties of the GLM equations
are discussed in Andrews \& McIntyre [1978a]. Many of these
properties arise from general mathematical structures that are shared
by all exact nonlinear ideal fluid theories. With the help of the
thermodynamic identity (\ref{1st-law-thermo}) for
$d\,\overline{e(D^\xi,\bar{s}^L)}$ we shall recast the GLM fluid
motion equation (\ref{GLM-eqns}) as an {\bf Euler-Poincar\'e (EP)
equation},
\boxeq3{
\begin{equation}\label{EP-GLM-eqn}
\frac{\partial}{\partial t}
\frac{\delta \bar\ell}{\delta \bar{u}^L_i}
+\,
\frac{\partial}{\partial x_k}
\Big(\frac{\delta \bar\ell}{\delta \bar{u}^L_i}\bar{u}^L_k\Big)
+\,
\frac{\delta \bar\ell}{\delta \bar{u}^L_k}
\frac{\partial \bar{u}^L_k}{\partial x_i}
=
\tilde{D}\frac{\partial}{\partial x_i}
\frac{\delta \bar\ell}{\delta \tilde{D}}
-
\frac{\delta \bar\ell}{\delta \bar{s}^L}
\frac{\partial \bar{s}^L}{\partial x_i}
\,,
\end{equation}
}
\medskip

\noindent
expressed in terms of variational derivatives of an averaged
Lagrangian,  $\bar\ell(\bar\mathbf{u}^L,\tilde{D},\bar{s}^L)$.  See
Holm, Marsden \& Ratiu [1998a,b] for an exposition of the
mathematical structures that arise in the EP theory of
ideal fluids that possess advected quantities such as heat and mass.
For GLM the Eulerian expression of the averaged Lagrangian is
\boxeq5{
\begin{eqnarray}\label{Lag-mean-Lag}
\bar\ell(\bar\mathbf{u}^L,\tilde{D},\bar{s}^L)
&=&
\int
d^3x\,\tilde{D}
\bigg[
\frac{1}{2}
\overline{
\Big|\bar\mathbf{u}^L
+
\frac{D^L\xi}{Dt}\Big|^2}
\,
+
\overline{
(\Omega\times\mathbf{x}^\xi)
\cdot\Big(\bar\mathbf{u}^L
+
\frac{D^L\xi}{Dt}\Big)}
\nonumber\\
&&\hspace{2cm}
-\
\overline{\Phi(\mathbf{x}^\xi)}
-\,
\overline{e(D^\xi,\bar{s}^L)}
\bigg]
\,.
\end{eqnarray}
}
\medskip

\noindent
The mean Lagrangian $\bar\ell\equiv\int\bar{\cal  L}
(\bar\mathbf{u}^L,\tilde{D},\bar{s}^L;\xi)d^3x$ is a straight
transcription of the standard Lagrangian for adiabatic fluids into
the GLM formalism, followed by taking the Eulerian mean. If desired,
the rotation frequency can be allowed to depend on position by
replacing
$(\Omega\times\mathbf{x}^\xi)
\to \mathbf{R}(\mathbf{x}^\xi)$, in which case 
$2\Omega\to({\rm curl}\,\mathbf{R})^\xi$. The variational derivatives
of $\bar\ell$ are given by
\begin{eqnarray}\label{GLM-Lag-der}
\delta\bar\ell
&=&
\int
d^3x\,\bigg[
\tilde{D}\Big( \bar\mathbf{u}^L 
- \bar\mathbf{p}
+
\Omega\times\mathbf{x}
\Big)
\cdot\delta\bar\mathbf{u}^L
-\
\tilde{D}\,\bar{T}^L\,\delta\bar{s}^L
-\,\Pi\,
\delta\tilde{D}
\\
&&
+\,\tilde{D}\,\overline{[\,u^\ell_k+(\Omega\times\xi)_k]
(\partial_t\,\delta\xi^k
+
\bar\mathbf{u}^L\cdot\nabla\delta\xi^k)}
+
\overline{p^\xi K^j_k(\partial\,\delta\xi^k/\partial x^j)}
\,\bigg]
\,,
\nonumber
\end{eqnarray}
where $K^j_k=J\partial\,x^j/\partial (x^k+\xi^k)$ is the cofactor
that arises from the thermodynamic identity (\ref{1st-law-thermo}).
Thus, the pseudomomentum $\bar\mathbf{p}$ defined in
(\ref{pseudomom-def}), the Lagrangian-mean temperature
$\bar{T}^L=\partial\overline{e(D^\xi,\bar{s}^L)}/\partial\bar{s}^L$
and the potential $\Pi$ in (\ref{Pi-def}) all arise naturally in the
variational derivatives of the Lagrangian $\bar\ell$ in
(\ref{Lag-mean-Lag}) with respect to the mean fluid quantities.

One may verify that the GLM motion equation (\ref{GLM-eqns}) for the
mean fluid motion is now recovered by substituting the variational
derivatives of $\bar\ell$ in $\bar\mathbf{u}^L$, $\tilde{D}$ and
$\bar{s}^L$ into the EP equation (\ref{EP-GLM-eqn}). This 
computation places the GLM theory into the EP framework for the
averaged Lagrangian (\ref{Lag-mean-Lag}) and, thus, directly proves
the following. 

\comment{
\begin{lemma} \label{GLM/EP}
{\bf GLM adiabatic fluids satisfy EP equations}
The GLM motion equation (\ref{GLM-eqns}) for a compressible
adiabatic fluid results when the Euler-Poincar\'e equation
(\ref{EP-GLM-eqn}) is applied to the averaged Lagrangian
(\ref{Lag-mean-Lag}). 
\end{lemma}
}

This result suggests that a much broader principle is operating,
namely,

\comment{

\begin{lemma} \label{EPA-lemma}
{\bf EP Averaging Lemma}
GLM-averaging preserves the four equivalence relations of the EP
theorem. 
\end{lemma}
}

Andrews \& McIntyre's exposition of the GLM theory was published in
1978, while the EP theory for advective fluid dynamics was only
recently developed in Holm, Marsden and Ratiu [1998a]. Therefore, it
may be more natural to state the EP Averaging Lemma from the
viewpoint of a variational principle. Namely, 

\comment{
\begin{corollary}\label{var-red-thm}
{\bf Variational Reduction Property}
The GLM equations follow from a GLM-averaged EP variational
principle.
\end{corollary} 
}

\noindent
The Variational Reduction Property is summarized by the
following commutative diagram.

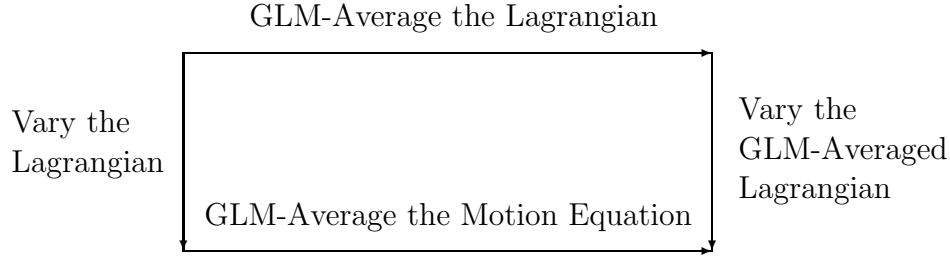
\begin{picture}(150,100)(-60,0)
\put(0,75){\vector(1,0){200} }
\put(25,85){GLM-Average the Lagrangian}
\put(200,75){\vector(0,-1){75} }
\put(210,50){Vary the}
\put(210,35){GLM-Averaged}
\put(210,20){Lagrangian}
\put(0,0){\vector(1,0){200} }
\put(8,10){GLM-Average the Motion Equation}
\put(0,75){\vector(0,-1){75} }
\put(-65,45){Vary the}
\put(-65,30){Lagrangian}
\end{picture}
\vspace{5mm}

\noindent
{\bf Sketches of Proofs.}
Lemma \ref{EPA-lemma} follows by using the definition of GLM
averaging, in combination with the four equivalence relations in the
first EP theorem \ref{rarl} and the Kelvin-Noether circulation result
of the second EP theorem
\ref{KelvinNoetherthm}  in Appendix \#1 (section \ref{EP-appendix}).
Three observations are relevant in sketching the proof of Lemma  
\ref{EPA-lemma}. First, one observes that the GLM average of a
right-invariant Lagrangian is still right-invariant, so the
Lagrange-to-Euler reduction and GLM averaging are compatible in the EP
theorem. (This observation takes us along the top and down the right
side of the diagram.) Second, the GLM average of the motion equation
preserves the transport structure of the Kelvin circulation theorem,
which is also implied by the second EP theorem. (Recall the GLM
Kelvin circulation loop analysis in section
\ref{Kel-transport-structure-sec}.) Third, the GLM average preserves
the form of the advection relations, as discussed in section
\ref{GLM-advection-sec}. Finally, to identify the averages of
thermodynamic derivatives that appear in the averaged motion equation
and, thus, complete the proof, one uses commutation of
exterior derivatives and GLM averaging in the definitions of these
average dual variables. For example, the average temperature of a
fluid parcel is correctly defined from the GLM average of the First
Law, since thermodynamic relations are applied in ideal fluid
theories for each fluid parcel as a {\it closed system}. Moreover,
the {\it same} definitions are used in the variational derivatives of
the averaged Lagrangian. These observations are sufficient to prove
Lemma \ref{EPA-lemma} -- the EP Averaging Lemma. For
more details, see Holm [2001], who proves that the 
EP Averaging Lemma is the front face of a cube of six interlocking
equivalence relations and commutative diagrams representing the
Lagrangian Averaged Euler-Poincar\'e (LAEP) Theorem.

Corollary \ref{var-red-thm} follows immediately from Lemma
\ref{EPA-lemma} for any Euler fluid equation that is also an EP
equation before the averaging is applied. However, Corollary
\ref{var-red-thm} can also be proven independently, by, say, using
the Clebsch procedure to place the GLM averaged equations and their
average advection relations directly into the EP variational
framework. Descriptions of the Clebsch procedure in this context are
given in Marsden and Weinstein [1983] and in Holm and Kupershmidt
[1983]. 

The EP Averaging Lemma and its corollary the Variational Reduction
Property allow extension of the exact nonlinear GLM theory to
include, for example, the continuum theory applications of the EP
theorem considered in Holm, Marsden \& Ratiu [1998a,b], and the
geophysical fluids applications considered in Allen, Holm \&
Newberger [2001] and in  Holm, Marsden \& Ratiu [2001].  The
remainder of this paper will be devoted to exploring some of the
applications of the EP Averaging Lemma in the small disturbance
approximation.


\subsection{GLM results arising in the EP framework}

The EP framework instills several fundamental
properties, including some that the GLM theory is already known to
possess. These known properties include the Kelvin circulation
theorem, the balance laws for energy and momentum, and the potential
vorticity conservation law for GLM. These properties are
briefly expressed in the EP framework, as follows. For
more details and the original development of the
EP theory with advected parameters, 
see Holm, Marsden \& Ratiu [1998a,b].


\subsubsection*{EP Kelvin circulation theorem for adiabatic GLM}

The EP motion equation (\ref{EP-GLM-eqn}) can be rewritten in
Lie-derivative form, as
\begin{equation}\label{Lie-der-EP-eqn}
\Big(\frac{\partial}{\partial t}+\pounds_{\bar\mathbf{u}^L}\Big)
\Big(\frac{1}{\tilde{D}}
\frac{\delta \bar\ell}{\delta \mathbf{u}^L}
\cdot{d}\mathbf{x}\Big)
=
d\frac{\delta \bar\ell}{\delta \tilde{D}}
-
\frac{1}{\tilde{D}}
\frac{\delta \bar\ell}{\delta \bar{s}^L}d\bar{s}^L
\,,
\nonumber
\end{equation}
where $\pounds_{\bar\mathbf{u}^L}$ is the Lie derivative with respect
to the Lagrangian mean velocity, $\bar\mathbf{u}^L$.
Integrating this form of the EP motion equation around a loop
$c(\bar\mathbf{u}^L)$  moving with the average motion of the fluid
provides the {\bf Kelvin-Noether theorem} in the EP framework for
adiabatic compressible fluids, as
\begin{equation}\label{KN-circ-GLM}
\frac{d}{dt}\oint_{c(\bar\mathbf{u}^L)}
\frac{1}{\tilde{D}}
\frac{\delta \bar\ell}{\delta \bar\mathbf{u}^L}
\cdot
d\mathbf{x}
=
-\,\oint_{c(\bar\mathbf{u}^L)}
\frac{1}{\tilde{D}}
\frac{\delta \bar\ell}{\delta \bar{s}^L}
\,d\bar{s}^L
\,.
\nonumber
\end{equation}
Hence, for adiabatic compressible GLM flow, from equation
(\ref{GLM-Lag-der}) for the required variational derivatives one
recovers equation (\ref{GLM-comp-Kel-thm}) as, 
\begin{equation}
\frac{d}{dt}\oint_{c(\bar\mathbf{u}^L)}
\Big( \bar\mathbf{u}^L 
- \bar\mathbf{p}
+
\Omega\times\mathbf{x}
\Big)
\cdot
d\mathbf{x}
=
\oint_{c(\bar\mathbf{u}^L)}
\bar{T}^L\,d\bar{s}^L
\,.
\nonumber
\end{equation}
%


\subsubsection*{Energy balance for adiabatic GLM}

Legendre transforming in $\bar\mathbf{u}^L$ yields,
\begin{eqnarray}
\bar{E}
&=&
\int \frac{\delta \bar\ell}{\delta \bar\mathbf{u}^L}
\cdot \bar\mathbf{u}^L
\ d^3x\
-\
\bar\ell
\nonumber\\
&=&
\int 
\tilde{D}\,\bigg[
\tfrac{1}{2}|\bar\mathbf{u}^L|^2
+
\tfrac{1}{2}\overline{|\mathbf{u}^\ell|^2}
+
\bar{\Phi}^L(\mathbf{x})
+
\overline{e(D^\xi,\bar{s}^L)}
\nonumber\\
&&\hspace{1cm}
-\
\overline{
(\mathbf{u}^\ell
+
\Omega\times\xi)\cdot\partial_t\,\xi}
\bigg]
d^3x
\,.
\nonumber
\end{eqnarray}
Except for the last term, this is the total mean 
energy of the adiabatic GLM theory. We notice the
``pseudoenergy'' 
\begin{equation}\label{pseudo-erg-def}
\bar{e}
\equiv
\overline{[u^\ell_k+(\Omega\times\xi)_k]\partial_t\,\xi^k}
\,,
\end{equation}
appearing as the last term in the energy quantity $\bar{E}$. This
term is independent of the internal energy and has a common factor
with the pseudomomentum defined earlier,
\begin{equation}\label{pseudo-mom-redef}
-\,\bar\mathbf{p}
\equiv
\overline{[u^\ell_k+(\Omega\times\xi)_k]\nabla\xi^k}
\,.
\nonumber
\end{equation}
In fact, these two quantities may be expressed equivalently as
\begin{equation}\label{pseudo-mom/erg-rel}
\tilde{D}\,\bar{e}
=
\overline{\pi_k\partial_t\,\xi^k}
\quad\hbox{and}\quad
\tilde{D}\,\bar\mathbf{p}
=
-\,\overline{\pi_k\nabla\xi^k}
\,,
\nonumber
\end{equation}
where 
$\pi_k
\equiv
\delta{\ell}/\delta(\partial_t\,\xi^k)
=
\tilde{D}\,[u^\ell_k+(\Omega\times\xi)_k]
$
is the momentum density canonically conjugate to $\xi^k$, {\bf
before} the Eulerian mean is taken in the Lagrangian $\bar{\ell}$.

The spatially integrated pseudoenergy is given by
\[
\langle\,\bar{e}\,\rangle 
= \int \tilde{D}\,\bar{e}\, d^3x 
= \int\overline{\pi_k\,\partial_t\,\xi^k}\,
d^3x
\,.
\]
This  term would have {\bf cancelled}, had we performed the complete
the Legendre transformation,
\begin{equation}
\bar{\cal E}
=
\int 
\bigg(\frac{\delta \bar\ell}{\delta \bar\mathbf{u}^L}
\cdot \bar\mathbf{u}^L
+
\overline{
\frac{\delta \bar\ell}{\delta (\partial_t\,\xi)}\cdot\partial_t\,\xi
}
\
\bigg)d^3x\
-\
\bar\ell
=
\bar{E}
+
\int
\overline{
\pi\cdot\partial_t\,\xi
}
\,d^3x
\,.
\nonumber
\end{equation}
in both fluid and wave properties.
The pseudoenergy $\bar{e}$ in equation (\ref{pseudo-erg-def}) is thus
understood to be the mean classical-mechanical action per unit mass
of the fluctuating Lagrangian displacement field. 
This complete Legendre transformation yields the expected result
for the conserved total mean energy for a self-consistent theory,
\begin{equation}
\bar{\cal E}
=
\int 
\tilde{D}\,\Big[
\tfrac{1}{2}|\bar\mathbf{u}^L|^2
+
\tfrac{1}{2}\overline{|\mathbf{u}^\ell|^2}
+
\bar{\Phi}^L(\mathbf{x})
+
\overline{e(D^\xi,\bar{s}^L)}
\,\Big]\,d^3x
\,.
\nonumber
\end{equation}
Hence, we find that $d\bar{E}/dt
=
-\,\frac{d}{dt}\int\overline{\pi\cdot\partial_t\,\xi}\,d^3x
=-\,\frac{d}{dt}\int\tilde{D}\bar{e}\,d^3x$,
since the total mean energy $\bar{\cal E}$ must be conserved for a
theory with no sources or sinks of energy.

\paragraph{Remark about averaging and conservation laws.} 
Before averaging, the integrated {\it instantaneous} pseudomomentum
is defined as,
\[
\langle\mathbf{p}\rangle = \int \tilde{D}\,\mathbf{p}\, d^3x =
- \int\pi_k\nabla\xi^k\, d^3x
\,,
\]
The spatially integrated quantity $\langle\mathbf{p}\rangle$
generates infinitesimal Eulerian spatial shifts of the wave
properties as {\bf canonical transformations}. That is,
\begin{equation}\label{pseudo-mom/shift}
\{\langle\mathbf{p}\rangle, \xi\}
=
\nabla\xi
\quad\hbox{and}\quad
\{\langle\mathbf{p}\rangle, \pi\}
=
\nabla\pi
\,,
\nonumber
\end{equation}
where  $\{F,H\}$ is the canonical Poisson bracket with
$\{\xi(\mathbf{x}^{\,\prime}),\pi(\mathbf{x})\}
= 
\delta(\mathbf{x}-\mathbf{x}^{\,\prime})$.

Under this canonical Poisson bracket, one may verify the formulas 
\[
\mathsf{A}=-\int\pi\cdot\partial_a\xi\,d^3x
\quad\Rightarrow\quad
\{\mathsf{A}, \xi\}
=
\partial_a\xi
\quad\hbox{and}\quad
\{\mathsf{A}, \pi\}
=
\partial_a\pi
\,.\]
That is, the functional $\mathsf{A}$ generates a translation in phase
space for any parameter $a$ that admits integration by parts. If the
solutions in phase space $(\pi,\xi)$ are averaged over such a
parameter, then the averaged generator of the the translations,
$\bar\mathsf{A}=-\int\overline{\pi\cdot\partial_a\xi}\,d^3x$,
will be conserved. For example, the $i^{th}$ component $\bar{p}_i$ of
the pseudomomentum would be conserved, if the solutions $(\pi,\xi)$
were averaged over space in the $i^{th}$ direction.

\paragraph{Remark -- the relation between GLM and WMFI.}

The GLM and WMFI theories are closely related. For example, the
WMFI wave action density has the same character as the GLM
quantities, pseudomomentum and pseudoenergy, which may also be aptly
expressed in terms of a single-frequency WKB wave packet. By varying
the wave properties $\xi$ in the averaged Lagrangian as well as the
mean fluid properties, Gjaja \&  Holm [1996] constructed a
{\it  self-consistent} Lagrangian mean WMFI theory. This WMFI theory
reduces to GLM theory when the statistics of $\xi$ are {\it 
prescribed}. See Appendix \#3 (section \ref{WMFI-appendix}) for a
brief discussion. 

To explain how the wave action density of the WMFI theory is
related to the GLM pseudomomentum, we make the following
pre-canonical transformation,
\[
\tilde{D}\,\bar\mathbf{p}\cdot d\mathbf{x}
=
-\,\overline{\pi_k\cdot\nabla\xi^k}\cdot d\mathbf{x}
=
-\,\overline{\pi\cdot d\xi}
\,.\]
If $\xi$ and $\pi$ depend on a phase parameter $\phi$, we may write
the phase-averaged differential relation as
\[
-\,\overline{\pi\cdot d\xi}
=
- \overline{\pi_k\partial_\phi\xi^k} \, d\phi
=
Nd\phi
=
N\mathbf{k}\cdot d\mathbf{x}
\,,\]
where the wavevector $\mathbf{k}$ is defined by $d\phi =
\nabla\phi\cdot d\mathbf{x} =
\mathbf{k}\cdot d\mathbf{x}$. 

Thus, we obtain the wave action
density $N =-\overline{\pi_k\partial_\phi\xi^k}$, which is 
related to the GLM pseudomomentum by 
$\tilde{D}\bar\mathbf{p}=N\mathbf{k}$. For the
WKB wavepacket 
$\xi = \tfrac{1}{2}(\mathbf{a}e^{i\phi/\epsilon}
+
\mathbf{a}^*e^{-i\phi/\epsilon})$, 
one finds the formula,
\[
\frac{N}{\tilde{D}}
=
-\,\overline{\Big[
\frac{D^L\xi}{Dt}+(\Omega\times\xi)\Big]\cdot\partial_\phi\xi}
=
2\tilde{\omega}|\mathbf{a}|^2
+
2i\Omega\cdot\mathbf{a}\times\mathbf{a}^*
+
2\Im\Big(\mathbf{a}\cdot\frac{D^L\mathbf{a}^*}{Dt}\Big)
\,,
\]
in which
$\tilde{\omega}=-D^L\phi/Dt
=\omega-\mathbf{k}\cdot\bar\mathbf{u}^L$
is the Doppler-shifted wave frequency. This formula is in agreement
with the wave action density $N$ appearing in WMFI studies such as
that of Gjaja \&  Holm [1996]. As a result of the symmetry under
translations in $\phi$ introduced by phase-averaging the Lagrangian,
we have 
\begin{equation}\label{Wave-action-cons}
0
=
-\,
\frac{\partial }{\partial t}
\frac{\partial \bar{\cal L}}{\partial (\partial_t\phi)}
-\,
{\rm div}\,
\frac{\partial \bar{\cal L}}{\partial (\nabla\phi)}
=
\frac{\partial N}{\partial t}
+
\frac{\partial }{\partial x^j}
\Big(N\bar{u}^{L\,j}
-
\overline{p^\xi K^j_i \partial_\phi\,\xi^i}\, \Big)
\,,
\nonumber
\end{equation}
upon using the variational derivatives in equation
(\ref{GLM-Lag-der}). Andrews \& McIntyre [1978b] obtain the same
conservation law by directly manipulating the GLM motion equation
(\ref{GLM-eqns}). This equivalence, of course, is guaranteed by the
EP Averaging Lemma \ref{EPA-lemma}.

We recover this conservation law as a result of {\bf Noether's
theorem} for the averaged Lagrangian. Thus, we have the following.
%
\comment{\begin{lemma}
When averaging introduces an ignorable coordinate, the average of the
corresponding canonically conjugate momentum is conserved. In this
case, the conserved wave action density $N$ is the phase-averaged
generator of phase shifts.
\end{lemma}
}

\noindent
{\bf Remarks.}
\begin{description}
\item$\quad\bullet$
The GLM pseudoenergy is related to $N$ by $\tilde{D}\bar{e}=N\omega$,
which again identifies $\bar{e}$ as an action variable.

\item$\quad\bullet$
The self-consistent WMFI theory is closed by writing the
pseudomomentum as $\tilde{D}\bar\mathbf{p}=N\mathbf{k}$ and
using the {\it conservation of waves} relation, 
$
\partial_t\,\mathbf{k} = \nabla\omega
\,.
$
In this equation, the frequency variable $\omega$ must still be
determined. Until this point, no small-amplitude assumption has
been made. Introducing a small amplitude approximation allows the
frequency $\omega$ to be determined from its dispersion relation in
terms of fluid and wave mean properties. See Gjaja \& Holm [1996]
for more details, including the Lie-Poisson Hamiltonian structure of
the self-consistent WMFI theory, which is reminiscent of the Landau
two-fluid model of superfluid Helium.  

\end{description}


\subsubsection*{EP momentum balance for adiabatic GLM}

The momentum conservation law for the EP theory is,
\begin{equation}\label{EP-mom-cons}
{\partial_t}\,\bar{m}^L_i
+\,
{\partial_j}\,\bar{T}_i^j
=
 \frac{\partial \bar{\cal L}}{\partial x^i}\bigg|_{exp}
\,,
\end{equation}
where 
$\bar\mathbf{m}^L =
{\delta \bar\ell}/{\delta \bar\mathbf{u}^L}
$ 
is the Lagrangian-mean momentum density, the stress tensor
$\bar{T}_i^j$ is given by
\begin{eqnarray}\label{EP-stress-def}
\bar{T}_i^j
&=&
\bar{m}_i\bar{u}^{L\,j}
+
\delta_i^j\Big(\bar{\cal L} 
- 
\tilde{D}
\frac{\partial \bar{\cal L}}{\partial \tilde{D}}
\Big)
\nonumber
\end{eqnarray}
and $\partial \bar{\cal L}/\partial x^i\big|_{exp}$ denotes the
derivative with respect to the 
explicit spatial dependence that arises in the mean
Lagrangian $\bar\ell$ in (\ref{Lag-mean-Lag}) after
averaging over the statistics of $\xi$. For the adiabatic GLM theory,
this stress tensor is given by
\begin{eqnarray}
\bar{T}_i^j
&=&
\tilde{D}\Big( \bar{u}_i^L 
- \bar{p}_i
+
(\Omega\times\mathbf{x})_i
\Big)\bar{u}^{L\,j}
+
\delta_i^j\,\tilde{D}\,\overline{(p^\xi/D^\xi)}
\,.
\end{eqnarray}
The momentum balance law for adiabatic GLM is specified,
only after ${\partial \bar{\cal L}/\partial x^i}\big|_{exp}$ is
specified, by giving the spatial dependence in (\ref{Lag-mean-Lag})
of the wave properties and external potential in the Lagrangian
density  $\bar{\cal L}$. This is the requirement for obtaining closure
in the GLM theory.


\subsubsection*{Local EP potential vorticity conservation for
adiabatic GLM}

Invariance of the Lagrangian under diffeomorphisms
(interpreted physically as Lagrangian particle relabeling) implies
the local conservation law for  EP potential vorticity, 
\begin{equation}
\frac{D^L}{Dt}\bar{q}^L
=
0
\,,\quad\hbox{where}\quad
\bar{q}^L
=
\frac{1}{\tilde{D}}\nabla\bar{s}^L\cdot
{\rm curl}\,
\Big(
\frac{1}{\tilde{D}}
\frac{\delta \bar\ell}{\delta \bar\mathbf{u}^L}\Big)
\,.
\nonumber
\end{equation}
For the adiabatic GLM case, the potential vorticity is given
explicitly as
\begin{equation}
\bar{q}^L
=
\frac{1}{\tilde{D}}\nabla\bar{s}^L\cdot
{\rm curl}\,
\Big( \bar\mathbf{u}^L 
- \bar\mathbf{p}
+
\Omega\times\mathbf{x}
\Big)\,.
\nonumber
\end{equation}
Note the relation of the potential vorticity to the Kelvin
circulation theorem. This is particularly apparent when the Kelvin
theorem for adiabatic GLM theory is re-cast in terms of surface
integrals using Stokes theorem, as
\[
\frac{d}{dt}\int\!\!\!\!\int_{A}
{\rm curl}\,
\Big( \bar\mathbf{u}^L 
- \bar\mathbf{p}
+
\Omega\times\mathbf{x}
\Big)\cdot\boldsymbol{\hat{n}}\,dA
=
\int\!\!\!\!\int_{A}
\nabla\bar{T}^L\times\nabla\bar{s}^L
\cdot\boldsymbol{\hat{n}}\,dA
\,,\]
where the boundary of the surface $A$ is the fluid loop, 
$\partial{A}=c(\mathbf{u}^L)$.


\subsubsection*{GLM helicity}

The EP helicity is given by
\begin{equation}
\bar{\Lambda}^L
=
\int
\Big(
\frac{1}{\tilde{D}}
\frac{\delta \bar\ell}{\delta \bar\mathbf{u}^L}\Big)
\cdot
{\rm curl}\,
\Big(
\frac{1}{\tilde{D}}
\frac{\delta \bar\ell}{\delta \bar\mathbf{u}^L}\Big)
d^3x
\,.
\nonumber
\end{equation}
The corresponding GLM helicity is not conserved in the adiabatic
case, although it is conserved in the GLM theory for the three
dimensional barotropic case. (The same is true, before averaging.)


\subsection{EP results for the GLM Boussinesq stratified fluid}

The Eulerian expression of the averaged Lagrangian for a Boussinesq
stratified fluid is 
\boxeq5{
\begin{eqnarray}\label{Lag-mean-Bouss-Lag}
\bar\ell(\bar\mathbf{u}^L,\tilde{D},\bar{\theta}^L)
&=&
\int\bigg\{
\tilde{D}
\bigg[
\frac{1}{2}
\overline{
\Big|\bar\mathbf{u}^L
+
\frac{D^L\xi}{Dt}\Big|^2}
\,
+
\overline{
(\bar\mathbf{R}^L+\mathbf{R}^\ell)
\cdot\Big(\bar\mathbf{u}^L
+
\frac{D^L\xi}{Dt}\Big)}
\nonumber\\
&&\hspace{1cm}
-\
\overline{\Phi(\mathbf{x}^\xi)}
-\,
g\,z\,\bar{\theta}^L
\bigg]
-
\overline{p^\xi
\bigg(
\tilde{D}
-
{\cal J}
\bigg)}
\bigg\}
d^3x
\,.
\end{eqnarray}
}
This mean Lagrangian $\bar\ell\equiv\int\bar{\cal  L}
(\bar\mathbf{u}^L,\tilde{D},\bar{\theta}^L)d^3x$ is a straight
GLM decomposition of the standard Lagrangian for Boussinesq
stratified fluids, followed by taking the Eulerian mean. The relative
buoyancy $\theta$ is advected as a scalar in the Boussinesq
approximation, 
\[
\partial_t\,\theta+\mathbf{u}\cdot\nabla\theta=0
\,,
\]
so we have already substituted
$\theta^\xi=\bar{\theta}^L$. The rotation frequency $\Omega$ depends
on position and is given by
$2\Omega(\mathbf{x}^\xi) = ({\rm curl}\,\mathbf{R})^\xi$. The
rotation potential is decomposed in standard GLM fashion as 
$\mathbf{R}^\xi=\bar\mathbf{R}^L + \mathbf{R}^\ell$. Finally, the
pressure $p^\xi$ is a Lagrange multiplier that imposes the
the constraint relation defining the conserved
GLM density $\tilde{D}=D^\xi{\cal J}$, for $D^\xi=1$.

\begin{quote}
{\bf Remark.}
The kinetic energy is the same here as in equation
(\ref{Lag-mean-Lag}) for the adiabatic compressible fluid, and
the relative buoyancy is perfectly analogous to the entropy per unit
mass. Moreover, the pressure constraint is also analogous to
internal energy. So, one should expect no substantial difference
to occur in passing from the adiabatic GLM case to the Boussinesq GLM
equations.
\end{quote}


\subsubsection*{Variational derivatives and EP equation for GLM
Boussinesq stratified fluid}
 
The variational derivatives required for the EP equation
(\ref{EP-GLM-eqn}) -- with entropy $\bar{s}^L$ replaced by
buoyancy $\bar{\theta}^L$ -- are obtained from (ignoring variational
derivatives in $\xi$ now and henceforth)
\begin{eqnarray}\label{Lag-Bouss-var-der}
\delta\bar\ell
=
\int
d^3x\,\bigg[
\tilde{D}\Big( \bar\mathbf{u}^L 
- \bar\mathbf{p}
+ \bar\mathbf{R}^L
\Big)
\cdot\delta\bar\mathbf{u}^L
-\
\tilde{D}\,gz\,\delta\bar{\theta}^L
-\,\Pi^B\,
\delta\tilde{D}
\bigg]
\,.
\end{eqnarray}
Here, the pseudomomentum is defined by 
$\bar\mathbf{p}=-\overline{(u^\ell_j+R^\ell_j)\nabla\xi^j}$ and the
Boussinesq potential $\Pi^B$ is defined by
\begin{equation}\label{Pi-B-def}
\Pi^B
=
\pi^B
+
gz\,\bar{\theta}^L
+
\bar\mathbf{u}^L\cdot\bar\mathbf{R}^L
\,,\nonumber
\end{equation}
where
\begin{equation}\label{pi-B-def}
\pi^B
=
\bar{p}^L
+
\bar\Phi^L(\mathbf{x})
-
\frac{1}{2}
\overline{\mathbf{u}^\xi\cdot(\mathbf{u}^\xi+2\mathbf{R}^\ell)}
\,.\nonumber
\end{equation}
Here $\bar{p}^L=\overline{p^\xi}$ is the Lagrangian mean pressure,
cf. equation (\ref{Pi-def}) for the potential in the adiabatic
compressible case. We substitute these variational derivatives into
the  Euler-Poincar\'e (EP) equation (\ref{EP-GLM-eqn}),  with the
analogous replacement $\bar{s}^L\to\bar{\theta}^L$, to find the
following GLM motion equation for a stratified Boussinesq fluid in
Cartesian coordinates, \\
\boxeq2{
\begin{eqnarray}\label{GLM-Bouss-eqns}
\Big[
\frac{D^L}{Dt}\big(
\bar\mathbf{u}^L
-
\bar\mathbf{p}
\big)
+
\big(
\bar{u}^L_k
-
\bar{p}_k
\big)
\nabla
\bar{u}^L_k
\Big]
+
2\Omega\times\bar\mathbf{u}^L
+
\nabla\,\pi^B
+
g\bar{\theta}^L\boldsymbol{\hat{z}}
=
0
\,.
\end{eqnarray}
}

\noindent
{\bf Remarks.}
\begin{description}
\item$\quad\bullet$
Stratified Boussinesq fluids and adiabatic compressible fluids
admit very similar forms of the EP Averaging Lemma. 

\item$\quad\bullet$
The GLM Boussinesq motion equation (\ref{GLM-Bouss-eqns}) is very
similar to the corresponding adiabatic compressible equation
(\ref{GLM-eqns}). However, it is different in an important way from
the corresponding equations (8.7a) and (9.1) of Andrews \& McIntyre
[1978a], which both contain only $D^L\bar\mathbf{u}^L/Dt$, instead of
the combination of four terms in square brackets written here. (This
error was not repeated in B\"uhler and McIntyre [1998] equation
(9.4).) Without this correct combination of four terms, the Kelvin
circulation theorem cannot be satisfied properly.
\end{description}


\subsubsection*{EP Kelvin circulation theorem for GLM Boussinesq
stratified fluid}

The EP framework provides the {\bf Kelvin-Noether
theorem} for Boussinesq stratified fluid, in the form
\begin{equation}
\frac{d}{dt}\oint_{c(\bar\mathbf{u}^L)}
\frac{1}{\tilde{D}}
\frac{\delta \bar\ell}{\delta \bar\mathbf{u}^L}
\cdot
d\mathbf{x}
=
-\,\oint_{c(\bar\mathbf{u}^L)}
\frac{1}{\tilde{D}}
\frac{\delta \bar\ell}{\delta \bar{\theta}^L}
\,d\bar{\theta}^L
\,.
\nonumber
\end{equation}
Hence, for the  GLM Boussinesq stratified fluid one has,
\begin{equation}
\frac{d}{dt}\oint_{c(\bar\mathbf{u}^L)}
\Big( \bar\mathbf{u}^L 
- \bar\mathbf{p}
+
\bar\mathbf{R}^L(\mathbf{x})
\Big)
\cdot
d\mathbf{x}
=
\oint_{c(\bar\mathbf{u}^L)}
gz\,d\bar{\theta}^L
\,,
\nonumber
\end{equation}
where ${\rm curl}\,\bar\mathbf{R}^L(\mathbf{x})
= 2\Omega(\mathbf{x})$.
If the loop $c(\bar\mathbf{u}^L)$ moving with the Lagrangian mean
flow lies entirely on a level surface of $\bar{\theta}^L$, then the
right hand side vanishes, and one recovers for this case the
``generalized Charney-Drazin theorem'' for transient Boussinesq
internal waves, in analogy to the discussion in Andrews \& McIntyre
[1978a] for the adiabatic compressible case.


\subsubsection*{Momentum balance for GLM Boussinesq stratified
fluid}
 
For a mean Lagrangian density $\bar{\cal L}$, the EP theory yields
the momentum balance,
\begin{equation}
{\partial_t}\,\bar{m}_i
+
{\partial_j}\,\bar{T}_i^j
=
\frac{\partial\bar{\cal L}}{\partial x^i}\bigg|_{exp}
\,,\nonumber
\end{equation}
with terms defined in analogy with the compressible GLM case.


\subsubsection*{Local potential vorticity conservation for GLM
Boussinesq stratified fluid}

Invariance of the Lagrangian under diffeomorphisms
(interpreted physically as Lagrangian particle relabeling) implies
the local conservation law for EP potential vorticity, 
\begin{equation}
\frac{D^L}{Dt}\bar{q}^L
=
0
\,,\quad\hbox{where}\quad
\bar{q}^L
=
\frac{1}{\tilde{D}}\nabla\bar{\theta}^L\cdot
{\rm curl}\,
\Big(
\frac{1}{\tilde{D}}
\frac{\delta \bar\ell}{\delta \bar\mathbf{u}^L}\Big)
\,.
\nonumber
\end{equation}
For the GLM case, the potential vorticity is given explicitly as
\begin{equation}
\bar{q}^L
=
\frac{1}{\tilde{D}}\nabla\bar{\theta}^L\cdot
{\rm curl}\,
\Big( \bar\mathbf{u}^L 
- \bar\mathbf{p}
+
\bar\mathbf{R}^L(\mathbf{x})
\Big)\,.
\nonumber
\end{equation}
Again, the EP framework explains the relation of the potential
vorticity to the Kelvin circulation theorem.

Other considerations in the EP framework for the GLM
Boussinesq stratified fluid closely follow the developments for the
GLM adiabatic fluid, modulo simple adjustments for replacing
$\bar{s}^L\to\bar{\theta}^L$, in the momentum and energy balance
laws, for example.


\subsection{Section summary}

This section shows that passing from the Euler equations for ideal 
compressible and incompressible fluids to the GLM equations admits
the {\bf EP Averaging Lemma}. Namely, when the Lagrangian mean is
used for averaging, the EP equations for the averaged Lagrangian are
identical to the averaged EP equations. Because of the EP Averaging
Lemma, one finds that the Kelvin circulation theorem, the balances
for energy and momentum and the local conservation law for potential
vorticity  arise as general features of all GLM-averaged EP fluid
theories. Concepts in GLM theory such as pseudomomentum and wave
action density also arise naturally as general features in the EP 
context.

Thus, the EP Averaging Lemma places the exact nonlinear GLM theory
into the realm of {\bf averaged Lagrangians} for Eulerian fluid
mechanics in the EP framework. This framework allows further
structure-preserving approximations of the GLM equations to be made
using the EP variational formulation.

Being derivable in the EP framework, the GLM theory also possesses
other fundamental structure that is shared by all ideal fluid
theories in the EP framework. In particular, the  EP framework leads
to the Lie-Poisson Hamiltonian formulation for GLM theory, as well as
to the potential-vorticity Casimirs associated with this Lie-Poisson
bracket. In turn, this structure leads to the energy-Casimir method
for characterizing equilibrium solutions of the GLM equations for
ideal fluids as critical points of a constrained energy and for
establishing their nonlinear Liapunov stability conditions. For an
explanation of this additional structure and many applications in
fluids and plasmas, see Holm, Marsden, Ratiu \& Weinstein [1985]. 

All of these additional features are now available to the GLM theory
of fluid dynamics. However, we shall forego investigating these other
implications here and pass to the formulation of an approximate set
of Eulerian mean equations based on a small-amplitude approximation
of the GLM theory.  We refer to Holm, Marsden \& Ratiu [1998a,b] for
detailed descriptions, derivations and basic references to other
works concerning the underlying geometry associated with the EP
framework for ideal fluids with advected quantities.

\vfill

\paragraph{Remarks.}
\begin{description}
\item$\bullet\quad$
The GLM equations may also be obtained by averaging in Hamilton's
principle at constant fluid parcel label in the Lagrangian
description, then transforming the result to the Eulerian description
and again using the EP theory. This approach was taken in Gjaja \& 
Holm [1996] in developing a self-consistent WMFI theory for the
Boussinesq stratified case. The same approach was taken by Holm
[1999] in developing nonlinear Taylor hypothesis closures (THC)  for
both compressible and incompressible flows. The equivalence of these
other approaches to the present approach is proven by the LAEP Theorem
in Holm [2001].

\item$\bullet\quad$
Regarding stability of the GLM solutions, see Andrews \& McIntyre
[1978b] for discussion of a variational principle for linear
evolution of small disturbances of a Lagrangian-mean flow. In this
regard, see also the classical works mentioned earlier on Lagrangian
fluid stability analysis and WMFI theory. 
\end{description}


\section{Linearized Eulerian/Lagrangian fluctuation relations}
\label{Lin-fluct-rel}

In principle, the GLM theory is more accurate than an Eulerian mean
theory, because its scalar advection relations hold exactly, and it
preserves the Euler-Poincar\'e (EP) structure of the original
unapproximated equations. That is, being an EP theory,
GLM preserves the standard ideal fluid relations for energy, momentum
and potential vorticity, as well as possessing a Kelvin circulation
theorem. However, the results of any Lagrangian mean theory are often
difficult to interpret accurately in an Eulerian setting. In
addition, the Lagrangian mean statistics themselves are affected by
the mean motion at finite-amplitude Lagrangian displacement and,
thus, cannot be taken as prescribed quantities. Therefore, one sees
the need for an Eulerian mean counterpart to the GLM theory in the
small-amplitude approximation. A theory of this type was recently
initiated in B\"uhler \& McIntyre [1998] in the context of the
gravity wave parameterization problem.

In preparation for producing a variational complement to the
small-amplitude GLM theory, we shall first discuss the linearized
Eulerian/Lagrangian fluctuation relations. 


\subsection{Taylor series approximations of Eulerian fluctuations at
linear order in the Lagrangian displacement $\xi$}

In the GLM theory, the displaced fluid velocity is given by
\begin{equation}
\mathbf{u}(\mathbf{x} + \xi,t)
=
\bar\mathbf{u}^L(\mathbf{x},t)
+
\mathbf{u}^\ell(\mathbf{x},t)
\,,
\nonumber
\end{equation}
where
\begin{equation}
\mathbf{u}^\ell(\mathbf{x},t)
=
\frac{\partial}{\partial t}\xi
+
\bar\mathbf{u}^L\cdot\nabla\xi
\equiv
\frac{D^L\xi}{D t}
\,.
\nonumber
\end{equation}
A Taylor series approximation shows that the Eulerian velocity
fluctuation $\mathbf{u}^{\,\prime}$ is related to the Lagrangian
disturbance velocity $\mathbf{u}^\ell$, as well as the fluctuating
displacement $\xi$ and the Eulerian mean velocity
$\bar\mathbf{u}$ at linear order in $\xi$ by
\[
\mathbf{u}^\ell
=
\mathbf{u}^{\,\prime} 
+ 
\xi\cdot\nabla\bar\mathbf{u}
\,.
\]
Therefore, we find the important relation at linear order,
\begin{equation}\label{u-prime}
\mathbf{u}^{\,\prime}(\mathbf{x},t)
=
\frac{\partial}{\partial t}\xi
+
\bar\mathbf{u}\cdot\nabla\xi
-
\xi\cdot\nabla\bar\mathbf{u}
\,.
\hspace{17mm} \hbox{\fbox{$\mathbf{u}^{\,\prime}-$equation}}
\end{equation}
Likewise,  for a scalar quantity $\chi$, we have the 
linear-order relation,
$\chi^\ell=\chi^{\,\prime}
+ 
\xi\cdot\nabla\bar{\chi}
\,.$
Consequently, we find,
\begin{equation}\label{chi-prime}
\chi^{\,\prime}
=-\,
\xi\cdot\nabla\bar\chi
\,,
\hspace{4cm} \hbox{\fbox{$\chi^{\,\prime}-$equation}}
\end{equation}
for an {\bf advected scalar} $\chi$
(since $\chi^\ell=0$ for an advected scalar).
For a conserved density, $D$, 
the linear-order Taylor approximation
is 
\begin{equation}\label{D-ell}
D^\ell=D^{\,\prime} + 
\xi\cdot\nabla\bar{D} = -\bar{D}{\rm div}\xi
\,.
\nonumber
\end{equation}
Consequently, the Eulerian density fluctuation ${D}^{\,\prime}$ and
Eulerian mean density $\bar{D}$ are related to the Lagrangian
fluctuating displacement at linear order in $\xi$ for a conserved
density $D$ by
\begin{equation}\label{D-prime}
{D}^{\,\prime}
=
-\,{\rm div}\,(\bar{D}\xi)
\,.
\hspace{3cm} \hbox{\fbox{$D^{\,\prime}-$equation}}
\end{equation}

The $\mathbf{u}^{\,\prime}$ and ${D}^{\,\prime}$ equations imply 
\begin{equation}
(\bar{D}\mathbf{u}^{\,\prime} + {D}^{\,\prime}\bar\mathbf{u})
=
\partial_t\,(\bar{D}\xi)
-
{\rm curl}\,(\bar\mathbf{u}\times\bar{D}\xi)
\,.
\nonumber
\end{equation}
Taking the divergence of this relation and using the
${D}^{\,\prime}$ equation then implies the 
{\bf linearized continuity equation},
\begin{equation}\label{lin-cont-eqn}
\partial_t\,{D}^{\,\prime}
=
-\,{\rm div}\,
(\bar{D}\mathbf{u}^{\,\prime} + {D}^{\,\prime}\bar\mathbf{u})
\,.
\end{equation}
%

\paragraph{Remarks.}
\begin{description}
\item$\bullet\quad$
The $\mathbf{u}^{\,\prime}$, $\chi^{\,\prime}$ and ${D}^{\,\prime}$
equations are {\bf standard} in Lagrangian stability analysis. See
Friedman \& Schutz [1978a] for a historical survey of the use of
these linearized fluctuation relations, especially in astrophysics.
\item$\bullet\quad$
As a consequence of the linearized continuity equation, the Eulerian
mean density $\bar{D}$ satisfies the usual 
{\bf continuity equation}
\begin{equation}\label{usual-cont}
\partial_t\,\bar{D}
+
{\rm div}\bar{D}\bar\mathbf{u}
=
0
\,,
\end{equation}
in terms of the Eulerian mean velocity $\bar\mathbf{u}$.
\item$\bullet\quad$
The $\mathbf{u}^{\,\prime}$ equation (\ref{u-prime}) may also be
expressed in geometrical language in terms of the ad-operator defined
on the Lie algebra of vector fields. Namely, as the linear relation
$\mathbf{u}^{\,\prime}=\partial_t\,\xi+{\rm ad}_{\bar\mathbf{u}}\xi$. In
this expression, the ad-operator is defined in terms of the
commutator operation for vector fields $[\cdot\,,\cdot]$ by
\begin{equation}
{\rm ad}_{\bar\mathbf{u}}\xi
=
[\bar\mathbf{u},\xi]
=
\bar\mathbf{u}\cdot\nabla\xi
-
\xi\cdot\nabla\bar\mathbf{u}
=
-\,\pounds_\xi\,\bar\mathbf{u}^{\,\sharp}
\,,
\nonumber
\end{equation}
where superscript $(\,\cdot\,)^{\,\sharp}$ denotes a vector field.
\item$\bullet\quad$
In geometrical language, the
Eulerian fluctuating component of any advected quantity $a$ is given
at linear approximation in $\xi$ by 
\[
a^{\,\prime} = -\,\pounds_\xi\, \bar{a}
\,,\]
where $\bar{a}$ is the Eulerian mean and $\pounds_\xi$ denotes the Lie
derivative corresponding to the diffeomorphism generated by the
vector field $\xi$, the fluctuating displacement of the Lagrangian
trajectory away from its mean position $\mathbf{x}$.
\item$\bullet\quad$
In Appendix \#2 (section \ref{GEOM-appendix}) we explain in more
detail the geometric interpretations of these linear relations for the
Eulerian fluctuations of a scalar, a density and the fluid velocity
in terms of the Lagrangian fluctuation displacement.
\item$\bullet\quad$
A strong connection exists between the present approach and
the GLM approach to WMFI discussed in Gjaja \& Holm [1996]. In that
paper, attention concentrated on modeling the Lagrangian fluid
trajectory displacement fluctuation $\xi(\mathbf{x},t)$ as a WKB wave
packet. See Appendix \#3 (section \ref{WMFI-appendix}) for a list of
available WMFI theories based on asymptotic expansions and the WKB
approximation in Hamilton's principle. Here we shall use the linear
relations for
${D}^{\,\prime}$ and $\mathbf{u}^{\,\prime}$ to derive Eulerian mean
fluid equations that approximate the GLM fluid motion equations at
second order in the fluctuation displacement $\xi$. 
\end{description}


\paragraph{Frozen-in Lagrangian fluctuations.}
Seen as equations for $\xi$, the linearized fluctuation relations for
${D}^{\,\prime}$, $\chi^{\,\prime}$ and $\mathbf{u}^{\,\prime}$ only
determine the Lagrangian fluctuating displacement up to adding
an appropriate homogeneous solution determined by the initial
conditions. Such homogeneous solutions are ``frozen'' into the mean
motion, and for them ${D}^{\,\prime}$,
$\chi^{\,\prime}$ and
$\mathbf{u}^{\,\prime}$ all vanish. For example,
$\bar{D}\xi=\nabla\bar{\chi}\times\nabla\psi$ implies
${D}^{\,\prime}=0$ and $\chi^{\,\prime}=0$ for any function $\psi$.
If, in addition, the function $\psi$ satisfies the advection relation
for the mean flow, $\partial_t\,\psi+\bar\mathbf{u}\cdot\nabla\psi=0$,
then $\mathbf{u}^{\,\prime}=0$, as well. These homogeneous solutions
are known from the traditional theory of Lagrangian stability analysis
for fluids. See, e.g., Friedman \& Schutz [1978a] for a
clear discussion of them in this context.

Thus, homogeneous solutions of the linearized fluctuation
relations (the ${D}^{\,\prime}$, $\chi^{\,\prime}$ and
$\mathbf{u}^{\,\prime}$ equations) cause no changes in the
corresponding Eulerian mean fluid quantities $\bar{D}$, $\bar{\chi}$
and $\bar\mathbf{u}$. That is, they generate symmetries of the
Eulerian mean fluid quantities. In this way, the linearized
description distinguishes between {\bf passive} Lagrangian
fluctuations that are frozen-in (i.e., move with the mean flow) and
{\bf active} Lagrangian fluctuations that propagate through the fluid
in a Lagrangian sense (i.e., they move through the fluid from one
fluid element to the next). We shall assume that we may choose initial
conditions for which the frozen-in homogeneous contributions may be
set equal to zero. Being zero initially, they will remain so.
Physically, this means we are choosing the mean and current positions
of each Lagrangian fluid element to refer to the same Lagrangian
coordinate. Fortunately, this has been our convention from the
outset, so no further changes are needed. Henceforth, the fluctuating
Lagrangian displacement will be understood as the inhomogeneous
component of the solution. This inhomogeneous (propagating)
Lagrangian fluctuation component $\xi$ will be determined by a
{\bf Green's function method} that obtains the inhomogeneous
solution for $\xi$ uniquely from the $\mathbf{u}^{\,\prime}-$equation.
This solution will then imply ${D}^{\,\prime}$ and $\chi^{\,\prime}$
from the other linearized fluctuation relations. We shall give an
explicit example next.


\subsection{Green's function relation 
$\xi=G*\mathbf{u}^{\,\prime}$}

This subsection discusses the Green's function relation
$\xi=G*\mathbf{u}^{\,\prime}$ and derives the corresponding 
{\bf DuHamel formula} explicitly for the special case of affine
motion. 

The inhomogeneous solution of the linear velocity fluctuation
relation  (\ref{u-prime})
\[
\mathbf{u}^{\,\prime}
=
\partial_t\,\xi+{\rm ad}_{\bar\mathbf{u}}\,\xi
=
\partial_t\,\xi+\bar\mathbf{u}\cdot\nabla\xi 
- \xi\cdot\nabla\bar\mathbf{u}
\,,
\]
may be solved in terms of a Green's function $G$ as
\[\xi=G*\mathbf{u}^{\,\prime}\,,\] 
where $G*\mathbf{u}^{\,\prime}$ denotes the componentwise Green's
function convolution, 
\begin{equation}
(G*\mathbf{u}^{\,\prime\,})_i
=
\int G_i(\mathbf{x}-\mathbf{y},t-\tau)
{u}_i^{\,\prime}(\mathbf{y},\tau)\,d^3y\,d\tau
\,,\quad\hbox{(no sum on }i)
\,.
\nonumber
\end{equation}
%

\begin{quote}
{\bf Remark.}
The vector Green's function with components
$G_i$ introduces memory (depending on the  history of
$\bar\mathbf{u}$) into the solution for $\xi$ from
$\mathbf{u}^{\,\prime\,}$. Note that $\mathbf{G}$ is a 
{\it deterministic} quantity.
\end{quote}


The components of the vector Green's function
$G_i$, $i=1,2,3$, satisfy the {\bf deterministic dual equation},
\begin{equation}
\partial_t\, G_i+({\rm ad}^*_{\bar\mathbf{u}}\, G)_i
=
\delta(\mathbf{x}-\mathbf{y})\delta(t-\tau)
\,.
\nonumber
\end{equation}
In tensor index notation, this dual equation is written as
\begin{equation}
\partial_t\, G_i
+
\partial_j(G_i\bar{u}^j)
+
G_j\partial_i\bar{u}^j
=
\delta(\mathbf{x}-\mathbf{y})\delta(t-\tau)
\,,
\quad
\forall
\quad
i=1,2,3
\,.
\nonumber
\end{equation}
In vector notation this equation may also be expressed as 
\begin{equation}
\Big[\partial_t\, \mathbf{G}
+
\mathbf{G}\,{\rm div}\,\bar\mathbf{u}
\,-\,
\bar\mathbf{u}\times
{\rm curl}\,\mathbf{G}
+
 \nabla(\mathbf{G}\cdot\bar\mathbf{u})
\Big]_i
=
\delta(\mathbf{x}-\mathbf{y})\delta(t-\tau)
\,.
\nonumber
\end{equation}
The linear expression on the left side contains several elements
that are familiar from the motion equation for fluids. 

\begin{quote}
{\bf Remark.}
Geometrically, the operation ad$^*$ is defined as 
being the dual to the operation $-$ad under the $L_2$ pairing, namely,
\begin{equation}
\Big\langle \mathbf{G},{\rm ad}_{\bar\mathbf{u}}\,\xi\Big\rangle
=
-\,
\Big\langle{\rm ad}^*_{\bar\mathbf{u}}\, \mathbf{G},\xi\Big\rangle
=
-\,
\Big\langle\pounds_{\bar\mathbf{u}}\, \mathbf{G},\xi\Big\rangle
\,.
\nonumber
\end{equation}
Thus, $\mathbf{G}\in\mathfrak{g}^*$ is a one-form density -- which
is the geometrical object dual under $L_2$ pairing to vector fields.
That is, the Green's functions $G_i$ are components of a one-form
density (physically, $\mathbf{G}$ is a momentum density, if $\xi$ is
a velocity). In Cartesian coordinates, 
$({\rm ad}^*_{\bar\mathbf{u}}\, \mathbf{G})_i$ is written as
\begin{equation}
({\rm ad}^*_{\bar\mathbf{u}}\, \mathbf{G})_i
=
\partial_j( G_i\bar{u}^j)
+
 G_j\partial_i\bar{u}^j
\,.
\nonumber
\end{equation}
We shall find these formulas convenient for interpreting the
geometrical meaning of the expressions arising in the manipulations
to follow.
\end{quote}


\subsubsection*{Example: Green's function for affine mean motion} 

We shall derive the explicit Green's function relation 
$\xi=G*\mathbf{u}^{\,\prime}$ for the inhomogeneous
solution of the $\mathbf{u}^{\,\prime}-$equation in the
special case that the mean Lagrangian trajectory is given by the
affine relation 
\[
\mathbf{x}(t,\mathbf{x}_0)=F(t)\cdot\mathbf{x}_0+ \mathbf{b}(t)
\]
so that $\nabla\bar\mathbf{u}=\dot{F}(t)F^{-1}(t)$. In this special
case, the ${u}_i^{\,\prime}-$equation can be written with an
integrating factor as 
\begin{equation}
F(t)\cdot\frac{d}{dt}\Big|_{\mathbf{x}_0}
\Big(F^{-1}(t)\cdot\xi\Big)
=
\frac{d\xi}{dt}\Big|_{\mathbf{x}_0}
-
\dot{F}F^{-1}(t)\cdot\xi
=
\mathbf{u}^{\,\prime}
\,.\nonumber
\end{equation}
This implies an explicit {\bf DuHamel formula}
(implying memory, or history dependence) in which the
inhomogeneous solution is given by
\begin{equation}
\xi
=
G*\mathbf{u}^{\,\prime}
=
F(t)\cdot\int_0^t
F^{-1}(\tau)\cdot\mathbf{u}^{\,\prime}
(\mathbf{x}(\tau,\mathbf{x}_0),\tau)
\,d\tau
\,.
\nonumber
\end{equation}
This formula is reminiscent of the exponential integrating factor
for $\nabla\bar\mathbf{u}=const$ considered in Ristorcelli [2001] for
deriving tensor eddy viscosities in a turbulence model.

The full solution for the fluctuation displacement is  
\[\xi=G*\mathbf{u}^{\,\prime} + \xi^{(0)}\,,\]
where $\xi^{(0)}$ is the homogeneous solution
of the $\mathbf{u}^{\,\prime}-$equation. 
The homogeneous solution for the affine case is simply
\[\xi^{(0)}(\mathbf{x}(t,\mathbf{x}_0),t)
=
F(t)\cdot\xi^{(0)}(\mathbf{x}_0)
\,,
\]
as follows from 
\[
\frac{d\xi^{(0)}}{dt}\Big|_{\mathbf{x}_0}
=
\dot{F}(t)\cdot\xi^{(0)}(\mathbf{x}_0)
=
\dot{F}F^{-1}(t)\cdot\xi^{(0)}(\mathbf{x},t)
\,,
\]
where $\xi^{(0)}(\mathbf{x}_0)$ is the initial value of
$\xi^{(0)}(\mathbf{x},t)$. This initial value may be set to zero,
if one assumes the current and mean positions of a Lagrangian
trajectory refer to the same Lagrangian coordinate.

In fact, the homogeneous solution of the
$\mathbf{u}^{\,\prime}-$equation in the {\bf general case} is the
same as Cauchy's solution for the vorticity equation of an
incompressible ideal fluid, namely, 
\[\xi^{(0)}(\mathbf{x},t)
=
F(t,\mathbf{x}_0)\cdot\xi^{(0)}(\mathbf{x}_0)
\,,
\]
where $F(t,\mathbf{x}_0) 
=
\partial\mathbf{x}(t,\mathbf{x}_0)/\partial\mathbf{x}_0$ 
is the deformation gradient, or Jacobian, of the Lagrange-to-Euler
map $(t,\mathbf{x}_0)\to\mathbf{x}$.


\section{Deriving $g\ell{m}$ -- the order $O(|\xi|^2)$ GLM
equations}
\label{glm-der}

In this section, we shall obtain a set of Eulerian-mean
equations that approximate the GLM equations at second order in the
displacement $\xi$. Following ideas familiar in Lagrangian fluid
stability analysis, we shall derive these approximate equations from
a variational principle based on first taking the Eulerian mean of the
second-variation of the GLM Lagrangian and then using the
EP formulation. Our strategy for developing this order
$O(|\xi|^2)$ approximate Eulerian mean counterpart for GLM is as
follows.

We base the structure-preserving approximations of the GLM theory
implemented here in the EP framework on a two-step procedure. The
first step linearizes the Eulerian/Lagrangian fluctuation relation.
(This linearization describes how small fluctuations of a given fluid
quantity around its Eulerian mean are related to the fluctuating
displacement of a Lagrangian fluid parcel trajectory around its mean
position.) The second step substitutes these linearized fluctuation
relations into the second-variation of the Lagrangian in Hamilton's
principle. The Eulerian mean is then applied. This produces an
averaged  second-variation Hamilton's principle whose coefficients
are Eulerian mean second moments of fluctuating Lagrangian
displacements. Finally, variations are taken with respect to Eulerian
mean fluid quantities and thereby one obtains the averaged motion
equation in the EP framework. Thus, the first step of this procedure
is reminiscent of the traditional approach in linear Lagrangian
stability analysis for fluids. This traditional approach also invokes
the second-variation of a fluid action principle with respect to
Lagrangian displacements. However, there is a difference -- the
procedure here involves fluctuating linear displacements from a mean
solution, not deterministic linear displacements from a steady
solution as in the traditional stability analysis. 

The second-variation Lagrangian contains quadratic terms in $\xi$,
whose coefficients depend on the Eulerian mean fluid quantities. The
second step of our procedure begins by taking the Eulerian mean of
the second-variation Lagrangian over the quadratic moments of these
fluctuating displacements, $\xi$. We then take variations of this
averaged Lagrangian, with respect to the Eulerian mean fluid
quantities, and use the EP framework. (In Lagrangian stability
analysis, first, one does not average and, second, one takes
variations with respect to the Lagrangian displacements, not the
steady solutions.) The resulting EP equations express the
back-reaction effects on the Eulerian mean motion equation of the
fluctuating displacements of the Lagrangian trajectories in terms of
their Eulerian second moments. This two-step procedure is performed
within the EP framework for right-invariant Lagrangians that are
defined on the tangent space of a group. For fluids, this is the
group of diffeomorphisms representing the fluid motions, including
the Lagrangian fluctuating displacements themselves. See Appendices
\#1 and \#2 for further discussions of this point.


\paragraph{We summarize this two-step procedure in symbols,
as follows.}
\begin{description}

\item \underline{Step 1}
           \begin{description}

           \item$\bullet\quad$
           Linearize the fluctuation relations to find
equations (\ref{u-prime}) - (\ref{D-prime}),
\[
D^{\,\prime}=-\,{\rm div} \bar{D}\xi
\,,\
\chi^{\,\prime}=-\,\xi\cdot\nabla\bar{\chi}
\,,\
\mathbf{u}^{\,\prime}
=
\partial_t\,\xi+\bar\mathbf{u}\cdot\nabla\xi 
- \xi\cdot\nabla\bar\mathbf{u}
\,.\]
           \item$\bullet\quad$
           Substitute these relations into the second-variation
Lagrangian, to form
\[
\ell^{\,\prime\,\prime}
=
\int\Big[
\partial_t\,\xi\cdot{A}\cdot\partial_t\,\xi
+
\partial_t\,\xi\cdot{B}\cdot{\xi}
+
{\xi}\cdot{C}\cdot{\xi}
\Big]\,
d^3x
\,,\]
where $A$, $B$, $C$ are matrix operators involving the mean fluid
quantities {\bf and their gradients}, i.e., the
set $\{\bar{\mathbf{u}},
\bar{D}, \nabla\bar{\mathbf{u}}, \nabla\bar{D}\}$.

           \end{description}

\item \underline{Step 2}
           \begin{description}

           \item$\bullet\quad$
           Take the Eulerian mean to form the total mean Lagrangian
\[
\bar{\ell} 
= \bar{\ell}_0 
+ \tfrac{1}{2}\overline{\ell^{\,\prime\,\prime}}
\,.\]

           \item$\bullet\quad$
           Derive the $g\ell{m}$ motion equation for barotropic
           compressible fluids by 
           computing the EP equation 
\begin{equation}
\frac{d}{dt}
\frac{1}{\bar{D}}
\frac{\delta \bar\ell}{\delta \bar\mathbf{u}}
+
\frac{1}{\bar{D}}
\frac{\delta \bar\ell}{\delta \bar{u}^{\,j}}
\nabla\bar{u}^{\,j}
=
\nabla
\frac{\delta \bar\ell}{\delta \bar{D}}
\,,\nonumber
\end{equation}
           for the total mean Lagrangian $\bar{\ell}$
           by taking its variations
\[
\delta\bar{\ell}
=
\int \Big[
\Big(\frac{\delta\bar{\ell}}{\delta\bar{\mathbf{u}}}\Big)
\cdot\delta\bar{\mathbf{u}}
 +
\Big(\frac{\delta\bar{\ell}}{\delta\bar{D}}\Big)\delta\bar{D}
\
\Big] d^3x
\,.\]
           \end{description}
These variational derivatives involve Eulerian means of 
quadratic combinations of the Lagrangian fluctuating displacement
$\xi$, and its derivatives $\partial_t\,\xi$ and $\nabla\xi$. For
example, one combination that appears is 
$\overline{\pi_j\nabla\xi^j}$,
where
$\pi
=
\tfrac{1}{2}
\delta\ell^{\,\prime\,\prime}/\delta(\partial_t\,\xi)
=
A\cdot\partial_t\,\xi
+
B\cdot\xi$ 
is the momentum canonically conjugate to $\xi$. These Lagrangian
quadratic statistical moments are {\bf unknown parameters} in the
$g\ell{m}$ equations that must be independently specified, or
modeled, in closing the equations. Thus, a number of modeling
decisions must be made in closing any $g\ell{m}$ model.

In section \ref{alpha}, we shall discuss the various modeling
parameters required to produce a closed $g\ell{m}$ model. This
will be done in the context of simplifying them and constructing a
more manageable class of closed equations -- the alpha models --
obtained by using closures based on Taylor's hypothesis of frozen-in
turbulence.

\end{description}

The equations derived from this two-step procedure -- being the
small-amplitude approximation of the GLM equations -- 
are called $g\ell{m}$ equations. They are derived within the
EP framework. These new equations describe the dynamics of Eulerian
mean fluid quantities influenced by small amplitude fluctuations.
Being EP equations, they still retain the properties that result from
particle relabeling symmetry. In particular, the $g\ell{m}$
equations retain the Kelvin-Noether circulation theorem and its
associated local conservation law for potential vorticity.


\subsection{Expanding the Lagrangian $\ell(\mathbf{u}, D)$
in Hamilton's principle for barotropic fluids}

Into the Lagrangian $\ell$ we substitute  
$\mathbf{u}=\bar\mathbf{u}+\epsilon\mathbf{u}^{\,\prime}$ and
$D=\bar{D}+\epsilon{D}^{\,\prime}$, then truncate at quadratic order
in $\mathbf{u}^{\,\prime\,}$ and ${D}^{\,\prime\,}$ to find
\begin{equation}
\ell = \ell_0 
+ 
\epsilon\ell^{\,\prime} 
+ 
\frac{\epsilon^2}{2}\ell^{\,\prime\prime}
\,,\quad\hbox{with}\quad
^{\,\prime} = \frac{d}{d\epsilon}\Big|_{\epsilon=0}
\,.\nonumber
\end{equation}
Variations of $\ell$ are given by
\begin{eqnarray}
\ell^{\,\prime}
&=&
\frac{d}{d\epsilon}\Big|_{\epsilon=0}
\ell(\bar\mathbf{u}+\epsilon\mathbf{u}^{\,\prime},
\bar{D}+\epsilon{D}^{\,\prime})
\nonumber\\
&=&
\Big\langle
\frac{\delta \ell}{\delta \bar\mathbf{u}}
\,,\,
\mathbf{u}^{\,\prime\,}
\Big\rangle
+
\Big\langle
\frac{\delta \ell}{\delta \bar{D}}
\,,\,
D^{\,\prime\,}
\Big\rangle
\,,\nonumber
\end{eqnarray}
where
$
\langle{f\,,\,g}\rangle=\int f\,g\, d^3x
$,
is the $L_2$ pairing.
The quadratic functional $\ell^{\,\prime\prime}$ is the second
variation of the Lagrangian $\ell$ in the basis
$\mathbf{u}^{\,\prime\,}$ and
${D}^{\,\prime\,}$. That is, 
\begin{equation}\label{ell-prime-prime}
\ell^{\,\prime\prime}
=
\Big\langle
(\mathbf{u}^{\,\prime\,},{D}^{\,\prime\,})
\,,\,
D^2\ell(\bar\mathbf{u},\bar{D})
\cdot
(\mathbf{u}^{\,\prime\,},{D}^{\,\prime\,})
\Big\rangle
\,.
\end{equation}
(This is the connection to linear Lagrangian fluid stability theory.) 
Note that we treat $\ell^{\,\prime\prime}$ genuinely as a
second variation; so there are no double-prime terms, such
as $\mathbf{u}^{\,\prime\prime}$.
\bigskip

The {\bf averaged Lagrangian} at second order is then
\begin{equation}
\bar\ell
= 
\bar\ell_0 
+ 
\tfrac{\epsilon^2}{2}\,
\overline{\ell^{\,\prime\prime}}
\,,\quad\hbox{since }
\overline{\ell^{\,\prime}}
=
0
\,,\hbox{ for }
\overline{u^{\,\prime}}
=
0
=
\overline{D^{\,\prime}}
\,.
\nonumber
\end{equation}
Recall that the Eulerian mean satisfies the
projection property, $\bar{\bar\mathbf{u}} =
\bar\mathbf{u}$, and it commutes with the
spatial gradient, $\overline{\nabla\mathbf{u}} =
\nabla\bar\mathbf{u}$. On substituting the linearized fluctuation
relations for
${D}^{\,\prime}$, $\chi^{\,\prime}$ and $\mathbf{u}^{\,\prime}$ 
into $\overline{\ell^{\,\prime\prime}}$, we find the expected
quadratic form,
\[
\overline{\ell^{\,\prime\prime}}
=
\int\Big[
\overline{\partial_t\,\xi\cdot{A}\cdot\partial_t\,\xi}
+
\overline{ \partial_t\,\xi\cdot{B}\cdot{\xi} }
+
\overline{ {\xi}\cdot{C}\cdot{\xi} }
\Big]\,
d^3x
\,.\]
The $A$, $B$, $C$ in this quadratic form are matrix operators
involving the mean fluid quantities {\bf and their gradients}, i.e.,
the set $\{\bar{\mathbf{u}}, \bar{D}, \nabla\bar{\mathbf{u}},
\nabla\bar{D}\}$. Consequently, after taking variations, the
contribution from the mean fluctuation Lagrangian
$\overline{\ell^{\,\prime\prime}}$ to the mean momentum in the
corresponding EP equation will depend on {\bf second-gradients} of the
mean fluid quantities. The
Lagrangian $\overline{\ell^{\,\prime\prime}}$ is a functional of the
Eulerian mean quadratic moments of the Lagrangian fluctuation
displacements. Consequently, the resulting EP equation will also
depend parametrically on the second-order statistics of the
Lagrangian fluctuations.

\begin{quote}
{\bf Summary.}
Thus, in the EP framework for these $g\ell{m}$ equations we must
expand the Lagrangian to second order in $\xi$, take its Eulerian
mean, vary it with respect to $\bar\mathbf{u}$ and $\bar{D}$, and then
model the second-order statistics of $\xi$ in the resulting EP motion
equation for $\bar\mathbf{u}$.

\end{quote}


\subsection*{Our next steps are:}

\begin{enumerate}

\item 
Compute the mean momentum of the fluctuations,
\[
\overline{\mathbf{m}^{\,\prime\prime}}
=
\frac{\delta}{\delta \bar\mathbf{u}}
\,\Big(\tfrac{1}{2}\,\overline{\ell^{\,\prime\prime}}\Big)
\,.\]
\item
Write the EP $g\ell{m}$ equations for total
momentum
\[
\bar\mathbf{m}=\frac{\delta \overline{\ell}}{\delta \bar\mathbf{u}}
\,,\quad
\bar\ell
= 
\bar\ell_0 
+ 
\Big(\tfrac{1}{2}\,\overline{\ell^{\,\prime\prime}}\Big)
\,.\]
\item
Obtain a Kelvin circulation theorem for $g\ell{m}$ equations from their
corresponding EP equations and the Kelvin-Noether
theorem for these equations.
\item
Derive the $g\ell{m}$ energy balance by Legendre transforming
$\bar\ell$, the averaged Lagrangian.
\item Derive the $g\ell{m}$ stress tensor $\bar{T}_i^j$ in the $g\ell{m}$ momentum
balance law,
$
{\partial_t}\,\bar{m}_i
+{\partial_j}\,\bar{T}_i^j
=
\partial{\cal L}/\partial x^i\big|_{exp}
\,,
$
including the ``fluctuation stresses'' by invoking Noether's theorem
again.
\item
Use the result in section \ref{alpha} to interpret the Euler$-\alpha$
model stresses,  circulation and momentum in $g\ell{m}$ terms.
\end{enumerate}


\subsection{The $g\ell{m}$ approximations for a barotropic compressible
fluid}

We shall now drop any dependence of the fluid internal energy on
specific entropy, here and in what follows. Thus, we shall treat only
the case of a {\bf barotropic, or isentropic, compressible fluid}.  We
shall evaluate the necessary variational derivatives of $\bar\ell$
with respect to $\bar\mathbf{u}$ and $\bar{D}$ by using the relations
(definitions) for $\mathbf{u}^{\,\prime}$ and $D^{\,\prime}$
re-derived in Appendix \#2 in terms of the infinitesimal generator
$\xi(\mathbf{x},t)$. 


\subsubsection*{Eulerian-mean Lagrangian at order $O(|\xi|^2)$ } 

To second order, the Eulerian-mean of the Lagrangian for a barotropic
(isentropic) compressible fluid is given by
\boxeq5{
\begin{eqnarray}\label{lag-bar}
\bar\ell(\bar\mathbf{u},\bar{D})
&=&
\bar\ell_0 
+ 
\tfrac{1}{2}\,\overline{\ell^{\,\prime\prime}}
=
\int \Big[
\tfrac{1}{2}\bar{D}|\bar\mathbf{u}|^2
-
\bar{D}e(\bar{D})
\Big]\,d^3x
\\
&&\hspace{5mm}
+
\int \Big[
\tfrac{1}{2}\bar{D}\overline{|\mathbf{u}^{\,\prime}|^2}
+
\overline{D^{\,\prime}\mathbf{u}^{\,\prime}}
\cdot\bar\mathbf{u}
-
\frac{c^2(\bar{D})}{2\bar{D}}
\,\overline{D^{\,\prime\,2}}
\Big]\,d^3x
\,.
\nonumber
\end{eqnarray}
}

\noindent
For such a fluid, the equation of state defines $c^2(\bar{D})$ via 
\begin{equation}
\frac{\partial^2}{\partial \bar{D}^2}
\big(\bar{D}e(\bar{D})\big)
= 
\frac{\partial}{\partial \bar{D}}
h(\bar{D})
=
\frac{ c^2(\bar{D})}{\bar{D}}
\,.
\nonumber
\end{equation}
Note, before averaging, $\ell^{\,\prime\prime}$ in equation
(\ref{ell-prime-prime}) is the second variation of the Lagrangian
$\ell(\mathbf{u},{D})$ with respect to the Eulerian mean velocity and
density, evaluated at the mean fluid values, $\bar\mathbf{u}$ and
$\bar{D}$. 

The variational derivatives of the mean fluctuational parts of
$\bar\ell$ are given by
\begin{eqnarray}
\delta(\tfrac{1}{2}\,\overline{\ell^{\,\prime\prime\,}})
&=&
\int 
\delta\bar\mathbf{u}\cdot
\Big[
\overline{D^{\,\prime}\mathbf{u}^{\,\prime}}
+
\overline{
{\rm ad}^*_\xi
(\bar{D}\mathbf{u}^{\,\prime} 
+ 
D^{\,\prime}\bar\mathbf{u})
}
\Big]
\nonumber
\\
&&
+\
\delta\bar{D}
\bigg[
\tfrac{1}{2}\overline{|\mathbf{u}^{\,\prime}|^2}
+
\overline{\xi\cdot\nabla(\mathbf{u}^{\,\prime}\cdot\bar\mathbf{u})}
\\
&&
-\
\overline{D^{\,\prime\,2}}
\,
\frac{\partial}{\partial\bar{D}}
\Big(\frac{c^2(\bar{D})}{2\bar{D}}\Big)
-
\overline{\xi\cdot\nabla
\Big(
D^{\,\prime}
\,
\frac{c^2(\bar{D})}{\bar{D}}
\Big)}\
\bigg]\,d^3x
\,.
\nonumber
\end{eqnarray}
In these formulas, recall from (\ref{u-prime}) and 
(\ref{D-prime}) that
$D^{\,\prime}
=-\,
{\rm div}(\bar{D}\xi)
$
 and
\begin{equation}
\mathbf{u}^{\,\prime}(\mathbf{x},t)
=
\frac{\partial {\xi}}{\partial t}
+
\bar\mathbf{u}\cdot\nabla\xi - \xi\cdot\nabla\bar\mathbf{u}
=
\partial_t\,\xi-\,{\rm ad}_\xi \bar\mathbf{u}
\,.
\nonumber
\end{equation}
%
\paragraph{Remarks.}
\begin{description}
\item$\bullet\quad$
{\bf Boundary conditions} are 
$\hat\mathbf{n}\cdot\bar\mathbf{u}=0$
and
$\hat\mathbf{n}\cdot\xi=0$
on the boundary.
\item$\bullet\quad$
Recall this is the same $\xi$ as in the GLM theory, so
we will be able to make direct comparisons between $g\ell{m}$ and GLM after
assembling the EP equations for the order $O(|\xi|^2)$
approximate theory. 

\item$\bullet\quad$
Note that after substituting the
linearized approximations for the fluctuations, the mean Lagrangian
and its variational derivatives now also depend on the {\bf
gradients} of mean fluid properties.

\end{description}


\subsection{The mean fluctuation momentum} 

Using the geometrical notation of Appendices \#1 and \#2, we express
the mean fluctuational momentum in various equivalent forms as
\begin{eqnarray}
\overline{\mathbf{m}^{\,\prime\prime}}
&=&
\frac{\delta}{\delta \bar\mathbf{u}}
\,\Big(\tfrac{1}{2}\,\overline{\ell^{\,\prime\prime}}\Big)
\nonumber\\
&=&
\overline{D^{\,\prime}\mathbf{u}^{\,\prime}}
+
\overline{
{\rm ad}^*_\xi
(\bar{D}\mathbf{u}^{\,\prime} 
+ 
D^{\,\prime}\,\bar\mathbf{u})
}
\nonumber\\
&=&
\bar{D}\big(
\overline{\xi\cdot\nabla\mathbf{u}^{\,\prime}}
+
\overline{{u}^{\,\prime}_j\nabla\xi^j}
\big)
+
\overline{ {\rm ad}^*_\xi {D}^{\,\prime}\,\bar\mathbf{u} }
\nonumber\\
&=&
\bar{D}\overline{\pounds_\xi(\mathbf{u}^{\,\prime})^\flat}
+
\overline{ {\rm ad}^*_\xi {D}^{\,\prime}\,\bar\mathbf{u} }
\nonumber\\
&\equiv&
\bar{D}\big(
\bar\mathbf{u}^S
-
\bar\mathbf{p}
\big)
+
\overline{ {\rm ad}^*_\xi {D}^{\,\prime}\,\bar\mathbf{u} }
\,,
\nonumber
\end{eqnarray}
where superscript `flat' $(\,\cdot\,)^{\,\flat}$ denotes a one-form
and one defines 
\begin{equation}
\bar\mathbf{u}^S
\equiv
\overline{\xi\cdot\nabla\mathbf{u}^{\,\prime}}
\quad\hbox{is \fbox{Stokes mean drift velocity}}
\nonumber
\end{equation}
\begin{equation}
\bar\mathbf{p}
\equiv
-\,
\overline{{u}^{\,\prime}_j\nabla\xi^j}
\quad\hbox{is \fbox{GLM pseudomentum}}
\nonumber
\end{equation}
In Cartesian components the geometrical combinations
$\overline{\pounds_\xi(\mathbf{u}^{\,\prime})^\flat}$ and
$\overline{{\rm ad}^*_\xi {D}^{\,\prime}\,\bar\mathbf{u} }$ are
expressed as 
\begin{equation}
\big(\,\overline{\pounds_\xi(\mathbf{u}^{\,\prime})^\flat}\,\big)_i 
=
\overline{\xi^j{u}^{\,\prime}_{i\,,j}}
+
\overline{{u}^{\,\prime}_j\xi^j_{\,,i}}
=
\big(
\bar\mathbf{u}^S
-
\bar\mathbf{p}
\big)_i
\nonumber
\end{equation}
and
\begin{equation}
\big(\,\overline{ {\rm ad}^*_\xi {D}^{\,\prime}\,\bar\mathbf{u}}
\,\big)_i 
=
\partial_j\big(\bar{u}_i\overline{{D}^{\,\prime}\xi^j}\,\big)
+
\bar{u}_j
\,
\overline{ {D}^{\,\prime}\partial_i\xi^j}
\,.\nonumber
\end{equation}
These are recurring combinations of terms, reappearing throughout
the $g\ell{m}$ theory.


\subsection*{$g\ell{m}$ pseudomomentum} 

Before the second-variation Lagrangian for the $g\ell{m}$
theory is averaged, one finds the momentum canonically conjugate to
$\xi$, given by, cf. the linearized continuity
equation (\ref{lin-cont-eqn}),
\[
\pi
=
\delta\bar\ell/\delta(\partial_t\,\xi)
=
(\bar{D}\mathbf{u}^{\,\prime} 
+ 
D^{\,\prime}\,\bar\mathbf{u})
\,.
\]
The corresponding pseudomomentum for the $g\ell{m}$ theory is
then given by
\[
\tilde{\mathbf{p}}
=
-\overline{ \pi_k\nabla\xi^k}
=
-\overline{(\bar{D}{u}_k^{\,\prime} 
+ 
D^{\,\prime}\,\bar{u}_k)
\nabla\xi^k}
\,.
\]
Thus, our earlier discussion indicates that a WMFI version of
$g\ell{m}$ theory would possess a conserved wave action density
given by 
\[
N
=
-\overline{ \pi_k\partial_\phi\xi^k}
=
-\overline{(\bar{D}{u}_k^{\,\prime} 
+ 
D^{\,\prime}\,\bar{u}_k)
\partial_\phi\xi^k}
\,.
\]


\subsubsection*{Mean fluctuation momentum -- incompressible case} 

The mean fluctuation momentum takes a simpler form in the
incompressible case. Recall $D^{\,\prime} = -\,{\rm div}(\bar{D}\xi)$.
Consequently, ${\rm div}\mathbf{u}=0$ (which implies
${\rm div}\bar\mathbf{u}=0={\rm div}\mathbf{u}^{\,\prime}$)
is consistent with setting $\bar{D}=1$ in the mean continuity
equation
\[
\partial_t\, \bar{D}
=-\,
{\rm div}(\bar{D}\bar\mathbf{u})
\,.
\]
Also setting $\bar{D}=1$ in the density fluctuation gives,
\[
D^{\,\prime}\Big|_{\bar{D}=1}
= -\,
{\rm div}\xi
\,.
\]
Taking the divergence of the $\mathbf{u}^{\,\prime}$ equation
(\ref{u-prime}) then yields
\[
{\rm div}\mathbf{u}^{\,\prime}
=
0
=
\partial_t\,({\rm div}\xi)
+
\bar\mathbf{u}\cdot\nabla({\rm div}\xi)
\,.
\]
So ${\rm div}\,\xi=0$ is preserved, which means we may choose
{\bf initial conditions} so that $D^{\,\prime}=0$.  Thus,
$D^{\,\prime}$ vanishes (after taking variations) in the
incompressible case, upon invoking the preserved initial conditions
$\bar{D}=1$ and div$\,\xi=0$.

Upon setting $D^{\,\prime}=0$ in the formulas for the
incompressible case, the mean momentum may be expressed equivalently
as 
%
\boxeq6{
\begin{eqnarray}\label{glm-mom-incomp}
\bar\mathbf{m}
=
\frac{\delta \overline{\ell}}{\delta \bar\mathbf{u}}
\Big|_{\bar{D}=1}
&=&
\bar\mathbf{u}
+
\overline{
{\rm ad}^*_\xi\mathbf{u}^{\,\prime} 
}
=
\bar\mathbf{u}
+
\overline{
\pounds_\xi(\mathbf{u}^{\,\prime} )^{\,\flat}
}
\nonumber\\
&=&
\bar\mathbf{u}
+
\overline{\xi\cdot\nabla\mathbf{u}^{\,\prime}}
+
\overline{{u}^{\,\prime}_j\nabla\xi^j}
\nonumber\\
&=&
\bar\mathbf{u}
+
\bar\mathbf{u}^S
-
\bar\mathbf{p}
\nonumber\\
&=&
\bar\mathbf{u}
-\,
\overline{\xi\times
{\rm curl}\,\mathbf{u}^{\,\prime}}
+
\nabla\overline{(\xi\cdot\mathbf{u}^{\,\prime}\,)}
\,.
\end{eqnarray}
}
Here the quantities  $\bar\mathbf{u}^S$ and $\bar\mathbf{p}$ are the
same as in the GLM theory, when rotation is absent.


\subsubsection*{Simplifications in the $g\ell{m}$ Lagrangian for
incompressible mean flow} 

For incompressible mean flow, the second order Eulerian mean
$g\ell{m}$ Lagrangian (\ref{lag-bar}) reduces to 
\begin{equation}\label{lag-bar-incomp}
\bar\ell(\bar\mathbf{u},\bar{D})
=
\int \Big[
\tfrac{1}{2}\bar{D}
\Big(
|\bar\mathbf{u}|^2
+
\overline{|\mathbf{u}^{\,\prime}|^2}
\Big)
+
\bar{p}(1 - \bar{D})
\Big]\,d^3x
\,.
\end{equation}
Here we have used the GLM density equation (\ref{GLM-density}) and
truncated at second order, by enforcing $\bar{D}=1$ with the pressure
constraint. As a result, the Eulerian mean velocity $\bar\mathbf{u}$
satisfying the usual continuity equation (\ref{usual-cont}) is
incompressible. Thus, in the Lagrangian $\bar\ell$ in this case, the
fluctuations contribute only to the mean $g\ell{m}$ kinetic energy
and the Eulerian mean flow is incompressible.


\subsection{$g\ell{m}$ results arising in the EP
framework}


\subsection*{The motion equation for barotropic $g\ell{m}$ } 

For the $g\ell{m}$ theory in which
$\bar\ell \equiv \int \bar{\cal
L}(\bar\mathbf{u}, \nabla\bar\mathbf{u}, \bar{D},
\nabla\bar{D}; \xi(\mathbf{x},t))\,d^3x$, the EP framework yields the
equations of motion,
\boxeq2{
\begin{equation}
\partial_t\,\bar{m}_i
+
\partial_j(\bar{m}_i\bar{u}^j)
+
\bar{m}_j\partial_i\bar{u}^j
=
\bar{D}
\frac{\partial}{\partial {x}^i}
\frac{\delta\bar\ell}{\delta\bar{D}}
\quad\hbox{and}\quad
\partial_t\,\bar{D}
+
{\rm div}\bar{D}\bar\mathbf{u}
=
0
\,.\nonumber
\end{equation}
}
%
Here the total mean momentum $\bar{m}_i$ for $g\ell{m}$ is
defined by
%
\begin{eqnarray}\label{glm-mom-def}
\bar{m}_i
&=&
\frac{\delta \bar\ell}{\delta \bar{u}^i}
=
\frac{\partial \bar{\cal L}}{\partial \bar{u}^i}
-
{\partial_k}\,
\frac{\partial \bar{\cal L}}{\partial \bar{u}^i_{,k}}
\nonumber\\
&=&
\bar{D}\big(
\bar{u}_i
+
\bar{u}^S_{i}
-
\bar{p}_i
\big)
+
\big(\overline{ {\rm ad}^*_\xi {D}^{\,\prime}\,\bar\mathbf{u} }\big)_i
%
\\
&=&
\bar{D}
\bar{u}_i
+
\overline{{D}^{\,\prime}\,{u}_i^{\,\prime}}
+
\big(
{\rm ad}^*_\xi
(
\bar{D}\mathbf{u}^{\,\prime}
+
{D}^{\,\prime}\,\bar\mathbf{u}
)
\big)_i
\,,
\nonumber
\end{eqnarray}
%
where $\bar{u}^S_{i}$ is the GLM Stokes correction and $\bar{p}_i$ is
the GLM pseudomomentum given earlier. The variational derivative with
respect to mean density is obtained from
\begin{equation}
\frac{\delta\bar\ell}{\delta\bar{D}}
=
\frac{\partial \bar{\cal L}}{\partial \bar{D}}
-
{\partial_k}\,
\frac{\partial \bar{\cal L}}{\partial \bar{D}_{,k}}
\,.\nonumber
\end{equation}

\paragraph{Remarks:}
\begin{description}
\item$\bullet\quad$
Note, the {\bf linear} fluctuation relations modify the
$g\ell{m}$ total mean momentum $\bar{m}_i$, which, however, 
appears in the {\bf nonlinearity} of the EP motion equation
for the $g\ell{m}$ theory.
\item$\bullet\quad$
The combination of {\it Lagrangian mean velocity} $\bar\mathbf{u}^L$ 
and pseudomomentum $\bar\mathbf{p}$ appearing as
$
\bar\mathbf{u}
+
\bar\mathbf{u}^S
-
\bar\mathbf{p}
=
\bar\mathbf{u}^L
-
\bar\mathbf{p}
$
in the total mean momentum for $g\ell{m}$ also 
appears in the same way in the GLM theory. Here, however,
the combination also adds to 
$\bar{D}^{-1}
\,
\overline{ {\rm ad}^*_\xi {D}^{\,\prime}\,\bar\mathbf{u} }
$.

\end{description}


\subsubsection*{Mean Kelvin circulation theorem for barotropic
$g\ell{m}$ }
\begin{equation}
\frac{d}{dt}\oint_{c(\bar\mathbf{u})}
\frac{1}{\bar{D}}
\frac{\delta \bar\ell}{\delta \bar\mathbf{u}}
=
\oint_{c(\bar\mathbf{u})}
\nabla
\frac{\delta \bar\ell}{\delta \bar{D}}
\cdot
d\mathbf{x}
=
0
\nonumber
\end{equation}
where for $g\ell{m}$ one has, in the compressible barotropic
case,
\begin{equation}
\frac{1}{\bar{D}}
\frac{\delta \bar\ell}{\delta \bar\mathbf{u}}
=
\bar\mathbf{u}
+
\bar\mathbf{u}^S
-
\bar\mathbf{p}
+
\frac{1}{\bar{D}}
\,
\overline{ {\rm ad}^*_\xi {D}^{\,\prime}\,\bar\mathbf{u} }
\,.
\nonumber
\end{equation}
In contrast, for the incompressible case, div$\bar\mathbf{u}=0$ and
one again sets $\bar{D}=1$ and ${D}^{\,\prime}=0$  in this formula
(after taking variations) thereby dropping the last term. Thus, in the
incompressible case, the contributions to the circulation integrands
are the {\bf same} for both $g\ell{m}$ and GLM theories. However, the
velocities of the fluid loops in the Kelvin circulation theorems are
{\bf different}. They are $\bar\mathbf{u}$ for $g\ell{m}$ and
$\bar\mathbf{u}^L$ for GLM.

\paragraph{Remarks:}
\begin{description}
\item$\bullet\quad$
When curl$\,(\bar\mathbf{u}^S - \bar\mathbf{p}
+
\bar{D}^{-1}
\,
\overline{ {\rm ad}^*_\xi {D}^{\,\prime}\,\bar\mathbf{u} }
)$ vanishes, this is the Charney-Drazin ``nonacceleration theorem''
for $g\ell{m}$ for barotropic compressible fluids. See
Andrews \& McIntyre [1978a] for their discussion of the GLM case.
\item$\bullet\quad$
The {\it Eulerian mean vorticity} due to the fluctuations in the
{\it incompressible} case is
\begin{eqnarray}
{\rm curl}\,(\bar\mathbf{u}^S-\bar\mathbf{p})
&=&
-\,{\rm curl}\,\overline{(\xi\times\omega^{\,\prime\,})}
\,,\quad\hbox{where}\quad
\omega^{\,\prime}
=
{\rm curl}\,\mathbf{u}^{\,\prime}
\nonumber\\
&=&
\overline{\xi\cdot\nabla\omega^{\,\prime}}
-
\overline{\omega^{\,\prime}\cdot\nabla\xi}
\nonumber\\
&=&
\overline{{\rm ad}_\xi\,\omega^{\,\prime} }
\,.
\end{eqnarray}
For potential fluctuations, one sets $\omega^{\,\prime}=0$.

\end{description}


\subsubsection*{Momentum balance for barotropic $g\ell{m}$ }

For a mean Lagrangian density $\bar{\cal L}$, the EP theory yields
the momentum balance,
\begin{equation}
{\partial_t}\,\bar{m}_i
+
{\partial_j}\,\bar{T}_i^j
=
\frac{\partial\bar{\cal L}}{\partial x^i}\bigg|_{exp}
\,,
\nonumber
\end{equation}
%
where the total mean momentum $\bar{m}_i$ for
barotropic $g\ell{m}$ is evaluated in equation
(\ref{glm-mom-def}). The stress tensor is defined in the EP theory
for this class of Lagrangians as 
%
\begin{equation}
\hspace{-1mm}
\bar{T}_i^j
=
\bar{m}_i\bar{u}^j
+
\delta_i^j\Big(
\bar{\cal L} 
- 
\bar{D}\frac{\partial \bar{\cal L}}{\partial \bar{D}}
+
\bar{D}{\partial_k}\,
\frac{\partial \bar{\cal L}}{\partial \bar{D}_{,k}}
\Big)
-
\bar{u}^k_{,i}\,
\frac{\partial \bar{\cal L}}{\partial \bar{u}^k_{,j}}
-
\bar{D}_{,i}\,
\frac{\partial \bar{\cal L}}{\partial \bar{D}_{,j}}
\,.
\end{equation}
%
Explicitly evaluating the partial derivatives
of $\bar\ell$ for $g\ell{m}$ gives, 
%
\boxeq7{
\begin{eqnarray}\label{glm-stress-tensor-comp}
\bar{T}_i^j
&=&
\bar{m}_i\bar{u}^j
+
\delta_i^j\Big[
p(\bar{D}) 
+ 
\overline{D^{\,\prime}\mathbf{u}^{\,\prime}}\cdot\bar\mathbf{u}
-\,
\frac{c^2(\bar{D})}{\bar{D}}
\,
\overline{D^{\,\prime\,2}}
\Big]
\\
&&\hspace{-15mm}
+\
\bar{D}
\delta_i^j
\Big[
-
\overline{\xi\cdot\nabla(\mathbf{u}^{\,\prime}\cdot\bar\mathbf{u})}
+\
\overline{D^{\,\prime\,2}}
\,
\frac{\partial}{\partial\bar{D}}
\Big(\frac{c^2(\bar{D})}{2\bar{D}}\Big)
+
\overline{\xi\cdot\nabla
\Big(
D^{\,\prime}
\,
\frac{c^2(\bar{D})}{\bar{D}}
\Big)}\
\Big]
\nonumber
\\
&&
+\
\bar{u}^k_{,i}\,
\overline{(D^{\,\prime}\bar{u}_k+\bar{D}u_k^{\,\prime})\xi^j}
+\
\bar{D}_{,i}\,\Big(
\overline{\xi^j\mathbf{u}^{\,\prime}}\cdot\bar\mathbf{u}
-\,
\frac{c^2(\bar{D})}{\bar{D}}
\,
\overline{D^{\,\prime}\xi^j}
\Big)
\,.\nonumber
\end{eqnarray}
}

\noindent
The momentum balance law is specified, only
after ${\partial\bar{\cal L}/\partial x^i}\big|_{exp}$ is  known.
This requires specifying the explicit spatial dependence in
(\ref{lag-bar}) of the wave properties and external potential in the
Lagrangian density $\bar{\cal L}$ for $g\ell{m}$. 

\begin{quote}
{\bf Remark.}
Thus, the form of the theory is fixed -- it is the
EP theory. However, its manifestations and channels for expressing
energy exchange are many. Even in the barotropic case, for example,
there are many different contributions to the stress tensor from the
Lagrangian fluctuations $\xi$. These contributions are
primarily isotropic, including the term 
$\partial\bar{\cal L}/\partial x^i\big|_{exp}$.
\end{quote}


\subsubsection*{Energy balance for barotropic $g\ell{m}$ }

A Legendre transformation gives the energy 
quantity for the $g\ell{m}$ fluid flow, namely,
\begin{eqnarray}
\bar{E}
&=&
\Big\langle{\frac{\delta \bar\ell}{\delta \bar\mathbf{u}}
\,\cdot\,\bar\mathbf{u}}\Big\rangle
-
\bar\ell(\bar\mathbf{u},\bar{D},\xi)
\nonumber
\\
&=&
\int \Big[
\tfrac{1}{2}\,\bar{D}|\bar\mathbf{u}|^2
+
\bar{D}e(\bar{D})
+
\tfrac{1}{2}\,\bar{D}
\overline{|\mathbf{u}^{\,\prime}|^2 }
+
\overline{D^{\,\prime}\mathbf{u}^{\,\prime}}
\cdot\bar\mathbf{u}
+
\frac{c^2(\bar{D})}{2\bar{D}}
\,\overline{D^{\,\prime\,2}}
\Big]
d^3x
\nonumber
\\
&&
-
\int \Big(
\bar{D}\overline{|\mathbf{u}^{\,\prime}|^2}
+
\overline{D^{\,\prime}\mathbf{u}^{\,\prime}}
\cdot\bar\mathbf{u}
\
-\
\big(\,
\overline{{\rm ad}^*_\xi
(
\bar{D}\mathbf{u}^{\,\prime}
+
{D}^{\,\prime}\,\bar\mathbf{u}
)}\,
\big)\cdot\bar\mathbf{u}
\Big)
d^3x
\nonumber
\\
&=&
\Big[
E(\bar\mathbf{u},\bar{D})
\
+\
\tfrac{1}{2}
\overline{E^{\,\prime\,\prime}}\
\Big]\
-\
\int \overline{\bigg(\
\frac{\delta\bar{\ell}}{\delta(\partial_t\,\xi)}
\cdot\partial_t\,\xi\
\bigg)}d^3x
\,.
\nonumber
\end{eqnarray}
We recognize the last integral term as
$\int\overline{\pi\cdot\partial_t\,\xi}\,d^3x$, 
the total ``pseudoenergy'' for the $g\ell{m}$ theory. 
Hence, just as for the GLM theory, but now with
correspondingly different definitions of terms, we find that
$d\bar{E}/dt =
-\,\frac{d}{dt}\int\overline{\pi\cdot\partial_t\,\xi}\,d^3x$,
in $g\ell{m}$ theory, since the mean total energy must be
conserved for a self-consistently coupled theory. 

\begin{quote}
{\bf Remark.}
The quantity
$\tfrac{1}{2}\overline{E^{\,\prime\,\prime}}$ is the same as the
approximately conserved expression from acoustics due to Blokhintsev
[1945],
\begin{equation}\label{Blok-erg}
\tfrac{1}{2}\overline{E^{\,\prime\,\prime}}
=
\int \Big[
\tfrac{1}{2}\,\bar{D}
\overline{|\mathbf{u}^{\,\prime}|^2 }
+
\overline{D^{\,\prime}\mathbf{u}^{\,\prime}}
\cdot\bar\mathbf{u}
+
\frac{c^2(\bar{D})}{2\bar{D}}
\,\overline{D^{\,\prime\,2}}
\Big]
d^3x
\nonumber
\end{equation}
as discussed in Andrews \& McIntyre [1978b]. Of course, this
quantity is not the pseudoenergy for barotropic $g\ell{m}$
theory.
\end{quote}


\subsection{Remarks about $g\ell{m}$ closure and rapid distortion
theory}

\begin{description}

\item$\quad\bullet$
The $g\ell{m}$ theory linearizes the
$\mathbf{u}^{\,\prime}$ equation, so it neglects the nonlinear
term 
$
{\rm div}\big(
\mathbf{u}^{\,\prime}\mathbf{u}^{\,\prime}
-
\overline{\mathbf{u}^{\,\prime}\mathbf{u}^{\,\prime}}\big)
$
that appears in the $\mathbf{u}^{\,\prime}$ equation for Reynolds
turbulence closure in the Eulerian mean setting.

\item$\quad\bullet$
Based on ideas from Lagrangian stability analysis
and closely related to ideas from WMFI theory, the $g\ell{m}$
equations are also related to ideas from {\bf rapid
distortion theory}. See Cambon and Scott [1999] for an interesting
discussion of the close connections between rapid distortion theory
and  WKB stability theory. Some of the relations between WKB
stability and WMFI theory are also briefly discussed in the EP
framework in Appendix \#3 (section \ref{WMFI-appendix}).

\item$\quad\bullet$
In principle, the Lagrangian statistics for the coefficients
in the nonlinear $g\ell{m}$ equations may be closed at second moments,
since the fluctuations $\xi$ and $\mathbf{u}^{\,\prime}$ are both
taken to satisfy {\it linear} equations. Thus, the linearity of the
$g\ell{m}$ equations would allow one to derive a set of equations for
second moments such as
$\overline{\xi\times{\rm curl}\,\mathbf{u}^{\,\prime}}\,$ in the
incompressible case and treat the combined system for the motion and
the Lagrangian statistics as an initial value problem. This could be
done by computing 
$\partial_t(\overline{\xi\times{\rm curl}\,\mathbf{u}^{\,\prime}})\,$
using the linearized Euler motion equation for evolving 
$\mathbf{u}^{\,\prime}$ and using the $\mathbf{u}^{\,\prime}-$equation
for the evolution of $\xi$. 

Of course, such a linear closure would not produce only linear
effects in the mean motion equation. The $g\ell{m}$ effects arise
from second moments. The $g\ell{m}$ effects appear multiplicatively 
in the stress tensor and additively in the definition of the total
mean momentum. The latter appears also in the nonlinearity of the
$g\ell{m}$ equations. Thus, although the the mean advection relations
are enforced only at {\it linear} order, the contributions of the
fluctuations to the
$g\ell{m}$ motion equation are both linear and nonlinear.

\item$\quad\bullet$
We note that the $g\ell{m}$ theory expresses wave properties in
terms of Lagrangian displacement statistics and gradients of mean
flow properties. This idea suggests we may consider substituting
aspects of these relations between wave properties and mean
gradients, before taking variations in Hamilton's principle, by
regarding these wave, mean flow relations as a type of {\bf Taylor
hypothesis}. We shall follow this idea further in section
\ref{alpha}. 

\item$\quad\bullet$
The Green's function relation $\xi=G*\mathbf{u}^{\,\prime}$
implies an explicit DuHamel formula
(with its memory, or history dependence) and suggests that
$\overline{\xi\xi}$ might be usefully computed as a {\bf diagnostic}
in direct numerical simulations. We shall follow this suggestion in
the next subsection.

\end{description}


\subsection{
Determining Lagrangian fluctuation statistics in DNS}

In assessing and benchmarking the $g\ell{m}$ model,  one may
obtain $\overline{\xi\xi}$ by measuring $\mathbf{u}^{\,\prime}$ and
$D^{\,\prime}$ and then {\bf inverting} their Eulerian relations with 
displacement fluctuation $\xi$ in direct numerical
simulations. For example, in compressible simulations, the density 
fluctuations satisfy equation (\ref{D-prime})
\begin{equation}
D^{\,\prime}(\mathbf{x},t)
=-\,
{\rm div}(\bar{D}\xi)
\,.\nonumber
\end{equation}
The Helmholtz decomposition
$\bar{D}\xi=\nabla\phi+{\rm curl}\,\mathbf{A}$ then implies the
curl-free part of $\bar{D}\xi$ as
\begin{equation}
\bar{D}\xi
=
-\,\nabla\Delta^{-1} D^{\,\prime}
\,.
\nonumber
\end{equation}
The divergence-free part of $\bar{D}\xi$ (the homogeneous
solution) is preserved and, so, it may be set to zero as an
initialization condition. For the velocity fluctuations, we have the
standard linearized relation (\ref{u-prime})
\begin{equation}
\mathbf{u}^{\,\prime}(\mathbf{x},t)
=
\frac{\partial {\xi}}{\partial t}
+
\bar\mathbf{u}\cdot\nabla\xi 
- 
\xi\cdot\nabla \bar\mathbf{u}
\,,\nonumber
\end{equation}
whose inhomogeneous solution is found using the Green's function
inversion, 
\[
\xi=G*\mathbf{u}^{\,\prime}
\,.
\]
As discussed earlier in section \ref{Lin-fluct-rel}, this also has a
homogeneous solution satisfying 
$d\xi^{(0)}/dt=\xi^{(0)}\cdot\nabla\bar\mathbf{u}$. When this
solution is initialized at zero, the remaining inhomogeneous
part of the solution for $\xi$ may be determined by the Green's
function method. Comparing the solutions for $\xi$ that are found from
the $D^{\,\prime}-$ and $\mathbf{u}^{\,\prime}-$equations may also be
used as a check of the accuracy of the method, provided the numerical
scheme does not unduly excite the homogeneous solutions. If this
excitation does occur, then perhaps an occasional re-initialization
may be required to suppress the homogeneous solutions for the
Lagrangian fluctuation,
$\xi$. This idea is reminiscent of the ``slow manifold'' concept,
introduced in Leith [1980] for atmospheric dynamics, which 
initializes gravity wave amplitudes to zero. These gravity waves may
need to be periodically reset to zero, especially when updating the
forecast with new data. However, the homogeneous solutions in the
present case do not grow unstably as the gravity waves do in
atmospheric dynamics, because these homogeneous solutions are not
unstable. In fact, being frozen into the motion, they cannot grow any
larger than their initial values.

In principle, performing this inversion would give us an Eulerian
diagnostic for determining the Lagrangian fluctuation statistics and
testing the basis for the derivation of the $g\ell{m}$
equations, as outlined below. 


\subsection{Protocol for using DNS to diagnose  
Lagrangian statistics input for $g\ell{m}$ 
in barotropic compressible fluid dynamics}

For using DNS to diagnose the  Lagrangian fluctuation statistics of
$g\ell{m}$ in barotropic compressible fluid dynamics (or,
shallow water dynamics), one could proceed as follows.
\begin{enumerate}
\item 
Measure $\bar\mathbf{u}$, $\mathbf{u}^{\,\prime}$, 
$\bar{D}$, $D^{\,\prime}$, with 
$\overline{D^{\,\prime}}=0=\overline{\mathbf{u}^{\,\prime}}$
in the DNS.
\item 
Recall 
$\mathbf{u}^{\,\prime}=\partial_t\,\xi-{\rm ad}_{\bar\mathbf{u}}\xi$, so
(componentwise)
\[
\xi_i=(G*\mathbf{u}^{\,\prime\,})_i
=
\int G_i(\mathbf{x}-\mathbf{y},t-\tau)
{u}_i^{\,\prime}(\mathbf{y},\tau)\,d^3y\,d\tau
\,,\]
with a {\bf deterministic} vector Green's function, whose components
$G_i$, $i=1,2,3$, each satisfy
\[
\partial_t\, G_i+({\rm ad}^*_{\bar\mathbf{u}} G)_i
=
\delta(\mathbf{x}-\mathbf{y})\delta(t-\tau)
\,.\]
\item 
In a local spatial domain, compute $G$,
then compute $\xi=G*\mathbf{u}^{\,\prime}$.
\item 
Compute $D^{\,\prime}=-{\rm div}(\bar{D}\xi)$ from the result and
compare with the measured $D^{\,\prime}$. This checks the
($\tilde{D}-$weighted)  curl-free part of the first computation of
$\xi$ against the DNS ``measurement.''
\item 
Compute $\bar{D}\xi=-\,\nabla\Delta^{-1}
D^{\,\prime}$.  This checks the two
computations of $\xi$ against each other.
\item 
The result $\bar{D}\xi=-\,\nabla\Delta^{-1}
D^{\,\prime}$ should be curl-free. Check this, by computing
curl$\bar{D}\xi$.
\item 
Compute $\overline{\xi\xi}(\mathbf{x},t)$.
Determine the spatial and temporal dependence of its isotropic and
anisotropic components.
\item 
Compute $\overline{\xi\mathbf{u}^{\,\prime}}(\mathbf{x},t)$.
This determines the relative dispersion tensor for Lagrangian
trajectories.
\item 
Compute 
$\overline{\mathbf{m}^{\,\prime\prime}}
=
\overline{D^{\,\prime}\mathbf{u}^{\,\prime}}
+
\overline{
{\rm ad}^*_\xi
(\bar{D}\mathbf{u}^{\,\prime} 
+ 
D^{\,\prime}\,\bar\mathbf{u})
}
$,
the mean momentum due to the fluctuations.
\item 
Compare 
${\bar\mathbf{m}}_{fluct}$
with $-\alpha^2\Delta\bar\mathbf{u}$. (This
would extract $\alpha^2$ in the isotropic case.
The original derivation of the alpha model 
in Holm, Marsden \& Ratiu [1998a,b] for the incompressible case 
assumes that $\overline{\xi\xi}$ is homogeneous and isotropic. See
section \ref{alpha} for more discussion of the alpha models.)
\end{enumerate}


\subsection{EP $g\ell{m}$ equations for incompressible mean flow}
\label{EP-inc-sec}

The variational derivatives of the $g\ell{m}$ Lagrangian
(\ref{lag-bar-incomp}) for incompressible flow
%
\begin{equation}\label{lag-bar-incomp-again}
\bar\ell(\bar\mathbf{u},\bar{D})
=
\int \Big[
\tfrac{1}{2}\bar{D}
\Big(
|\bar\mathbf{u}|^2
+
\overline{|\mathbf{u}^{\,\prime}|^2}
\Big)
+
\bar{p}(1 - \bar{D})
\Big]\,d^3x
\,,
\end{equation}
%
are given by, cf. equation (\ref{glm-mom-incomp}),
%
\begin{eqnarray}\label{lag-var-der-incomp}
\delta\bar\ell(\bar\mathbf{u},\bar{D})
&=&
\int \Big[
\delta\bar{D}
\Big(\tfrac{1}{2}\big(\,
|\bar\mathbf{u}|^2
+
\overline{|\mathbf{u}^{\,\prime}|^2}\,\big)
-
\bar{p}
\Big)
+
\delta\bar{p}\,(1 - \bar{D})
\nonumber\\
&&\hspace{-1cm}
+\
\bar{D}\,\delta\bar\mathbf{u}
\cdot
\Big(
\bar\mathbf{u}
-\,
\overline{\xi\times
{\rm curl}\,\mathbf{u}^{\,\prime}}
+
\nabla\overline{(\xi\cdot\mathbf{u}^{\,\prime}\,)}
\Big)
+
\delta\bar\mathbf{u}
\cdot
\overline{\mathbf{u}^{\,\prime}\,{\rm div}\,(\bar{D}\xi)}
\Big]\,d^3x
\,.
\nonumber
\end{eqnarray}
%
We define the $g\ell{m}$ circulation velocity as,
%
\begin{equation}\label{glm-vee-def}
\bar\mathbf{v}
\equiv
\bar\mathbf{u}
-\,
\overline{\xi\times
{\rm curl}\,\mathbf{u}^{\,\prime}}
+
\nabla\overline{(\xi\cdot\mathbf{u}^{\,\prime}\,)}
\,.\nonumber
\end{equation}
%
The corresponding EP motion equation (with
$\nabla\cdot\bar\mathbf{u}=0$) is expressed as,
%
\boxeq2{
\begin{equation}\label{glm-incomp1}
\frac{ \partial}{\partial t}\bar\mathbf{v}
+
\bar\mathbf{u}\cdot\nabla\bar\mathbf{v}
+
\bar{v}_j\nabla \bar{u}^j
+
\nabla\bar{p} = 0
\,.
\nonumber
\end{equation}
}
%
This is the EP equation for the Lagrangian
(\ref{lag-bar-incomp-again}). It also has the equivalent form,
%
\begin{equation}\label{glm-incomp2}
\frac{ \partial}{\partial t}\bar\mathbf{v}
-
\bar\mathbf{u}\times{\rm curl}\,\bar\mathbf{v}
+
\nabla\big(\bar\mathbf{v}\cdot\bar\mathbf{u} + \bar{p}\,\big)
 = 0
\,.
\nonumber
\end{equation}

The Kelvin circulation theorem for the incompressible
$g\ell{m}$ equations is simply,
\begin{equation}\label{glm-kel-circ}
\frac{d}{dt}\oint_{c(\bar\mathbf{u})}
\big(\bar\mathbf{u}
-\,
\overline{\xi\times
{\rm curl}\,\mathbf{u}^{\,\prime}}
\,\big)
\cdot
d\mathbf{x}
=
0
\,.
\end{equation}
%
\begin{quote}
{\bf Remark.}
We recall that 
$\mathbf{u}^{\,\prime}
=
\partial_t\,\xi+\bar\mathbf{u}\cdot\nabla\xi 
- \xi\cdot\nabla\bar\mathbf{u}$.
For the case that 
$\nabla\cdot\bar\mathbf{u}=0$
and $\nabla\cdot\xi=0$,
this becomes
$\mathbf{u}^{\,\prime}
=
\partial_t\,\xi
-
{\rm curl}\,\big(\bar\mathbf{u}\times\xi\big).
$
Thus, to {\bf close} the $g\ell{m}$ EP motion equation for
incompressible Eulerian mean flow, only {\bf one key element} from
the Lagrangian statistics must be specified. Namely, the quantity
\begin{equation}\label{glm-key-stat}\hspace{-3mm}
\overline{\xi\times\omega^{\,\prime}}
=
\overline{\xi\times
{\rm curl}\,\mathbf{u}^{\,\prime}}
=
\overline{\xi\times
{\rm curl}\,
\big(\partial_t\,\xi
-
{\rm curl}\,\big(\bar\mathbf{u}\times\xi\big)
\big)
}
\,,\!\!\!\!
\end{equation}
must be specified in terms of $\bar\mathbf{u}$,
$\nabla\bar\mathbf{u}$ and $\nabla\nabla\bar\mathbf{u}$.
This specification is one of the objectives of the discussions in
the next section.
\end{quote}


\section{Alpha models}
\label{alpha}


\subsection{Opening remarks}

We have seen that the use of Taylor expansions in the linearized
fluctuation relations summons gradients of Eulerian mean fluid
quantities into the mean second-variation Lagrangian. In turn, these
gradients summon {\bf second-order} spatial derivatives such as
$\nabla\nabla\bar\mathbf{u}$ into the $g\ell{m}$ motion equation
that results from the EP variational principle.

Among other things, the $g\ell{m}$ stress tensor
(\ref{glm-stress-tensor-comp}) for a compressible fluid shows the
variety of channels available for energy exchange to occur. These
channels arise through the various combinations of Eulerian mean
gradients that appear in the stress tensor for the
$g\ell{m}$ theory. The incompressible case is more
straightforward because it has fewer such channels. However, to
achieve closure, even the incompressible $g\ell{m}$ case still
requires an assumption to express the key element of the Lagrangian
statistics (\ref{glm-key-stat}) in terms of
$\bar\mathbf{u}$, $\nabla\bar\mathbf{u}$ and 
$\nabla\nabla\bar\mathbf{u}$. The linearized
fluctuation equations themselves (relating the Eulerian and
Lagrangian small disturbances) shall guide the formulation of such
approximate closure assumptions. 

\paragraph{Approach.}
The approach to the alpha-model equations is closely related to
the $g\ell{m}$ approach, but with one important difference.
Namely, the order is interchanged in the steps of making
approximations and varying the EP Lagrangian in Hamilton's principle.
        \begin{description}
\item$\qquad$\underline{To obtain the $g\ell{m}$ equations:} 
              \\We Expanded the Lagrangian, Took its Eulerian mean,
              then 
              \\Varied to obtain the equations of motion, 
              and \underline{finally} saw the need to Approximate the
              closure. This could be done, in principle, by using the
              $\mathbf{u}^{\,\prime}-$equation for
              the tendency of $\xi$
              and the linearization of the
              GLM equations for the tendency of 
              $\mathbf{u}^{\,\prime}$.
              We shall discuss a more direct approach to closure.
\item$\qquad$\underline{To obtain the $\alpha-$models:} 
              \\We shall Expand the Lagrangian, Take its Eulerian
              mean, 
              \\Approximate in the Lagrangian (by taking a 
              particular solution of
              the $\mathbf{u}^{\,\prime}-$equation as a {\bf Taylor
              hypothesis}),
              and then 
              \\Vary to find a closed set of EP motion equations.
        \end{description}

\begin{quote}
{\bf Remark.}
Because of the close relation between the approaches
used in deriving these two sets of equations, one might hope for a
bridge between them. For example, the $g\ell{m}$ equations could
potentially provide an Eulerian diagnostic for determining parameters
in the alpha model from DNS of the full Euler equations (or
Navier-Stokes equations). The $g\ell{m}$ equations form a
systematic approximation for the original GLM equations, within the EP
framework. Thus, perhaps the GLM equations could be used to help
answer questions that may arise at the other levels of approximation
in this framework, particularly, in the alpha models.
\end{quote}


\subsection{Taylor hypothesis closure (THC) approach}

We shall use partial, or particular, solutions of the
linearized velocity fluctuation equation (\ref{u-prime})
\[
\mathbf{u}^{\,\prime}
=\partial_t\,\xi 
+
\bar\mathbf{u}\cdot\nabla\xi
-
\xi\cdot\nabla\bar{\mathbf{u}}
\,,
\]
to guide certain choices of {\bf Taylor hypotheses}. The three Taylor
hypotheses we shall discuss for the $\mathbf{u}^{\,\prime}-$equation
are:%
\footnote{These three Taylor hypothesis closures are for the
$g\ell{m}$ equations. Later, we shall mention THC\#4 -- for the GLM
equations.}
%
\begin{description}

\item{\bf THC\#1}
Neglect space and time derivatives of $\xi$, or, set
$\partial_t\,\xi + \bar\mathbf{u}\cdot\nabla\xi = 0.$ In the
incompressible case, this yields the original Euler-alpha model of
Holm, Marsden \& Ratiu [1998a,b], in which one assumes
$\mathbf{u}^{\,\prime}=-\,\xi\cdot\nabla\bar\mathbf{u}$.

\item{\bf THC\#2}
Assume that $\xi$ is frozen-in as a one-form.  In the incompressible
case, this yields the anisotropic alpha model of Marsden
\& Shkoller [2001], with 
$\mathbf{u}^{\,\prime}
=-
2\xi\cdot\bar\mathsf{e}$, where 
$\bar\mathsf{e} = \tfrac{1}{2}(\nabla\bar\mathbf{u}
+\nabla\bar\mathbf{u}^T)$ is the mean strain rate tensor, and
$\partial_t\,\xi 
+
\bar\mathbf{u}\cdot\nabla\xi
=
-
\nabla\bar\mathbf{u}^T\cdot\xi$.

\item{\bf THC\#3}
Assume that $\xi$ is frozen-in as a two-form in three dimensions.
Hence, set
$\partial_t\,\xi  +
\bar\mathbf{u}\cdot\nabla\xi
=
\xi\cdot\nabla\bar\mathbf{u}
-
\xi\,{\rm div}\, \bar\mathbf{u}$.
This assumption implies
$\mathbf{u}^{\,\prime}
=
-\
\xi\,{\rm div}\,\bar\mathbf{u}$, 
which, of course, is only interesting in the compressible case. 
For compressible flows, this choice lead to similarities with the
Green-Naghdi equation for shallow water dynamics. The Green-Naghdi
equation is discussed in Green \& Naghdi [1976]. See, e.g.,
Camassa, Holm \& Levermore [1997] for more references and an
asymptotic treatment of these equations.
\end{description}
\begin{quote}
{\bf Remark about algebraic closure.}
Neglecting the partial time derivative in the linearized velocity
fluctuation equation for $\mathbf{u}^{\,\prime}$ leads to an
``algebraic closure relation,'' expressed as,
\[
\mathbf{u}^{\,\prime}
=
\bar\mathbf{u}\cdot\nabla\xi
-
\xi\cdot\nabla\bar{\mathbf{u}}
=
{\rm curl}\,(\xi\times\bar\mathbf{u})
\,,
\]
in the incompressible case.
We note that substituting this algebraic closure relation into the
$g\ell{m}$ Lagrangian (\ref{lag-bar}) yields the following
contribution to the mean fluctuational momentum, 
\begin{equation}\label{algebraic-close-hypoth}
\overline{\mathbf{m}^{\,\prime\prime}}
=
\frac{\delta}{\delta\bar\mathbf{u}}\int 
\tfrac{1}{2}\,\overline{|\mathbf{u}^{\,\prime}|^2}
\,d^3x
=
-\
\overline{\xi\times{\rm curl}\,
{\rm curl}\,(\xi\times\bar\mathbf{u})}
\,.
\nonumber
\end{equation}
So, in this incompressible case, the contribution of the fluctuations
to the mean total momentum (or pseudomomentum) keeps the same
$g\ell{m}$ form as in equations (\ref{glm-kel-circ}) and
(\ref{glm-key-stat}). However, this expression is not closed, because
it still remains to specify the evolution of the Lagrangian
statistics. This could perhaps be accomplished by imposing the
algebraic closure relation as a constraint in the Lagrangian.
The problem does not arise in using the Taylor hypotheses
THC\#1--\#3, above.

\end{quote}

The Taylor hypotheses THC\#1--\#3 are approximate relations
between Eulerian and Lagrangian statistics (namely, they are
relations for $\mathbf{u}^{\,\prime}$ as a function of $\xi$ and its
derivatives) that yield closures when substituted into the averaged
Lagrangian in Hamilton's principle. We shall first discuss the
incompressible case, which is simpler, and then we shall discuss the
barotropic compressible case.

In both cases we shall illustrate the {\bf Taylor hypothesis closure
technique} by substituting the first of these three Taylor hypotheses
into the $g\ell{m}$ Lagrangian, before taking its variations. This
approach results via the EP framework in closed equations based on the
$g\ell{m}$ equations that retain their Kelvin circulation theorem and
conservation properties. Among these closed equations for
the incompressible case are variants of the Euler-alpha model (or,
averaged Euler equations) that are also related to the theory of
second grade fluids and have been discussed as potential turbulence
closure models when Navier-Stokes viscous dissipation is introduced,
as in Chen et al. [1998], [1999a,b,c]. We shall show how this
approach also leads to a new generalization of the Euler-alpha model
that includes compressibility.

\subsection{A brief history of the alpha models}

The Euler-alpha equations for averaged incompressible ideal
fluid motion were first derived in Holm, Marsden \& Ratiu [1998a] in
the context of the Euler-Poincar\'e theory for fluid dynamics. That
derivation proceeded essentially by choosing the kinetic energy to be
the $H_1$ norm of the Eulerian fluid velocity, rather than the usual
$L_2$ norm. This choice generalized the unidirectional shallow water
equation of Camassa and Holm [1993] from one dimension to three
dimensions. The resulting n-dimensional Euler-alpha equation is (with
$\nabla\cdot\mathbf{u}=0$, 
$\mathbf{v}\equiv\mathbf{u}-\alpha^2\Delta\mathbf{u}$ and
constant length scale $\alpha$)
\begin{equation}\label{alpha-eqn}
\frac{ \partial}{\partial t}\mathbf{v}
+
\mathbf{u}\cdot\nabla\mathbf{v}
+
v_j\nabla u^j
+
\nabla{p} = 0
\,.
\end{equation}
This is the EP equation for the Lagrangian
\begin{equation}\label{Lag-alpha-incomp}
\ell=\tfrac{1}{2}\int
|\mathbf{u}|^2
+
\alpha^2|\nabla\mathbf{u}|^2
\
d^3x
\,,
\end{equation}
for a constant $\alpha$ and divergenceless fluid velocity
$\mathbf{u}$. Mathematically, this equation describes geodesic
motion on the volume-preserving diffeomorphism group of
$\mathbb{R}^3$ relative to the $H_1$ norm in a sense similar to the
work of Arnold [1966] and Ebin and Marsden [1970] in which the Euler
equations are shown to describe geodesic motion on the same
diffeomorphism group relative to the $L_2$ kinetic energy norm. 

Remarkably, the $H_1-$geodesic Euler-alpha equation was later
recognized as being identical to the well-known invisicid second
grade fluid equations introduced by Rivlin and Ericksen [1955],
although of course these equations were derived from a completely
different viewpoint. The differences in their derivations imply
corresponding differences in the interpretations of the solutions of
these equations in each of their contexts. In particular, the
constant parameter alpha (a length scale) is interpreted differently
in the two theories. In the Euler-alpha model, the parameter
alpha is associated with the flow regime and, in numerical
simulations, {\bf alpha separates active and passive degrees of
freedom}, as shown in Chen et al. [1999c]. (Physically, alpha is the
smallest active length scale participating in the nonlinear
interactions --  so scales smaller than alpha are swept along by the
larger ones.) In contrast, for the theory of second grade fluids,
alpha is a thermodynamic material parameter. 

\paragraph{Extensions of Euler$-\alpha$.} The works of Marsden \&
Shkoller [2001] and Marsden, Ratiu \& Shkoller [2001] used the EP
framework to introduce a certain type of filtering -- called
``fuzzying'' -- into the Lagrangian. Applying the EP reduction
theorem to the Lagrangian for ``fuzzy flow'' yielded an
alternative formal derivation of the incompressible Euler$-\alpha$
model, as well as an anisotropic variant of it and the extension of
that variant to Riemannian manifolds. These works also showed short
time existence for solutions of the Euler$-\alpha$ model and these 
extensions, by establishing for it the analog of the Ebin-Marsden
theorem for the incompressible Euler equations, proven in Ebin and
Marsden [1970]. See also Shkoller [1998] for the corresponding
existence result for the original Euler-alpha model of Holm, Marsden
\& Ratiu [1998a]. 

While these references do make use of the EP reduction theorem, they
do not show that the reduced equations obtained from it would also
result from applying the ``fuzzy flow'' averaging method directly to
Euler's equations. The EP Averaging Lemma {\it guarantees} this
result, however, when the GLM averaging method is applied.  For
example, to the extent that the fuzzy averaging used in Marsden and
Shkoller [2001] fails to possess the projection property, the
resulting EP equations will fail to coincide with fuzzy average of
the original equations. Investigations of the
relation between fuzzy averaging and GLM averaging are underway.

\paragraph{THC\#\,4, a nonlinear GLM Taylor hypothesis.} 

In Holm [1999], a nonlinear GLM Taylor hypothesis was introduced and
applied for both compressible and incompressible flows. This Taylor
hypothesis THC\#4 for GLM assumes that the Lagrangian displacement
fluctuation $\xi$ is frozen as a {\it Lagrangian vector field} into
the {\it nonlinear} GLM flow. Namely, (note $\bar\mathbf{u}^L$ rather
than
$\bar\mathbf{u}$)
\begin{equation}\label{THC4-def}
\mathbf{u}^\ell
\equiv
\partial_t\,\xi + \bar\mathbf{u}^L\cdot\nabla\xi
= 
\xi\cdot\nabla\bar\mathbf{u}^L
\,.
\end{equation}
This THC\#4 is substituted directly into the GLM averaged Lagrangian,
e.g., (\ref{Lag-mean-Lag}), or (\ref{Lag-mean-Bouss-Lag}), {\it
without} linearizing the fluctuation relations. One may then vary the
Lagrangian in the EP framework, {\it without} making the
small-amplitude approximation. This THC\#4 treats the fluctuating
Lagrangian displacement $\xi$ as a material property associated with
the frozen-in GLM motion of a ``cloud'' of fluid parcels initially
displaced from one another by the initial value of $\xi$, which is
{\it not} taken to vanish in this case. Under the GLM dynamics, the
assumed nonlinear frozen-in relation for THC\#4 implies additional
flow stresses as each fluid parcel convects this material property. 

The Taylor hypothesis closure THC\#4 may appear formally similar to
the others, especially to THC\#1. However, THC\#4 differs
fundamentally from the others by being imposed as a finite, rather
than a small-amplitude, approximation. Thus, THC\#4 couples to the
graients of the full Lagrangian mean velocity, rather than to the
gradients of its Eulerian mean small-amplitude approximation. Of
course, the other Taylor hypothesis closures THC\#1--\#3 could also
be made at the nonlinear GLM level, without first making the
linearized fluctuation hypotheses that lead to the $g\ell{m}$ theory.
It turns out that THC\#1 leads to a trivial result in this case, and
the other nonlinear Taylor hypotheses have not yet been analyzed at
the GLM level. The implications and physical interpretations of the
GLM results of THC\#4 are discussed in Holm [1999]. This includes
discussions of an interesting duality between the
Eulerian-mean and Lagrangian-mean fluid velocities that arises for
THC\#4 in the GLM theory.

\subsection{Barotropic $\overline{g\ell{m}}$, a closure model for 
barotropic $g\ell{m}$ }
\label{baro-comp-sec}

In seeking its variational closure, we shall start with the small
amplitude $g\ell{m}$ Lagrangian (\ref{lag-bar}) for barotropic
compressible flow at second order,
\begin{eqnarray}\label{lag-bar-baro-glm}
\bar\ell
&=&
\bar\ell_0 
+ 
\tfrac{1}{2}\,\overline{\ell^{\,\prime\prime}}
=
\int \Big[
\tfrac{1}{2}\bar{D}|\bar\mathbf{u}|^2
-
\bar{D}e(\bar{D})\Big)
\Big]\,d^3x
\\
&&\hspace{5mm}
+
\int \Big[
\tfrac{1}{2}\bar{D}\overline{|\mathbf{u}^{\,\prime}|^2}
+
\overline{D^{\,\prime}\mathbf{u}^{\,\prime}}
\cdot\bar\mathbf{u}
-
\tfrac{1}{2}\tfrac{c^2(\bar{D})}{\bar{D}}
\,\overline{D^{\,\prime\,2}}
\Big]\,d^3x
\,.
\nonumber
\end{eqnarray}
%
Into this $g\ell{m}$ Lagrangian we shall substitute the simplest
available hypothesis for closing the barotropic $g\ell{m}$
system, namely%
\footnote{For compressible flows, the other Taylor hypotheses THC\#2
\& THC\#3 lead to similar formulas to those given in this section. We
shall not discuss those other cases here. The implications of Taylor
hypothesis THC\#2 for {\it incompressible} flows are analyzed in
Marsden and Shkoller [2001].}
%
\begin{equation}\label{close-hypoth}
\mathbf{u}^{\,\prime}
=
-\,\xi\cdot\nabla\bar\mathbf{u}
\quad\hbox{and}\quad
{D}^{\,\prime}
=
-\,\xi\cdot\nabla\bar{D}
\,.
\end{equation}
This substitution yields the mean Lagrangian for the
closed barotropic $g\ell{m}$ system
(barotropic $\,\overline{g\ell{m}}\,$)
\begin{eqnarray}\label{lag-bar-glm-close}
\bar\ell
&=&
\bar\ell_0 
+ 
\tfrac{1}{2}\,\overline{\ell^{\,\prime\prime}}
=
\int \Big[
\tfrac{1}{2}\bar{D}|\bar\mathbf{u}|^2
-
\bar{D}e(\bar{D})\Big)
\Big]\,d^3x
\\
&&\hspace{5mm}
+
\int \Big[
\tfrac{1}{2}\bar{D}\, 
\overline{\xi^k\xi^l}\,
\bar\mathbf{u}_{,\,k}\!\cdot\!\bar\mathbf{u}_{,\,l}
+
\overline{\xi^k\xi^l}\,
\bar\mathbf{u}\!\cdot\!\bar\mathbf{u}_{,\,k}\,\bar{D}_{\!,\,l}
-
\tfrac{c^2(\bar{D})}{2\bar{D}}\,
\overline{\xi^k\xi^l}\,
\bar{D}_{\!,\,k}\bar{D}_{\!,\,l}
\Big]\,d^3x
\,.
\nonumber
\end{eqnarray}
With the $\overline{g\ell{m}}$ closure hypothesis
(\ref{close-hypoth}) no derivatives of the fluctuation statistics
appear in this mean Lagrangian.

Combining the closure hypothesis (\ref{close-hypoth}) with the
$\mathbf{u}^{\,\prime}-$equation (\ref{u-prime}) implies 
$(\partial_t+\bar\mathbf{u}\cdot\nabla)\xi=0$, i.e., componentwise
advection of $\xi$. Consequently, the components of the quadratic
Lagrangian moments are simply carried along with the Eulerian mean
flow, as
\begin{equation}\label{xixi-eqn}
(\partial_t+\bar\mathbf{u}\cdot\nabla)
\,\overline{\xi^k\xi^l}
=
0
\,.
\nonumber
\end{equation}
This equation admits the
isotropic solution 
\begin{equation}\label{lag-mom-iso}
\overline{\xi^k\xi^l}
=
\alpha^2\,\delta^{kl}
\,,
\nonumber
\end{equation}
where $\alpha$ is an advected scalar,
$(\partial_t+\bar\mathbf{u}\cdot\nabla)\alpha=0$, that has
dimensions of length. In turn, this advective relation for $\alpha$
also admits a constant solution, should we wish to
simplify the dynamics of the Lagrangian moments even further.

To develop the barotropic closure model $\overline{g\ell{m}}$, we
shall substitute the variational derivatives of $\bar\ell$ into the
following EP equation, cf. equation (\ref{EP-GLM-eqn}),
\boxeq3{
\begin{equation}\label{EP-glm-close-eqn}
\frac{\partial}{\partial t}
\frac{\delta \bar\ell}{\delta \bar{u}^i}
+\,
\frac{\partial}{\partial x^k}
\Big(\frac{\delta \bar\ell}{\delta \bar{u}^i}\bar{u}^k\Big)
+\,
\frac{\delta \bar\ell}{\delta \bar{u}^k}
\frac{\partial \bar{u}^k}{\partial x^i}
=
\bar{D}\frac{\partial}{\partial x^i}
\frac{\delta \bar\ell}{\delta \bar{D}}
-
\frac{\delta \bar\ell}{\delta \overline{\xi^k\xi^l}}
\frac{\partial \overline{\xi^k\xi^l}}{\partial x^i}
\,.
\end{equation}
}

\noindent
The contribution of the last term arises from the scalar advection of
the components of $\overline{\xi^k\xi^l}$. The necessary variational
derivatives may be obtained from
\begin{eqnarray}\label{Lag-var-der-glm-close}
\delta\bar\ell
&=&
\int
\bigg[
\Big(
(1-\hat\Delta)
(\bar{D}\bar\mathbf{u})
+
\overline{\xi^k\xi^l}\,
\bar\mathbf{u}_{,\,k}
\bar{D}_{,\,l}
\Big)
\cdot\delta\bar\mathbf{u}
\\
&&\hspace{3cm}+
\Gamma_{k\,l}\delta(\,\overline{\xi^k\xi^l}\,)
-
\bar\Pi^{g\ell{m}}\delta\bar{D}
\bigg]
d^3x
\,,
\nonumber
\end{eqnarray}
with homogeneous boundary conditions, 
\begin{equation}\label{bc-natural}
\hat\mathbf{n}\cdot\bar\mathbf{u}
=
0
\quad\hbox{and}\quad
\hat\mathbf{n}\cdot\overline{\xi\xi}
=
0
\quad\hbox{on the boundary.}
\nonumber
\end{equation}
These are the physically meaningful conditions at fixed
boundaries.  Weaker boundary conditions may also suffice in this case,
namely, 
\begin{equation}\label{bc-weaker}
\hat\mathbf{n}\cdot\bar\mathbf{u}
=
0
\quad\hbox{and}\quad
\hat\mathbf{n}\times\big(
\hat\mathbf{n}\cdot\overline{\xi\xi}
\cdot\nabla\big)\bar\mathbf{u}
=
0
\quad\hbox{on the boundary.}
\nonumber
\end{equation}

Here the generalized Laplacian operator $\hat\Delta$ is defined by
\begin{equation}
\hat\Delta 
=
 \partial_l\,\overline{\xi^k\xi^l}\,\partial_k
\,,
\end{equation}
and the $\overline{g\ell{m}}$ potential $\bar\Pi^{g\ell{m}}$ is defined by
\begin{equation}
\bar\Pi^{g\ell{m}} 
=
(1\!-\!\hat\Delta)\big(
\tfrac{1}{2}|\bar\mathbf{u}|^2
\!-\!
h(\bar{D})
\big)
+
\tfrac{1}{2}\, 
\overline{\xi^k\xi^l}\,
\bar\mathbf{u}_{,\,k}\!\cdot\!\bar\mathbf{u}_{,\,l}
-
\tfrac{1}{2}\, 
\overline{\xi^k\xi^l}\,
\bar{D}_{\!,\,k}\bar{D}_{\!,\,l}
h^{\,\prime\prime}(\bar{D})
\,,
\nonumber
\end{equation}
where $h^{\,\prime}(\bar{D})=c^2(\bar{D})/\bar{D}$.
The quantity $\Gamma_{k\,l}$ denotes the variational derivative
of $\bar\ell$ with respect to the mean Lagrangian statistical
moments. Namely, 
\begin{equation}
\Gamma_{k\,l}
=
\frac{\delta\bar\ell}
{\delta\,\overline{\xi^k\xi^l}}
=
\tfrac{1}{2}\, 
\bar{D}\,
\bar\mathbf{u}_{,\,k}\!\cdot\!\bar\mathbf{u}_{,\,l}
+
\bar\mathbf{u}\!\cdot\!\bar\mathbf{u}_{,\,k}\,\bar{D}_{\!,\,l}
-
\tfrac{1}{2}\, 
\bar{D}_{\!,\,k}\bar{D}_{\!,\,l}\,
h^{\,\prime}(\bar{D})
\,.
\nonumber
\end{equation}
The EP motion equation (\ref{EP-glm-close-eqn}) for barotropic
$\overline{g\ell{m}}$ is, thus,
\boxeq2{
\begin{equation}\label{glm-close-eqn}
\partial_t\,
\bar{m}_i
+
\partial_j(\bar{m}_i\bar{u}^j)
+
\bar{m}_j\partial_i\bar{u}^j
=
\bar{D}\,\partial_i\,\bar\Pi^{g\ell{m}} 
-
\Gamma_{k\,l}\,\partial_i\,\overline{\xi^k\xi^l}
\,,
\end{equation}
}

\noindent
where the total mean momentum for barotropic $\overline{g\ell{m}}$ is
given by
\begin{eqnarray}\label{glm-close-mom-def}
\bar\mathbf{m}
=
\frac{\delta \bar\ell}{\delta \bar\mathbf{u}}
&=&
(1-\hat\Delta)
(\bar{D}\bar\mathbf{u})
+
\overline{\xi^k\xi^l}\,
\bar\mathbf{u}_{,\,k}\,
\bar{D}_{,\,l}
\nonumber\\
&=&
\bar{D}\bar\mathbf{u}
-
\tfrac{1}{2}\, 
\big[
\hat\Delta(\bar{D}\bar\mathbf{u})
+
\bar{D}\hat\Delta\bar\mathbf{u}
+
\bar\mathbf{u}\,\hat\Delta\bar{D}
\big]
\,.
\end{eqnarray}
This momentum may be expressed as an operator acting on the mean
fluid velocity, 
\boxeq2{
\begin{equation}\label{glm-mom-def-op}
\bar\mathbf{m}
=
\big[
\big(
\bar{D}
-
\tfrac{1}{2}\,\hat\Delta\bar{D}
\big)
-
\tfrac{1}{2}\, 
\big(
\hat\Delta\bar{D}\!\cdot
+
\bar{D}\hat\Delta\!\cdot
\big)
\big]
\bar\mathbf{u}
\equiv
\hat{\cal O}\bar\mathbf{u}
\,,
\end{equation}
}

\noindent
which defines the operator $\hat{\cal O}$. (The first parenthesis in
the square brackets contains a multiplier and the second one
contains a symmetric operator.) 

To make a connection between the barotropic
$\overline{g\ell{m}}$ motion equation (\ref{glm-close-eqn}) and the
original GLM motion equation (\ref{GLM-eqns}), we shall define the
mean momentum as 
$\bar\mathbf{m}=\bar{D}(\bar\mathbf{u}-\bar\mathbf{p})$ 
with pseudomomentum density
\begin{equation}\label{glm-pseudomom}
\bar{D}\bar\mathbf{p}
\equiv
\tfrac{1}{2}\, 
\big[
\hat\Delta(\bar{D}\bar\mathbf{u})
+
\bar{D}\hat\Delta\bar\mathbf{u}
+
\bar\mathbf{u}\,\hat\Delta\bar{D}
\big]
\,.
\nonumber
\end{equation}
This definition of pseudomomentum and the continuity equation for
$\bar{D}$ allows equation (\ref{glm-close-eqn}) to be rewritten as,
cf. equation (\ref{GLM-eqns}),
\boxeq2{
\begin{eqnarray}\label{glm-close-mot-eqn}
(\partial_t+\bar\mathbf{u}\cdot\nabla)
\big(
\bar\mathbf{u}
-
\bar\mathbf{p}
\big)
+
\big(
\bar{u}_k
-
\bar{p}_k
\big)
\nabla
\bar{u}^k
-
\nabla\,\bar\Pi^{g\ell{m}}
+
\tfrac{1}{\bar{D}}
\Gamma_{kl}\nabla\overline{\xi^k\xi^l}
=
0
\,.
\nonumber
\end{eqnarray}
}

The closed barotropic $\overline{g\ell{m}}$ system consists of the EP
motion equation (\ref{glm-close-eqn}) and two auxiliary equations.
These are the continuity equation (\ref{usual-cont}) for $\bar{D}$ and
the advection equation (\ref{xixi-eqn}) for $\overline{\xi^k\xi^l}$,
recalled as
\begin{equation}\label{xixi-eqn-again}
\partial_t\,\bar{D}
+
{\rm div}\bar{D}\bar\mathbf{u}
=
0
\quad\hbox{and}\quad
(\partial_t+\bar\mathbf{u}\cdot\nabla)
\,\overline{\xi^k\xi^l}
=
0
\,.
\end{equation}
The dynamical properties of the closed barotropic
$\overline{g\ell{m}}$ system may be investigated using the EP
framework. For these equations, we have the Kelvin-Noether
circulation theorem, as well as conservation laws for potential
vorticity, momentum and energy. 


\subsubsection*{Kelvin-Noether circulation theorem for 
barotropic $\overline{g\ell{m}}$}
In the EP framework, the Kelvin-Noether theorem implies the
circulation relation, cf. equation (\ref{KN-circ-GLM}) for adiabatic
GLM,
\begin{equation}\label{KN-circ-glm}
\frac{d}{dt}\oint_{c(\bar\mathbf{u})}
\frac{1}{\bar{D}}
\bar\mathbf{m}
\cdot
d\mathbf{x}
=
-\,\oint_{c(\bar\mathbf{u})}
\frac{1}{\bar{D}}
\Gamma_{k\,l}
\,d\overline{\xi^k\xi^l}
\,.
\nonumber
\end{equation}
Thus, the advected Lagrangian statistical moments
$\overline{\xi^k\xi^l}$ play the same role that specific entropy and
relative buoyancy played in the adiabatic and stratified GLM cases
treated earlier. From Stokes theorem and scalar advection of
$\overline{\xi^k\xi^l}$, we also find {\bf local potential vorticity
conservation}
\begin{equation}\label{xixi-pv-cons}
\big(\partial_t+\bar\mathbf{u}\cdot\nabla\big)
q^{k\,l}
=
0
\,,
\quad
q^{k\,l}
=
\frac{1}{\bar{D}}
\nabla
\big(\,\overline{\xi^k\xi^l}\,\big)
\!\cdot
{\rm curl}
\Big(
\frac{1}{\bar{D}}
\bar\mathbf{m}\Big)
\,,
\quad
\forall\, k,l\,.
\nonumber
\end{equation}
%

\subsubsection*{Momentum conservation for 
barotropic $\overline{g\ell{m}}$}

Because the Lagrangian $\bar\ell$ in equation
(\ref{lag-bar-glm-close}) is invariant under translations, Noether's
theorem yields the momentum conservation law 
(\ref{EP-mom-cons}),
\begin{equation}\label{glm-mom-cons}
{\partial_t}\,\bar{m}_i
+\,
{\partial_j}\,\bar{T}_i^j
=
0
\,,\nonumber
\end{equation}
where 
$\bar\mathbf{m} =
{\delta \bar\ell}/{\delta \bar\mathbf{u}}
$ 
is the $\overline{g\ell{m}}$ Eulerian-mean momentum density in
equation (\ref{glm-mom-def-op}) and the Eulerian-mean stress tensor
$\bar{T}_i^j$ is written in equation (\ref{EP-stress-def}) in the form
\begin{eqnarray}
\bar{T}_i^j
&=&
\bar{m}_i\bar{u}^{\,j}
+
\delta_i^j\Big(\bar{\cal L} 
- 
\tilde{D}
\frac{\partial \bar{\cal L}}{\partial \tilde{D}}
\Big)
\,.\nonumber
\end{eqnarray}
For the $\overline{g\ell{m}}$ theory, this {\bf stress tensor} is
given in terms of mean fluid quantities by
\boxeq2{
\begin{eqnarray}
\bar{T}_i^j
&=&
\bar{m}_i\bar{u}^{\,j}
+
\delta_i^j\,{\cal P}
-
\bar{D}_{\!,\,i}\,
\overline{\xi^j\xi^k}\,
\big(\,
|\bar\mathbf{u}|^2
-
h(\bar{D})
\big)_{\!,\,k}
-
\bar{u}^{\,m}_{,\,i}\,
\overline{\xi^j\xi^k}\,
(
\bar{D}\bar{u}_{\,m}
)_{,\,k}
\,.
\nonumber
\end{eqnarray}
}

\noindent
Here ${\cal P}$ denotes the total $\overline{g\ell{m}}$ {\bf mean
pressure},
\begin{equation}\label{glm-tot-pressure}
{\cal P}
=
(1\!-\!\hat\Delta)p(\bar{D})
+
\tfrac{1}{2}\,\bar{D}\hat\Delta|\bar\mathbf{u}|^2
+
\tfrac{1}{2}\, 
\overline{\xi^k\xi^l}\,
\bar{D}_{\!,\,l}
\big(\,
|\bar\mathbf{u}|^2
+
c^2(\bar{D})
+
h(\bar{D})
\big)_{\!,\,k}
\,.\nonumber
\end{equation}
For an ideal $\gamma-$law gas, 
$c^2=\gamma\,{p}(\bar{D})/\bar{D}$ and
$c^2+h=\gamma c^2/(\gamma-1)$.


\subsubsection*{Energy conservation for 
barotropic $\overline{g\ell{m}}$}

The Legendre transformation of the mean Lagrangian
(\ref{lag-bar-glm-close}) yields the conserved mean energy,
also given by Noether's theorem, as
\boxeq5{
\begin{eqnarray}\label{glmc-erg}
\bar{\cal H}
&=&
\int{\frac{\delta \bar\ell}{\delta \bar\mathbf{u}}
\,\cdot\,\bar\mathbf{u}}\
d^3x
-
\bar\ell(\bar\mathbf{u},\bar{D},\overline{\xi^k\xi^l})
\nonumber
\\
&=&
\int \Big[
\tfrac{1}{2}\,\bar{D}|\bar\mathbf{u}|^2
+
\bar{D}e(\bar{D})
+
\tfrac{1}{2}\bar{D}\, 
\overline{\xi^k\xi^l}\,
\bar\mathbf{u}_{,\,k}\!\cdot\!\bar\mathbf{u}_{,\,l}
+
\frac{c^2(\bar{D})}{2\bar{D}}\,
\overline{\xi^k\xi^l}\,
\bar{D}_{\!,\,k}\bar{D}_{\!,\,l}
\Big]
d^3x
\nonumber
\end{eqnarray}
}
\smallskip

\noindent
We note that we may write the latter two terms in the mean conserved
energy $\bar{\cal H}$, i.e., those due only to fluctuations, as 
\begin{equation}
\tfrac{1}{2}\overline{{\cal H}^{\,\prime\,\prime}}=
\int \Big[
\tfrac{1}{2}\,\bar{D}
\overline{|\mathbf{u}^{\,\prime}|^2 }
+
\frac{c^2(\bar{D})}{2\bar{D}}
\,\overline{D^{\,\prime\,2}}
\Big]
d^3x
\ne
\tfrac{1}{2}\overline{E^{\,\prime\,\prime}}
\,.\nonumber
\end{equation}
This expression does not recover the result of Blokhintsev
[1945] mentioned earlier in equation (\ref{Blok-erg}). However, it
has the advantage of being a positive-definite mean
fluctuational energy for the closed $\overline{g\ell{m}}$ system.


\subsubsection*{Lie-Poisson Hamiltonian formulation of barotropic
$\overline{g\ell{m}}$}

Being an EP system, the barotropic $\overline{g\ell{m}}$ theory may
be transformed into Lie-Poisson Hamiltonian form, by following
the procedure explained in Holm, Marsden \& Ratiu [1998a]. This
Hamiltonian formulation begins by writing the Legendre-transformed
energy in terms of the momentum. We shall assume the operator
$\hat{\cal O}$ in momentum-velocity relation (\ref{glm-mom-def-op})
is invertible, so that one may solve for the velocity from the
momentum as 
$\bar\mathbf{u}=\hat{\cal O}^{-1}\bar\mathbf{m}$. Thus, the energy
Hamiltonian for 
\begin{eqnarray}
\bar{\cal H}
&=&
\int \Big[
\tfrac{1}{2}\,
\bar\mathbf{m}\cdot\hat{\cal O}^{-1}\bar\mathbf{m}
+
\bar{D}e(\bar{D})
+
\tfrac{c^2(\bar{D})}{2\bar{D}}\,
\overline{\xi^k\xi^l}\,
\bar{D}_{\!,\,k}\bar{D}_{\!,\,l}
\Big]
d^3x
\,.\nonumber
\end{eqnarray}
The ideal barotropic $\overline{g\ell{m}}$ equations may now be
treated in the Lie-Poisson Hamiltonian framework, if so desired. The
corresponding Lie-Poisson bracket is of the standard type, defined
on the dual of a certain semidirect-product Lie algebra, as
described, e.g., in Holm, Marsden, Ratiu \& Weinstein [1985].
See also Marsden \& Ratiu [1999] for an introduction to this
now-standard theory and references to the literature. 


\subsubsection*{Barotropic $\overline{g\ell{m}}-\alpha$, a
simplification of barotropic $\overline{g\ell{m}}$ for constant
isotropic Lagrangian statistics}

The scalar advection equation (\ref{xixi-eqn-again}) for the
Lagrangian statistical moments admits the constant isotropic solution
$\overline{\xi^k\xi^l}
=
\alpha^2\,\delta^{kl}
$
where $\alpha$ is a constant length scale. The $\overline{g\ell{m}}$
Lagrangian (\ref{lag-bar-glm-close}) in this case simplifies to, cf.
the Lagrangian (\ref{Lag-alpha-incomp}) for the incompressible
Euler-alpha model,
\boxeq4{
\begin{eqnarray}\label{lag-bar-glm-close-alpha}
\bar\ell
&=&
\int \Big[
\tfrac{1}{2}\bar{D}|\bar\mathbf{u}|^2
-
\bar{D}e(\bar{D})\Big)
\Big]\,d^3x
\\
&&\hspace{5mm}
+\
\alpha^2\!\!\int \Big[
\tfrac{1}{2}\bar{D}\, 
|\nabla\mathbf{u}|^2
+
\tfrac{1}{2}\nabla|\bar\mathbf{u}|^2\,
\cdot\nabla\bar{D}
-
\tfrac{c^2(\bar{D})}{2\bar{D}}\,
|\nabla\bar{D}|^2
\Big]\,d^3x
\,.
\nonumber
\end{eqnarray}
}
This is the Lagrangian for the compressible 
$\overline{g\ell{m}}-\alpha$ model with constant length scale
$\alpha$. For constant $\alpha$, the generalized Laplacian
$\hat\Delta$ in the previous equations reduces to
$\hat\Delta\to\alpha^2\Delta$, where $\Delta$ is the ordinary
Laplacian. The result is a {\bf compressible generalization of the
Euler-alpha model}. The equations of motion for this model are
(\ref{glm-close-eqn}), (\ref{glm-mom-def-op}) and
(\ref{xixi-eqn-again}) with
$\overline{\xi^k\xi^l} = \alpha^2\delta^{kl}$ and
$\hat\Delta\to\alpha^2\Delta$ for constant
$\alpha$.


\subsubsection*{Barotropic $g\ell{m}$ models in a rotating frame}

The EP setting is convenient for transforming the averaged Lagrangian
$\bar\ell$ into a rotating frame. One defines the rotation vector
potential $\mathbf{R}$ satisfying
curl$\,\mathbf{R}=2\Omega(\mathbf{x})$, for a spatially
dependent rotation frequency $\Omega(\mathbf{x})$.
The transformation begins by substituting
into the original Lagrangian the 
linearized relation
$\bar\mathbf{u}+\mathbf{u}^{\,\prime}
=
\bar\mathbf{u}^* + \mathbf{u}^{\,\prime\,*}
+
\bar\mathbf{R}+ \mathbf{R}^{\,\prime},$
with 
$\mathbf{R}^{\,\prime}
=
\mathbf{R}^\ell
-
\xi\cdot\nabla\bar\mathbf{R}.$
One then averages and finally drops
the asterisk in $(\,\cdot\,)^*$ to find, cf. equation
(\ref{lag-bar-baro-glm}),
\begin{eqnarray}\label{lag-bar-baro-glm-rot}
\bar\ell
&=&
\int \Big[ \tfrac{1}{2}\bar{D}
|\bar\mathbf{u}+\bar\mathbf{R}|^2
-
\bar{D}e(\bar{D})\Big)
\Big]\,d^3x
\\
&&\hspace{-1.3cm}
+\!\!
\int\! \bigg[
\tfrac{1}{2}\bar{D}\overline{|\mathbf{u}^{\,\prime}|^2}
+
\tfrac{1}{2}\bar{D}\overline{|\mathbf{R}^{\,\prime}|^2}
+
\overline{D^{\,\prime}
(\mathbf{u}^{\,\prime}+\mathbf{R}^{\,\prime})}
\cdot\big(\bar\mathbf{u}+\bar\mathbf{R}\big)
-
\frac{1}{2}\frac{c^2(\bar{D})}{\bar{D}}
\,\overline{D^{\,\prime\,2}}
\,\bigg]d^3x
\,.
\nonumber
\end{eqnarray}
For a {\it constant} rotation frequency,
$\bar\mathbf{R}=\Omega\times\mathbf{x}$
and $\mathbf{R}^{\,\prime}$
vanishes. In this Lagrangian, the velocities $\bar\mathbf{u}$ and
$\mathbf{u}^{\,\prime}$ are measured in the rotating frame. 
The analysis then proceeds as in the earlier sections in the
EP setting. See, e.g., equations (\ref{Lag-mean-Lag}),
(\ref{Lag-mean-Bouss-Lag}), (\ref{lag-bar}),
(\ref{lag-bar-glm-close}) and  (\ref{EP-glm-close-eqn}).
\vspace{-5mm}

\paragraph{Outlook.}
\begin{description}

\item$\bullet\quad$
We shall discuss the effects of rotation and its interaction with
fluctuations elsewhere, and compare with B\"uhler and McIntyre [1998]
in the Boussinesq stratified flow context. In the simpler shallow
water context, one may use the Green's function method of section
\ref{Lin-fluct-rel} to test the closure hypotheses in equation
(\ref{close-hypoth}). This will be reported in Holm, Tartakovsky,
Wingate and Winter [2001]. 

\item$\bullet\quad$
The analysis of the one-dimensional barotropic
$\overline{g\ell{m}}-\alpha$ equations (without rotation)
will also be studied elsewhere, in Holm, Lowrie and Wingate [2001].

\item$\bullet\quad$
The proper choice of dissipation for this system deserves further
investigation. We propose to add dissipation as {\bf shear viscosity}
in the form
\begin{equation}\label{glmc-dissip}
{\partial_t}\,\bar{m}_i
+
{\partial_j}\,\bar{T}_i^j
=
\frac{\partial}{\partial x^j}\Big(
\nu(\bar{D}-{\cal O})u_{i\,,j}\Big)
\,,\nonumber
\end{equation}
where ${\cal O}$ is the positive symmetric operator in equation
(\ref{glm-mom-def-op}). This choice assures monotonic decay of the
energy in (\ref{glmc-erg}).

\end{description}


\section{Conclusions}\label{conclusions}

\paragraph{Summary remarks.}
\begin{description}

\item$\bullet\quad$
{\bf Remarkable opportunity.} The EP Averaging Lemma implies that the
GLM-averaged Euler equations appear when the EP variational principle
is applied to the GLM-averaged Lagrangian. This is part of a general
theorem discussed in Holm [2001].

\item$\bullet\quad$
{\bf Theme.} We applied small-disturbance ideas that are also basic
for fluid stability analysis and rapid distortion theory in a
variational approach that uses Lagrangian displacements and GLM
averaging in the Euler-Poincar\'e (EP) framework.

\item$\bullet\quad$
{\bf Traditional stability analysis.} The Lagrangian
small-disturbance approach is fundamental and has a tradition in
stability analysis -- although the work here departs
from that tradition, because we also average the second-variation
Lagrangian and we vary with respect to different quantities than in
stability analysis.

\item$\bullet\quad$
{\bf Linearized fluctuation relations.} The approach used here
linearizes the expressions that relate the fluctuations of a given
fluid quantity around its Eulerian mean and the fluctuating
displacement of a Lagrangian fluid parcel trajectory around its mean
position. (These standard linearized fluctuation equations are
expressed geometrically in terms of Lie derivatives in Appendix \#2
in section
\ref{GEOM-appendix}.)

Use of the linearized fluctuation relations brings gradients of
Eulerian mean fluid quantities into the second-variation Lagrangian
for the EP variational principle that leads to the $g\ell{m}$
equations. We emphasized that the
linear $\mathbf{u}^{\,\prime}-$fluctuation relation provides a useful
ingredient for modeling the effects of fluctuations on the mean. The
effects of these fluctuations in the resulting EP equations, however, 
have both linear and nonlinear aspects. (We note that their linear
aspects may also be {\it nonlocal}.) 

\item$\bullet\quad$
{\bf Lagrangian statistics.} GLM and its small-amplitude
approximation $g\ell{m}$ both shift the closure problem from
Eulerian velocity correlations to the statistics of Lagrangian
displacements. However, these Lagrangian statistics are not readily
available or easily observable. We explored how these statistics
might be inferred by solving the linearized fluctuation equations
using a Green's function method. This Green's function method also
suggests some opportunities for diagnosing the Lagrangian
displacement statistics in a DNS, or, perhaps, in data assimilation.

Experiments for measuring Lagrangian statistics are under
development. See, e.g., E. Bodenschatz et al. [2000].

\item$\bullet\quad$
{\bf Green's function diagnostic.} The homogeneous and inhomogeneous
parts of the solution in the Green's function method 
correspond to  frozen-in, and propagating fluctuations.

We proposed an explicit method for using the Green's
function associated with the linearized fluctuation equations to
diagnose the propagating Lagrangian fluctuations in a direct numerical
simulation by using measurements of the Eulerian fluctuations, 
$\mathbf{u}^{\,\prime}$ and $D^{\,\prime}$, as input.

\item$\bullet\quad$
{\bf Variational closure ideas.} The GLM and $g\ell{m}$ theories
have the same advantages of the EP mathematical structure and also
the same disadvantages of not being closed. Three ideas for closure
assumptions are suggested by the second-order $g\ell{m}$ equations:
\begin{description}
\item
(1) Not unexpectedly, 
these closure approximations depend on gradients of Eulerian
mean fluid quantities. 

\item
(2) The $\mathbf{u}^{\,\prime}-$ velocity fluctuation relation 
plays a central role in modeling the necessary {\bf Taylor
hypotheses}.

\item
(3) The closure approximations may be introduced into the EP
variational principle for the $g\ell{m}$ equations, before its
variations are taken.

\end{description}

\item$\bullet\quad$
{\bf Alpha models.} Alpha models are seen as closure
models of these $g\ell{m}$ equations, obtained by ansatzes involving
terms from the linear $\mathbf{u}^{\,\prime}-$ and
$D^{\,\prime}-$ fluctuation equations. These ansatzes are introduced
into the EP variational principle for the $g\ell{m}$ equations,
before its variations are taken. Three of these models have simple
``frozen-in'' evolution equations for determining 
$\overline{\xi\xi}$. The correlation $\overline{\xi\xi}$, in turn,
appears in the nonlinearity of the motion equation for these alpha
models. 

\item$\bullet\quad$
{\bf Nonlinear feedback.} The linear approximation of the fluctuation
equations leads to a  nonlinear feedback in the closed motion 
equations that may tend to keep them within their range of validity.
In particular, the mean fluctuation energy contains gradients of the
mean fluid quantities. Because of this energy penalty, large mean
gradients will tend not to develop in the course of the mean dynamics.

\item$\bullet\quad$
{\bf Bridge.} 
The alpha model and some of its variants can also be a motivation for
further developing the $g\ell{m}$ equations. There are several
variants of alpha-models arising from ansatzes in a Taylor expansion
approximation of terms in an averaged Lagrangian. One could perhaps
use the Green's function method and $g\ell{m}$ theory to test
some of these hypotheses, since they tend to have the same
ingredients as in the $\mathbf{u}^{\,\prime}-$equation. In this sense,
the $g\ell{m}$ theory may be available as a ``bridge'' between
the alpha models and the exact nonlinear GLM theory.

Of course, much remains to be done in this regard. However, the
$g\ell{m}$ framework seems to offer a promising new opportunity for
modeling the nonlinearity of fluid turbulence from a Lagrangian
perspective.

\item$\bullet\quad$
{\bf Synopsis.} This paper connects the GLM equations to the Euler
alpha models through a new set of Eulerian mean $g\ell{m}$ fluid
equations that are derived in the small amplitude limit of the GLM
equations. These equations comprise a one-point closure approximation
that is second-order in the Lagrangian fluctuation statistics. In
principle, these equations may be closed by using the linearized
dynamics of the original equations in combination with the linearized
fluctuation relations. However, because of the complexity
still remaining even at the intermediate $g\ell{m}$ level, we
sought simpler closures by using these linearized fluctuation
relations to guide our choices among the various {\bf Taylor
hypotheses} for deriving variants of the Euler-alpha models and
related models in a variational closure procedure. Following the
original procedure discussed in Holm, Marsden \& Ratiu [1998a] for
deriving  the Euler-alpha model, one substitutes these linear
versions of the Taylor hypothesis into the Lagrangian before taking
its Eulerian mean and then its variations in the EP framework. This
procedure preserved the Kelvin-Noether circulation theorem, which we
regard as a basic geometrical property of all ideal fluid models.
This procedure also preserved the mean momentum and energy balances.
Finally, the procedure led to a barotropic compressible
generalization of the Euler-alpha models.

\end{description}

\subsection{Summary}

In this paper, we developed the geometric approach to dimension
reduction especially for models of turbulence in weakly compressible
fluids in the context of GLM averaging. Our approach concentrated on
reduction of the Lagrangian in Hamilton's principle for adiabatic
compressible fluid dynamics by using a combination of compatible
symmetries and averaging in the EP framework. This approach is
versatile enough to include ocean circulation models for global
climate modeling, as well as fundamental research in turbulence. The
present paper analyzes the basic equations in the framework of the EP
theory and thereby presents them in a unified geometrical context for
further application. 

The EP Averaging Lemma establishes the equivalence of modeling
using the GLM approach, either by directly averaging the equations of
motion, or by averaging the Lagrangian for these equations before
taking its variations.

We discussed EP formulations of both Lagrangian-mean,
and Eulerian-mean fluid equations for modeling turbulence.

\noindent
We used various elements of the classical theory of turbulence,
including:

$\bullet$ Reynolds decomposition(s), 

$\bullet$ Taylor hypothesis closures (THC), 

$\bullet$ Hamilton's principle, 

$\bullet$ Averaged Lagrangians and 

$\bullet$ Euler-Poincar\'e equations 

\noindent
to model and analyze the mean dynamical effects of fluctuations on
3D exact Lagrangian-mean and approximate second-order Eulerian-mean
fluid motion.

Our starting point was the exact nonlinear Generalized Lagrangian
Mean (GLM) equations of Andrews \& McIntyre [1978a] for a
compressible adiabatic fluid. We first recast the GLM equations as
EP equations resulting from the Lagrangian mean of
Hamilton's principle, written in the Eulerian fluid description.
This demonstrated the validity of the general principles underlying
the EP Averaging Lemma.  We then used the small-amplitude
approximation to linearize the relations between Lagrangian
disturbances and Eulerian fluctuations. We substituted these
linearizations into Hamilton's principle for the GLM equations and
kept terms up to quadratic order before taking the Eulerian mean. The
EP equations resulting from this approximate Eulerian-mean Lagrangian
produced a new set of
$g\ell{m}$ equations. These comprise a second-order (one-point,
weakly compressible) turbulence closure model that captures
some aspects of the influence of the small scale dynamics on the
large scale flow -- while preserving the mathematical structure of
the original Euler equations. 

We observed that $g\ell{m}$ theory relates certain combined
Eulerian and Lagrangian aspects of wave properties through
expressions also involving gradients of mean flow properties. This
observation suggested we consider closure schemes that
involve substituting approximations or truncated versions of these
relations between wave properties and mean gradients into Hamilton's
principle, before taking its variations. Thus, we regarded these
approximated, or truncated, relations as a type of {\bf Taylor
hypothesis closure}. We tried one of the simpler variants of this
idea and found a new compressible generalization of the Euler-alpha
models, the $\overline{g\ell{m}}$ closure, whose solution behavior
remains to be studied.

Thus, introducing such Taylor-hypotheses into the second-order
Eulerian-mean closure approximation for the $g\ell{m}$ theory
led to variants of the Euler-alpha models, and a framework for
exploring other options. This included finding new variants of them
for compressible flows that we discussed in section
\ref{baro-comp-sec}.

Being derivable in the EP framework, the GLM
theory, as well as its second-order Eulerian-mean closure
approximation, the new $g\ell{m}$ theory, and the new compressible
$\overline{g\ell{m}}$ generalization of the Euler-alpha equations,
all possess the same fundamental structure and underlying geometry
that are shared by all other ideal fluid theories in the EP
framework. This geometrical structure ensures that these fluid
theories (both exact and approximate ones) each retains its own
Kelvin circulation theorem and the associated conservation law for
potential vorticity arising from it by Noether's theorem, for
particle relabeling symmetry. The EP framework also implies balance
laws for momentum and energy exchanges between mean flow and wave
properties. 

The geometrical structure of the EP framework 
leads, in  addition, to the Lie-Poisson Hamiltonian formulation
for GLM theory, its Eulerian-mean closure approximation and the
variants of the Euler-alpha models. This Hamiltonian formulation
possesses potential-vorticity Casimirs associated with its
Lie-Poisson bracket. In turn, the Lie-Poisson Hamiltonian
structure leads to the energy-Casimir method for characterizing
equilibrium solutions as critical points of a constrained energy
and for establishing their nonlinear Liapunov stability
conditions. All of these additional features are now available,for
the GLM theory, for its Eulerian-mean closure approximation, the
$g\ell{m}$ theory, for the compressible $\overline{g\ell{m}}$ closure
model, and also for the alpha models and any new variants of them
that may arise in the future.

The $g\ell{m}$ theory provides a bridge that spans from the alpha
models to the exact nonlinear GLM theory. We hope this bridge will be
useful in answering questions that arise in the context of the alpha
models and other turbulence closure models.


\section*{Acknowledgements}
We are grateful for stimulating discussions of this topic with P.
Constantin, G. Eyink, J. Marsden, M. E. McIntyre and I. Mezic. Some
of these discussions took place at Cambridge University while the
author was a visiting professor at the Isaac Newton Institute for
Mathematical Science. We are also grateful to R. Lowrie, L.
Margolin, R. Ristorcelli, D. M. Tartakovsky, B. Wingate, L. Winter and
others in the Los Alamos Turbulence Working Group for many worthwhile
discussions.  This work was supported by the U.S. Department of Energy
under contracts W-7405-ENG-36 and the Applied Mathematical
Sciences Program KC-07-01-01.


\section*{References}

\begin{description}

\item
Allen, J.S., Holm, D. D., and Newberger, P.A. [2001] 
Toward an extended-geostrophic Euler--Poincar\'e
model for mesoscale oceanographic flow. 
In {\it Proceedings of the Isaac Newton Institute Programme on the
Mathematics of Atmospheric and Ocean Dynamics}, Cambridge University
Press, to appear.\\

\item
Andrews, D. G. \& McIntyre, M. E. [1978a],
An exact theory of nonlinear waves on a Lagrangian-mean flow.
{\it J. Fluid Mech.} {\bf 89}, 609--646.

\item
Andrews, D. G. \& McIntyre, M. E. [1978b],
On wave-action and its relatives.
{\it J. Fluid Mech.} {\bf 89}, 647--664, 
addendum {\it ibid} {\bf 95}, 796.

\item
Arnold, V. I. [1966],
Sur la g\'{e}ometrie differentielle des groupes de Lie de dimension
infinie et ses applications \`{a} l'hydrodynamique des fluides
parfaits.
{\it Ann. Inst. Fourier (Grenoble)} {\bf 16}, 319--361.

\item
Bayly, B. J., Holm, D. D. \& Lifschitz, A. [1996]
Three-dimensional stability of elliptical
vortex columns in external strain flows.
{\it Trans. Roy. Soc. London} {\bf 354}, 895-926.

\item
Bodenschatz, E. et al. [2000]
Lecture at May 2000 ITP turbulence meeting, UC Santa Barbara.\\
{http://online.itp.ucsb.edu/online/hydrot\_c00/bodenschatz/}

\item
Bretherton, F. P. 1971,
The general linearized theory of wave propagation. In {\it
Mathematical Problems in the Geophysical Sciences, Lect. Appl.
Math.}, {\bf 13}, pp 61--102. Providence, RI: Am. Math. Soc.

\item
Bernstein, I.B., Frieman, E.A., Kruskal, M.D. \& Kulsrud, R.M.
[1958] 
An energy principle for hydromagnetic stability theory.
{\it Proc. Roy. Soc. London A} {\bf244}, 17-40.

\item
Bernstein, I.B. [1983] The variational principle for problems of
ideal magnetohydrodynamic stability. In: {\it Handbook of Plasma
Physics, Vol. 1: Basic Plasma Physics I}, North-Holland, 421-449.

\item
Blokhintsev, D. I. [1945]
{\it Acoustics of a Nonhomogeneous Medium}
(In Russian) Moscow:Gostekhizdat. 
(English trans N. A. C. A. {\it Tech. Memo.} no. 1399.)

\item
B\"uhler, O. and McIntyre, M. E. [1998]
On non-dissipative wave-mean interactions in the atmosphere or oceans.
{\it J. Fluid Mech.} {\bf 354}, 301--343

\item
Cambon C. \&  Scott, J. F. 1999,
Linear and nonlinear models of anisotropic turbulence.
{\it Annu. Rev. Fluid Mech.} {\bf31}, 1Ð53.

\item
Camassa, R., Holm, D. D. \& Levermore, C. D., [1997]
Long-time shallow-water equations with a varying bottom.
{\it J. Fluid Mech.} {\bf 349} 173-189.

\item
Chandrasekhar, S. [1987] {\it Ellipsoidal Figures of
Equilibrium}. New York: Dover.

\item
Chen, S. Y., Foias, C., Holm, D. D., Olson, E.J., Titi, E.S. \&
Wynne, S. [1998]
The Camassa-Holm equations as a closure model for
turbulent channel and pipe flows.
{\it Phys. Rev. Lett.} {\bf 81}, 5338-5341.

\item
Chen, S. Y., Foias, C., Holm, D. D., Olson, E.J., Titi, E.S. \&
Wynne, S. [1999a]
The Camassa-Holm equations and turbulence in pipes and channels.
{\it Physica D} {\bf133}, 49-65.

\item
Chen, S. Y., Foias, C., Holm, D. D., Olson, E.J., Titi, E.S. \&
Wynne, S. [1999b]
A connection between the Camassa-Holm equations and turbulence
in pipes and channels. 
{\it Phys. Fluids} {\bf11}, 2343-2353.

\item
Chen, S. Y., Holm, D. D., Margolin, L. G. \& Zhang, R. [1999c]
Direct numerical simulations of the Navier-Stokes alpha model.
{\it Physica D}, {\bf133} 66-83.

\item
Dewar, R. L. [1970]
Interaction between hydromagnetic waves and a time-dependent
inhomogeneous medium.
{\it Phys. Fluids} {\bf 13}, 2710--2720.

\item
Ebin, D. \& Marsden, J. E.  [1970]
Groups of diffeomorphisms and the motion of an
incompressible fluid. 
{\it Ann. of Math.} {\bf92}, 102-163.

\item
Eckhart, C. [1963]
Some transformations of the hydrodynamic equations.
{\it Phys. Fluids} {\bf 6}, 1037--1041.

\item
Foias, C., Holm, D. D. \& Titi, E.S. [2001a]
The Three Dimensional Viscous Camassa--Holm Equations,
and Their Relation to the Navier--Stokes Equations and
Turbulence Theory.
{\it J. Diff. Eq.}, to appear.

\item
Foias, C., Holm, D. D. \& Titi, E.S. [2001b]
The Navier-Stokes-alpha model of fluid turbulence.
{\it Physica D}, to appear.

\item
Friedman, J. L. \& Schutz, B. F. [1978a]
Lagrangian perturbation theory of nonrelativistic fluids.
{\it Astrophys. J.} {\bf 221}, 937

\item
Friedman, J. L. \& Schutz, B. F. [1978b]
Secular stability of rotating Newtonian stars,
{\it Astrophys. J.} {\bf 222}, 281

\item
Frieman, E. \& Rotenberg, M. [1960]
On hydromagnetic stability of stationary equilibria.
{\it Rev. Mod. Phys.} {\bf32}, 898-902.

\item
Gjaja, I. \& Holm, D. D. [1996]
Self-consistent wave-mean flow interaction
dynamics and its Hamiltonian formulation for a rotating
stratified incompressible fluid.
{\it Physica D}, {\bf 98}, 343-378.

\item
Grappin, R., Cavillier, E. \& Velli, M. [1997]
Acoustic waves in isothermal winds in the vicinity of the sonic point.
{\it Astron. Astrophys.} {\bf322}, 659-670.

\item
Green, A. E. \& Naghdi, P. M. [1976]
A derivation of equations for wave propagation
in water of variable depth.
{\it J. Fluid Mech.} {\bf 78}, 237-246.

\item
Grimshaw, R. 1984,
Wave action and wave-mean flow interaction, with application
to stratified shear flows.
{\it Ann. Rev. Fluid Mech.} {\bf 16}, 11--44.

\item
Hameiri, E. [1998]
Variational principles for equilibrium states with plasma flow.
{\it Phys. Plasmas} {\bf5}, 3270-3281.

\item
Hayes, W. D. [1970]
Conservation of action and modal wave action.
{\it Proc. Roy. Soc. A} {\bf320}, 187-208.

\item
Holm, D. D. [1995]
The ideal Craik-Leibovich equations.
{\it Physica D} {\bf 98}, 415-441.

\item
Holm, D. D. [1999]
Fluctuation effects on 3D Lagrangian mean
and Eulerian mean fluid motion.
{\it Physica D} {\bf133}, 215-269.

\item
Holm, D. D. [2001]
Variational principles, geometry and topology of
Lagrangian-averaged fluid dynamics. In {\it Proceedings of the Isaac
Newton Institute Programme on the Geometry and Topology of Fluids},
Cambridge University Press, to appear.

\item
Holm, D. D. \& Kupershmidt, B. A. [1983]
Poisson brackets and Clebsch representations 
for magnetohydrodynamics, multifluid  plasmas and elasticity
{\it Physica D} {\bf6} 347-363. 

\item
Holm, D. D., Lowrie, R. \& Wingate, B. [2001]
In preparation.

\item
Holm, D. D., Marsden, J. E. \& Ratiu, T. S. [1998a]
The Euler--Poincar\'{e} equations and semidirect products
with applications to continuum theories.
{\it Adv. in Math.} {\bf 137}, 1-81.

\item
Holm, D. D., Marsden, J. E. \& Ratiu, T. S. [1998b]
Euler--Poincar\'e models of ideal fluids
with nonlinear dispersion,
{\it Phys. Rev. Lett.} {\bf 80}, 4173-4177.

\item
Holm, D. D., Marsden, J. E. \& Ratiu, T. S. [2001]
The Euler--Poincar\'{e} equations in geophysical fluid 
dynamics.
In {\it Proceedings of the Isaac Newton Institute Programme
on the Mathematics of Atmospheric and Ocean Dynamics}, Cambridge
University Press, to appear.

\item
Holm, D. D., Marsden, J. E., Ratiu, T. S. \&  Weinstein, A. [1985]
Nonlinear stability of fluid and plasma equilibria.
{\it Physics Reports} {\bf 123}, 1--116.

\item
Holm, D. D., Tartakovsky,D. M., Wingate, B. \& Winter, L. [2001]
In preparation.

\item
Jeffrey, A. \& Taniuti, T. [1966]
{\it Magnetohydrodynamic Stability and Thermonuclear Containment}.
Academic Press, New York.

\item
Leith, C. E. [1980]
Nonlinear normal mode initialization and quasi-geostrophic theory. 
{\it J. Atmos. Sci.} {\bf37}, 958-968.

\item
Lifschitz, A. [1994]
On the instability of certain motions of an ideal
incompressible fluid.
{\it Advances Appl. Math.} {\bf 15}, 404--436.

\item
Lochak, P. and Meunier, C. [1988]
{\it Multiphase averaging for classical systems}.
Springer-Verlag, New York.

\item
Marsden, J. E. \& Ratiu, T. S. [1999]
{\it Introduction to Mechanics and Symmetry} 
Springer: New York, 2nd Edition.

\item Marsden, J. E. \& Scheurle, J. [1993]
Lagrangian reduction and the double spherical pendulum.
{\it ZAMP} {\bf 44} 17-43. 

\item
Marsden, J. E.\& Shkoller, S. [2001] 
The anisotropic averaged Euler equations.
{\it J. Rat. Mech. Anal.} (To appear).

\item
Marsden, J. E., Ratiu, T. S. \& S. Shkoller [2001] 
The geometry and analysis of the averaged
Euler equations and a new diffeomorphism group. 
{\it Geom. Funct. Anal.} (to appear).

\item
Marsden, J. E. \& Weinstein, A. [1983] 
Coadjoint orbits, vortices and Clebsch variables for incompressible
fluids.
{\it Physica D} {\bf7} 305-323. 

\item
Newcomb, W. A. [1962]
Lagrangian and Hamiltonian methods in magnetohydrodynamics,
{\it Nucl. Fusion Suppl.} {part 2}, pp 451-463.

\item
Ristorcelli, J. R. Jr. [2001]
Lagrangian control volume coarse graining for 
for Navier Stokes. In preparation.

\item
Rivlin, R.S. and J.L. Erickson [1955] 
Stress-deformation relations for isotropic materials,
{\it J. Rat. Mech. Anal.} {\bf4}, 323-425.

\item
Sanders, J. A. and Verhulst, F. [1985]
{\it Averaging methods in nonlinear dynamical systems},
Springer-Verlag, New York.

\item
Shkoller, S. [1998]
Geometry and curvature of diffeomorphism groups
with $H_1$ metric and mean hydrodynamics.
{\it J. Funct. Anal.} {\bf160}, 337--365.

\item
Whitham, G. B. [1965]
A general approach to linear and nonlinear
dispersive waves using a Lagrangian.
{\it J. Fluid Mech.} {\bf 22}, 273--283.

\item
Whitham, G. B. [1970]
Two-timing, variational principles and waves.
{\it J. Fluid Mech.} {\bf44}, 373-395.

\end{description}

\vfill\eject


\section{Appendix \#1: The Euler-Poincar\'e theorems
for fluids with advected properties}
\label{EP-appendix}


\subsection{Mathematical setting and statement of the EP theorem}


The assumptions of the Euler-Poincar\'e
theorem from Holm, Marsden \& Ratiu [1998a] are briefly
listed below.
\begin{description}

\item$\quad\bullet$ There is a {\it right\/} representation of Lie
group $G$ on the vector space $V$ and $G$ acts in the natural way on
the {\it right\/} on $TG \times V^\ast$: $(v_g, a)h = (v_gh, ah)$.

\item$\quad\bullet$ Assume that the function $ L : T G \times V ^\ast
\rightarrow \mathbb{R}$ is right $G$--invariant.

\item$\quad\bullet$ In particular, if $a_0 \in V^\ast$, define the
Lagrangian $L_{a_0} : TG \rightarrow \mathbb{R}$ by
$L_{a_0}(v_g) = L(v_g, a_0)$. Then $L_{a_0}$ is right
invariant under the lift to $TG$ of the right action of
$G_{a_0}$ on $G$, where $G_{a_0}$ is the isotropy group of $a_0$.

\item$\quad\bullet$  Right $G$--invariance of $L$ permits one to define
$\ell: {\mathfrak{g}} \times V^\ast \rightarrow \mathbb{R}$ by
\begin{equation}
\ell(v_gg^{-1}, a_0g^{-1}) = L(v_g, a_0).
\nonumber
\end{equation}
Conversely,  this relation defines for any
$\ell: {\mathfrak{g}} \times V^\ast \rightarrow
\mathbb{R} $ a right $G$--invariant function
$ L : T G \times V ^\ast
\rightarrow \mathbb{R} $.

\item$\quad\bullet$ For a curve $g(t) \in G, $ let
\begin{equation}
u(t) \equiv \dot{g}(t) g(t)^{-1}\in TG/G\cong \mathfrak{g}
\nonumber
\end{equation}
and define the curve $a(t)$ as the unique solution of the linear
differential equation with time dependent coefficients 
\begin{equation}\label{advection-rel}
\dot a(t) = -a(t)u(t)
\end{equation}
where the action of $u\in\mathfrak{g}$ on the initial condition $a(0)
= a_0\in V^*$ is denoted by concatenation from the right. The
solution of (\ref{advection-rel}) can be written as the {\bf advective
transport relation},
\begin{equation}
a(t)
= a_0g(t)^{-1}
\,.\nonumber
\end{equation}

\end{description}

\begin{thm} [EP Theorem]\label{rarl}

The following are equivalent:
\begin{enumerate}

\item [{\bf i} ] Hamilton's variational principle
\begin{equation} \label{hamiltonprincipleright1}
\delta \int _{t_1} ^{t_2} L_{a_0}(g(t), \dot{g} (t)) dt = 0
\end{equation}
holds, for variations $\delta g(t)$
of $ g (t) $ vanishing at the endpoints.

\item [{\bf ii}  ] $g(t)$ satisfies the Euler--Lagrange
equations for $L_{a_0}$ on $G$.

\item [{\bf iii} ]  The constrained variational principle
\begin{equation} \label{variationalprincipleright1}
\delta \int _{t_1} ^{t_2}  \ell\,(u(t), a(t)) dt = 0
\end{equation}
holds on $\mathfrak{g} \times V ^\ast $, using variations of the form
\begin{equation} \label{variationsright1}
\delta u = \frac{ \partial \eta}{\partial t } + {\rm ad}_u\,\eta,
\quad
\delta a =  -a\,\eta ,
\end{equation}
where $\eta(t) \in \mathfrak{g}$ vanishes at the
endpoints.

\item [{\bf iv}] The Euler--Poincar\'{e} equations hold on
$\mathfrak{g} \times V^\ast$
\begin{equation} \label{eulerpoincareright1}
\frac{ \partial}{\partial t} \frac{\delta \ell}{\delta u} 
= 
-
 {\rm ad}_{u}^{\ast} \frac{ \delta \ell }{ \delta u}
+ 
\frac{\delta \ell}{\delta a} \diamond a.
\end{equation}
\end{enumerate}
\end{thm}


\noindent
{\bf Remarks.}
\begin{description}
\item$\quad\bullet$
The EP motion equation and advection relations may also be
written equivalently using Lie derivative notation as 
\begin{equation}
\Big(
\frac{ \partial}{\partial t}
+
\pounds_u
\Big)
\frac{ \delta \ell}{ \delta {u}}
-
\frac{ \delta \ell}{ \delta {a}}\diamond a
=
0
\,,\quad\hbox{and}\quad
\Big(
\frac{ \partial}{\partial t}
+
\pounds_u
\Big)
a
=
0
\,.
\nonumber
\end{equation}
The equivalence here of $\pounds_u$ and ad$^*_u$ arises
because $\delta\ell/\delta u$ is a one-form density and the equality 
ad$^*_u\mu=\pounds_u\mu$ holds for any one-form density $\mu$.

\item$\quad\bullet$
In the Lie derivative notation, one proves the {\bf Kelvin-Noether
circulation theorem} immediately as a corollary, by
\begin{equation}
\frac{d}{dt}
\oint_{c(u)}
\frac{1}{D}\
\frac{ \delta \ell}{ \delta {u}}
=
\oint_{c(u)}\!\!
\Big(
\frac{ \partial}{\partial t}
+
\pounds_u
\Big)
\frac{1}{D}\
\frac{ \delta \ell}{ \delta {u}}
=
\oint_{c(u)}
\frac{1}{D}\
\frac{ \delta \ell}{ \delta {a}}\diamond a
\,,
\nonumber
\end{equation}
for any closed curve $c(u)$ that moves with the fluid and advected
density $D$.

\end{description}



\subsection{Summary of the EP equations in 
geometric notation for a barotropic ideal fluid}

As proven in Holm, Marsden \& Ratiu [1998a], the Euler-Lagrange
equations of motion on $(TG\times V^*)$ for a Lagrangian that is
invariant under the group $G$ (where $G$ is the diffeomorphism group
for fluids) are equivalent to the {\bf EP equations},
expressed for a barotropic ideal fluid as
\begin{equation}
\frac{ \partial}{\partial t}
\frac{ \delta \ell}{ \delta {u}^i}
+ 
\underbrace{\,
\frac{ \partial}{\partial {x}^j}
\Big(
\frac{ \delta \ell}{ \delta {u}^i}{u}^j
\Big)
+
\frac{ \delta \ell}{  \delta u^j}
\frac{ \partial {u}^j}{\partial {x}^i}
}
_{\equiv\,\big({\rm ad}^*_u
\,
\frac{ \delta \ell}{ \delta {u}}\big)_i
\,}
-
\underbrace{\,
D
\frac{ \partial }{ \partial {x}^i}
\Big(
\frac{ \delta \ell}{ \delta {D}}
\Big)
\,}
_{\equiv\,\big(\frac{ \delta \ell}{ \delta {D}}\,\diamond \, D
\big)_i
}
=0
\,.
\nonumber
\end{equation}
Here the density $D\in V^*$ satisfies the advection relation,
\begin{equation}
\frac{ \partial D}{\partial t}
=
-\,{\rm div}(D\mathbf{u})
\,.
\nonumber
\end{equation}
In {\bf  EP geometric notation} the last two equations are written
equivalently as 
\begin{equation}
\Big(
\frac{ \partial}{\partial t}
+
{\rm ad}^*_u
\Big)
\frac{ \delta \ell}{ \delta {u}}
-
\frac{ \delta \ell}{ \delta {D}}\diamond D
=
0
\,,\quad\hbox{and}\quad
\frac{ \partial D}{\partial t}
=
-\,\pounds_u D
\,,
\nonumber
\end{equation}
where $\pounds_u$ denotes the Lie derivative with respect to velocity
${u}$, and the operations  ad${^*}$ and
$\boldsymbol{\diamond}$ are defined using the $L_2$ pairing 
$\langle{f,g}\rangle=\int fg\, d^3x$, as 
\begin{equation}
-\,
\Big\langle
{\rm ad}^*_u \frac{ \delta \ell}{ \delta {u}}, \xi
\Big\rangle
=
\Big\langle
\frac{ \delta \ell}{ \delta {u}}, {\rm ad}_u\xi
\Big\rangle
\nonumber
\end{equation}
\begin{equation}
{\rm ad}_u\xi 
= 
\xi u - u\xi 
= 
u\cdot\nabla\xi - \xi\cdot\nabla u
=
-\,{\rm ad}_\xi u
\nonumber
\end{equation}
\begin{equation}
-\,\Big\langle
\frac{ \delta \ell}{ \delta {D}}\diamond D,\xi
\Big\rangle
=
\Big\langle
\frac{ \delta \ell}{ \delta {D}},
\pounds_\xi D
\Big\rangle
=
\Big\langle
\frac{ \delta \ell}{ \delta {D}},
{\rm div}(D\xi) 
\Big\rangle
\,.
\nonumber
\end{equation}
%
%
\paragraph{Remark.}
\begin{description}
\item$\quad\bullet$
The EP equation may also be written using the Lie
derivative as 
\begin{equation}
\Big(
\frac{ \partial}{\partial t}
+
\pounds_u
\Big)
\frac{ \delta \ell}{ \delta {u}}
-
\frac{ \delta \ell}{ \delta {D}}\diamond D
=
0
\,,\quad\hbox{and}\quad
\Big(
\frac{ \partial}{\partial t}
+
\pounds_u
\Big)
D
=
0
\,.
\nonumber
\end{equation}
This equivalence using either $\pounds_u$ or ad$^*_u$ arises because
$\delta\ell/\delta u$ is a one-form density and the equality 
ad$^*_u\mu=\pounds_u\mu$ holds for any one-form density $\mu$.

\end{description}

See Appendix \#2 (section \ref{GEOM-appendix}) for more discussion of
how this underlying geometry expresses itself in the linearized
$g\ell{m}$ approximation.


\section{Appendix \#2: Geometric interpretations of the linearized
fluctuation formulas}
\label{GEOM-appendix}

In fluid dynamics the state space is the group of
diffeomorphisms -- the smooth invertible maps that take the reference
configuration of the fluid with coordinates $\mathbf{x}_0$ into its
current configuration $\mathbf{x}(t, \mathbf{x}_0)$ in the container.

\subsection{Lagrangian \& Eulerian pictures}

Lagrangian fluid trajectories are {\it orbits} under
the action of the diffeomorphism group $G=Diff$ parameterized by time
$t$, thus, 
\begin{equation}
\boldsymbol{x}(t,x_0) = g(t)\cdot x_0
\,,\quad
x_0 = g^{-1}(t)\cdot \boldsymbol{x}(t)
\,,\quad
g\in Diff
\,.\nonumber
\end{equation}
The corresponding {\it velocity relations} are
\begin{equation}
\boldsymbol{\dot{x}}(t,x_0) = \dot{g}(t)\cdot x_0
=
\dot{g}\,g^{-1}(t)\cdot \boldsymbol{x}
=
\boldsymbol{u}(\boldsymbol{x},t)
\,.\nonumber
\end{equation}
This formula relates the Lagrangian and Eulerian definitions of
velocity.

For {\it variations}, one introduces another parameter
$\epsilon$, and denotes,
\[g(t,\epsilon)
\,:\quad
\frac{\partial g}{\partial t}
=
\dot{g}(t,\epsilon)
\,,\quad
\frac{\partial g}{\partial \epsilon}
=
g^{\,\prime}(t,\epsilon)
\,.
\]
Thus, the corresponding {\it variational relations} are
\begin{equation}
\boldsymbol{x}^{\,\prime}(t,\epsilon,x_0) 
= 
g^{\,\prime}(t,\epsilon)\cdot x_0
=
g^{\,\prime}\,g^{-1}(t,\epsilon)\cdot \boldsymbol{x}
=
\boldsymbol{\xi}(\boldsymbol{x},t,\epsilon)
\,.
\nonumber
\end{equation}
This formula relates the Lagrangian and Eulerian definitions of
spatial trajectory fluctuations.

Thus, at spatial position $\boldsymbol{x}$ and time $t$, a given
Lagrangian trajectory has {\bf two tangent vectors}: the Eulerian
velocity, $\boldsymbol{u}=\dot{g}\,g^{-1}$, and the Eulerian
fluctuation/variation, $\boldsymbol{\xi}(\boldsymbol{x},t)
= g^{\,\prime}\,g^{-1}.$
\bigskip

These Lagrangian considerations lead to the following geometric
interpretations of the linearized relations for spatial trajectory
fluctuations in terms of the displacement fluctuation, $\xi$.


\subsection{The $\boldsymbol{u}^{\,\prime}-$ equation}

Equality of cross derivatives in the difference
\[
(\dot{g}\,g^{-1})^{\,\prime} 
- 
(g^{\,\prime}\,g^{-1})\boldsymbol{\,\dot{}}
\] 
gives
\begin{equation}
\boldsymbol{u}^{\,\prime}(\boldsymbol{x},t)
-
\frac{\partial {\xi}}{\partial t}
= 
\xi u - u\xi 
= 
u\cdot\nabla\xi - \xi\cdot\nabla u
=
-\,{\rm ad}_\xi u
\nonumber
\end{equation}
This is the $\boldsymbol{u}^{\,\prime}-$ equation,
\begin{equation}
\boldsymbol{u}^{\,\prime}(\boldsymbol{x},t)
=
\frac{\partial {\xi}}{\partial t}
+
u\cdot\nabla\xi - \xi\cdot\nabla u
=
\partial_t\,\xi-\,{\rm ad}_\xi u
\,,\nonumber
\end{equation}
which relates Eulerian velocity variations 
$\boldsymbol{u}^{\,\prime}(\boldsymbol{x},t)$
and Lagrangian trajectory variation tendencies
$\partial_t\,\xi(\boldsymbol{x},t)$.
This is a key equation for making the $g\ell{m}$ approximations
in the GLM Lagrangian. The particular solutions of the 
$\boldsymbol{u}^{\,\prime}-$ equation also play a role as sources of
inspiration for Taylor hypotheses in simplifying the
$g\ell{m}$ equations to obtain the alpha models in section
\ref{alpha}.

For a discussion of the linearized fluctuation relations
from the viewpoint of Lagrangian stability analysis with a similar
geometric viewpoint, see Friedman \& Schutz [1978a, 1978b].


\subsection{Advected quantities}

For {\bf advected quantities} the right action of the group, denoted
as
\[
a(t,\epsilon) = a_0 g^{-1}(t,\epsilon)
\,,\]
implies the {\it Eulerian advection} ($\,\boldsymbol{\dot{}}$ at
fixed $\boldsymbol{x}$)
\[
\dot{a}(\boldsymbol{x})
=
-\,
a_0 g^{-1}\dot{g}g^{-1}
=
-
a\,u
=
-
\pounds_u\,{a}
\,,\]
and the {\it Eulerian variation} (${}^{\,\prime}$ at fixed
$\boldsymbol{x}$)
\[
\delta a(\boldsymbol{x}) 
= 
a^{\,\prime}(\boldsymbol{x})
=
-\,
a_0 g^{-1}g^{\,\prime}g^{-1}
=
-
a\,\xi
=
-
\pounds_\xi\,{a}
\,.\]
%
\begin{quote}
{\bf Remark.}
This is the general result, examples of which were given
earlier. For an advected scalar $\chi$ one finds
$\chi^{\,\prime}
=
-
\pounds_\xi\,{\chi}
=
-\,\xi\cdot\nabla\chi$
and for an advected density $\bar{D}$ one finds
$D^{\,\prime}
=
-
\pounds_\xi\,\bar{D}
=
-\,{\rm div}\,\bar{D}\xi$.
We note that -- from their definitions in terms of Taylor series
approximations -- all of these linearized fluctuation relations
introduce gradients of Eulerian mean fluid quantities.

Thus, the smooth invertible $\epsilon-$dependence representing
variations in the $g\ell{m}$ and GLM theories is generated by the
displacement vector field $\xi$, just as the time dependence is
generated by the fluid velocity.
\end{quote}


\section{Appendix \#3: On WMFI Decomposition
\\(Lagrangian mean plus fluctuations)}
\label{WMFI-appendix}

The WKB wave packet form of $\xi$ is given by
(assuming $\epsilon\ll1$, $\alpha\ll1$)
$$
\mathbf{x}^\xi =\mathbf{x}+\alpha\,\xi(\mathbf{x},t) 
= \mathbf{x} + \alpha\,
\big(\mathbf{a} e^{i\phi/\epsilon} + cc\big), \quad
\mathbf{a}
=\mathbf{a}(\epsilon \mathbf{x},\epsilon t)
\,,\quad 
\phi=\phi(\epsilon \mathbf{x},\epsilon t)
$$

The {\bf WMFI Expansion of the Fluid Lagrangian} yields terms at
several orders of $\alpha$ and $\epsilon$, 
$$
{\cal L}^{(\alpha,\epsilon)}
= 
\underbrace{
{\cal L}^{(0,0)}
}_{mean flow}
+ 
\underbrace{
\alpha^2\Big[ {\cal L}^{(2,0)} 
+ 
\epsilon {\cal L}^{(2,1)}
+
\epsilon^2 {\cal L}^{(2,2)}
\Big]
}_{WMFI}
+\underbrace{
O(\alpha^4)
}_{Wave-Wave}
$$

Thus there are several levels of approximation available from this
asymptotic expansion of {\bf Hamilton's principle(s)}. We simply list
these as:

\begin{description}
\item
 (1) Set $\alpha=0$ and vary mean flow (MF) quantities
\\
$\Rightarrow$ 
Euler's equations for mean flow when $\delta{\cal L}^{(0,0)}=0$,
\\
as Euler-Lagrange equations in EP form

\item
 (2) Vary only {\it wave} quantities, 
evaluate MF at $\delta{\cal L}^{(0,0)}=0$
\\
$\Rightarrow$ Linearized spectral stability equations, 
\\
cf. Andrews \& McIntyre [1978b]

\item
 (3) Phase average, then vary only wave quantities, 
and evaluate MF at $\delta{\cal L}^{(0,0)}=0$
\\
$\Rightarrow$ 
Whitham's modulation equations, 
at order $O(\alpha^2,\alpha^2\epsilon)$
\\
$\Rightarrow$ WKB stability equations, 
at order $O(\alpha^2\epsilon^2)$, see Lifshitz [1994] and 
Bayly et al. [1996].

\item
 (4) Phase average at constant Lagrangian coordinate, then vary only
MF quantities,  and {\it prescribe} $\xi$
\\
$\Rightarrow$ GLM theory.

\item
 (5) Phase average, then vary both wave {\it and} MF quantities, 
\\
$\Rightarrow$ WMFI equations.

\end{description}

\begin{quote}
{\bf Remark.}
Details of the self-consistent theories of
Lagrangian-mean WMFI at these various levels of
approximation are given in Gjaja \& Holm [1996].
In particular, the self-consistent WMFI equations are shown in Gjaja
\& Holm [1996] to comprise a two-fluid theory, reminiscent of 
Landau's theory of superfluid Helium.
\end{quote}


\section{Appendix \#4: GLM, $g\ell{m}$, and $\alpha-$models of ideal
MHD}
\label{GLM-MHD-appendix}


\subsection*{GLM ideal MHD}

This appendix highlights some of the GLM, $g\ell{m}$, and
$\alpha-$models results in the rest of the paper by setting up the
corresponding results for ideal MHD (MagnetoHydroDynamics).

In ideal MHD one begins by including the magnetic energy potential
energy in the averaged Lagrangian,
$\bar\ell(\bar\mathbf{u}^L,\tilde{D},\bar{s}^L)$ in
(\ref{Lag-mean-Lag}),
\begin{equation}
P_{mag}
=
\int \tfrac{1}{2} \overline{|\mathbf{B}^\xi|^2}
\,d^3x
\,.\nonumber
\end{equation}
The variation of this GLM-averaged magnetic potential energy gives
\begin{equation}
\delta P_{mag}
=
\int \overline{ \mathbf{B}^\xi\cdot\delta\mathbf{B}^\xi }
\,d^3x
=
\int \overline{\mathbf{B}^\xi\cdot\mathsf{K}^{-1}}
\cdot\delta\tilde\mathbf{B}
\,d^3x
+
\hbox{terms in }\delta\xi
\,.\nonumber
\end{equation}
Here we recalled that 
$\tilde\mathbf{B}=\mathsf{K}\cdot\mathbf{B}^\xi$
is a mean quantity and used 
$\mathbf{B}^\xi=\mathsf{K}^{-1}\cdot\tilde\mathbf{B}$.

Hence, the effect of GLM averaging in MHD is to introduce a mean
{\it tensor permeability} due to the fluctuations as
\begin{equation}
\tilde\mathbf{H}
\equiv
\frac{\delta P_{mag}}{\delta \tilde\mathbf{B}}
=
\overline{\mathbf{B}^\xi\cdot\mathsf{K}^{-1}}
=
-\,
\frac{\delta \bar\ell}{\delta \tilde\mathbf{B}}
\,.\nonumber
\end{equation}
The EP equation for ideal {\it barotropic} MHD is given in Holm,
Marsden \& Ratiu [1998a] in the present notation as 
\begin{equation}\label{EP-GLM-MHD-eqn}
\frac{\partial}{\partial t}
\frac{\delta \bar\ell}{\delta \bar\mathbf{u}^L}
+\,
\frac{\partial}{\partial x_k}
\Big(\frac{\delta \bar\ell}{\delta \bar\mathbf{u}^L}
\bar{u}^{L\,k}\Big)
+\,
\frac{\delta \bar\ell}{\delta \bar{u}^{L\,k}}
\nabla
\bar{u}^{L\,k}
=
\tilde{D}\nabla
\frac{\delta \bar\ell}{\delta \tilde{D}}
+\,
\tilde\mathbf{B}\times
\,{\rm curl}\,
\frac{\delta \bar\ell}{\delta \tilde\mathbf{B}}
\,.
\nonumber
\end{equation}
The effect of the magnetic field is to add a 
$\tilde\mathbf{J}\times\tilde\mathbf{B}$ force, with
\[
\tilde\mathbf{J}={\rm curl}\,
\overline{\big(\mathbf{B}^\xi\cdot\mathsf{K}^{-1}\big)}
\,.
\]
The auxiliary equations are the continuity equation
(\ref{GLM-cont-eqn}) for $\tilde{D}$ and the frozen-in flux rule 
(\ref{GLM-mag-advect}) for $\tilde\mathbf{B}$. That is,
\begin{equation}
\partial_t\,\tilde{D}
+
{\rm div}\tilde{D}\bar\mathbf{u}^L
=
0
\,,\quad
\partial_t\, \tilde\mathbf{B}
=
{\rm curl}\,
(\bar\mathbf{u}^L\times\tilde\mathbf{B})
\quad\hbox{with}\quad
{\rm div}\,\tilde\mathbf{B} = 0
\,.\nonumber
\end{equation}
%

\paragraph{Remarks.}
\begin{description}
\item$\quad\bullet$
The GLM averaged EP variational principle yields the GLM averaged
equation for ideal isentropic MHD in Cartesian coordinates as
\boxeq3{
\begin{eqnarray}\label{GLM-eqns-mhd}
\frac{D^L}{Dt}\big(
\bar\mathbf{u}^L
-
\bar\mathbf{p}
\big)
+
\big(
\bar{u}^L_k
-
\bar{p}_k
\big)
\nabla
\bar{u}^L_k
+
\nabla\,\Pi
=
{\rm curl}\,
\overline{\big(\mathbf{B}^\xi\cdot\mathsf{K}^{-1}\big)}
\times
\tilde\mathbf{B}
\,.
\nonumber
\end{eqnarray}
}

\noindent
Here the {\bf pseudomomentum vector}, $\bar\mathbf{p}$, is defined
to be $\bar\mathbf{p} \equiv
-\,\overline{u^\ell_k\nabla\xi^k},$
as in equation (\ref{pseudomom-def}).
The mean potential $\Pi$ has the form of equation (\ref{Pi-def}),
\begin{equation}\label{Pi-def-mhd}
\Pi
=
\overline{e(D^\xi)}
+
\overline{(p^\xi/D^\xi)}
+
\bar{\Phi}^L(\mathbf{x})
-
\frac{1}{2}
\overline{
|\mathbf{u}^\xi|^2
}
\,.
\nonumber
\end{equation}
The computation of $\Pi$ uses $\tilde{D}=D^\xi{\cal J}$ from mass
conservation and the Eulerian mean of the thermodynamic First Law
following a fluid parcel in an isentropic compressible fluid, as in
equation (\ref{1st-law-thermo})
\boxeq2{
\begin{equation}\label{1st-law-thermo-mhd}
d\,\overline{e(D^\xi,\bar{s}^L)}
=
-\,\frac{1}{\tilde{D}}
\,\overline{(p^\xi \, d{\cal J})}
+
\frac{1}{\tilde{D}}
\,\overline{(p^\xi/D^\xi)}
\,d\tilde{D}
\,.
\nonumber
\end{equation}
}
\item$\quad\bullet$
The GLM averaging process preserves the transport structure of the
original ideal MHD equations. In particular, it also yields
preserved linking numbers -- {\bf  magnetic helicity and cross
helicity} -- involving the frozen-in averaged magnetic field and the
Lagrangian mean velocity. Thus, GLM averaging preserves not the
magnetic linking numbers themselves, but the {\it property} that the
GLM averaged dynamics has preserved linking numbers.
\end{description}


\subsection*{Ideal $g\ell{m}$ MHD}

The linearized Eulerian/Lagrangian fluctuation relation for a
magnetic field is
\begin{equation}\label{EP-glm-fluct-rel}
\mathbf{B}^{\,\prime}
=
-\,\pounds_\xi\,\bar\mathbf{B}
=
{\rm curl}\,(\,\xi\times\bar\mathbf{B})
=
-\xi\cdot\nabla\bar\mathbf{B}
+\bar\mathbf{B}\cdot\nabla\xi
-\bar\mathbf{B}\,{\rm div}\,\xi
\,.
\nonumber
\end{equation}
Hence, the $g\ell{m}$ ideal MHD energy variation may be computed as
\begin{equation}
\delta \int
\tfrac{1}{2} 
\overline{ |\mathbf{B}^{\,\prime}|^2 }
\,d^3x
=
-\,
\int 
\delta \bar\mathbf{B}\cdot
\overline{ \Big(\xi\times\,
{\rm curl}\,(\,\xi\times\bar\mathbf{B})\Big) }
\,d^3x
\,.\nonumber
\end{equation}
The motion equation for a $g\ell{m}$ theory of ideal MHD energy is
obtained by including this variational derivative in the
$\bar\mathbf{J}\times\bar\mathbf{B}$ force for the EP equation above,
where,
\begin{equation}
\bar\mathbf{J}={\rm curl}\,(\bar\mathbf{B}
-
\overline{ (\xi\times\mathbf{B}^{\,\prime})}
\,,
\quad\hbox{with}\quad
\mathbf{B}^{\,\prime}=
{\rm curl}\,(\,\xi\times\bar\mathbf{B})
\,.\nonumber
\end{equation}
The latter is a familiar combination from Lagrangian MHD stability
analysis, see, e.g., Bernstein et al. [1958]. (This was kindly
pointed out to the author by E. Caramana.)


\subsection*{EP $g\ell{m}$ equations for incompressible MHD}

The total mean Lagrangian for the incompressible $g\ell{m}$
MHD flow is 
%
\begin{equation}\label{lag-bar-incomp-MHD}
\bar\ell
=
\int \Big[
\tfrac{1}{2}\bar{D}
\Big(
|\bar\mathbf{u}|^2
+
\overline{|\mathbf{u}^{\,\prime}|^2}
\Big)
+
\bar{p}(1 - \bar{D})
-
\tfrac{1}{2}
\Big(
|\bar\mathbf{B}|^2
+
\overline{|\mathbf{B}^{\,\prime}|^2}
\Big)
\Big]\,d^3x
\,.
\end{equation}
%
The corresponding variational derivatives are given by, cf. equation
(\ref{lag-var-der-incomp}),
%
\begin{eqnarray}\label{lag-var-der-incomp-mhd}
\delta\bar\ell(\bar\mathbf{u},\bar{D},\bar\mathbf{B})
&=&
\int \Big[
\delta\bar{D}
\Big(\tfrac{1}{2}\big(\,
|\bar\mathbf{u}|^2
+
\overline{|\mathbf{u}^{\,\prime}|^2}\,\big)
-
\bar{p}
\Big)
+
\delta\bar{p}\,(1 - \bar{D})
\nonumber\\
&&\hspace{-1cm}
+\
\bar{D}\,\delta\bar\mathbf{u}
\cdot
\Big(
\bar\mathbf{u}
-\,
\overline{\xi\times
{\rm curl}\,\mathbf{u}^{\,\prime}}
+
\nabla\overline{(\xi\cdot\mathbf{u}^{\,\prime}\,)}
\Big)
+
\delta\bar\mathbf{u}
\cdot
\overline{\mathbf{u}^{\,\prime}\,{\rm div}\,(\bar{D}\xi)}
\nonumber\\
&&\hspace{1cm}
-\
\delta\bar\mathbf{B}
\cdot
\Big(
\bar\mathbf{B}
-\,
\overline{\xi\times
{\rm curl}\,\mathbf{B}^{\,\prime}}
\Big)
\Big]\,d^3x
\,.
\nonumber
\end{eqnarray}
%
On setting $\bar{D}=1$ and div$\xi=0$, we obtain the same   
$g\ell{m}$ circulation velocity as in subsection \ref{EP-inc-sec},
%
\begin{equation}\label{glm-vee-def-mhd}
\bar\mathbf{v}
\equiv
\bar\mathbf{u}
-\,
\overline{\xi\times
{\rm curl}\,\mathbf{u}^{\,\prime}}
+
\nabla\overline{(\xi\cdot\mathbf{u}^{\,\prime}\,)}
\,.\nonumber
\end{equation}
%
The corresponding EP motion equation (with
$\nabla\cdot\bar\mathbf{u}=0$) is expressed as,
%
\begin{equation}\label{glm-incomp-mhd}
\frac{ \partial}{\partial t}\bar\mathbf{v}
+
\bar\mathbf{u}\cdot\nabla\bar\mathbf{v}
+
\bar{v}_j\nabla \bar{u}^j
+
\nabla\bar{\pi} 
= 
(\bar\mathbf{J}+\overline{\mathbf{J}^{\,\prime\prime}})
\times\bar\mathbf{B}
\,,
\end{equation}
%
with 
$\overline{\mathbf{J}^{\,\prime\prime}}
=-{\rm curl}\,\overline{(\xi\times
{\rm curl}\,\mathbf{B}^{\,\prime})}$
and $\bar{\pi}=\bar{p}
-\tfrac{1}{2}\big(\,
|\bar\mathbf{u}|^2
+
\overline{|\mathbf{u}^{\,\prime}|^2}\,\big)$. This
is the EP equation for the $g\ell{m}$ MHD 
Lagrangian (\ref{lag-bar-incomp-MHD}). This equation also has the
equivalent form,
%
\begin{equation}\label{glm-incomp2-mhd}
\frac{ \partial}{\partial t}\bar\mathbf{v}
-
\bar\mathbf{u}\times{\rm curl}\,\bar\mathbf{v}
+
\nabla\big(\bar\mathbf{v}\cdot\bar\mathbf{u} + \bar{\pi}\,\big)
 = 
(\bar\mathbf{J}+\overline{\mathbf{J}^{\,\prime\prime}})
\times\bar\mathbf{B}
\,.
\nonumber
\end{equation}

The Kelvin circulation theorem for the incompressible
$g\ell{m}$ MHD equations is,
\begin{equation}\label{glm-kel-circ-mhd}
\frac{d}{dt}\oint_{c(\bar\mathbf{u})}
\big(\bar\mathbf{u}
-\,
\overline{\xi\times
{\rm curl}\,\mathbf{u}^{\,\prime}}
\,\big)
\cdot
d\mathbf{x}
=
\oint_{c(\bar\mathbf{u})}
\Big((\bar\mathbf{J}+\overline{\mathbf{J}^{\,\prime\prime}})
\times\bar\mathbf{B}\Big)
\cdot
d\mathbf{x}
\,.
\end{equation}

Thus, for incompressible $g\ell{m}$ MHD the additional statistical
element required for closure is
\begin{equation}\label{jay-prime2}
\overline{ \mathbf{J}^{\,\prime\prime} }
=
-\,{\rm curl}\,\overline{ \big(\xi\times
{\rm curl}\,(\,\xi\times\bar\mathbf{B})\big)}
=
-\,\overline{
{\rm ad}_\xi\big({\rm ad}_\xi\bar\mathbf{B}\big)
}
\,,
\end{equation}
where ad$_\xi\bar\mathbf{B}
=
\xi\cdot\nabla\bar\mathbf{B}
-\,\bar\mathbf{B}\cdot\nabla\xi
.$


\subsection*{EP equations for an incompressible MHD$-\alpha$
model}

Dropping derivatives of $\xi$ again in the linearized fluctuation
relations gives, cf. equation (\ref{close-hypoth}),
\begin{equation}\label{close-hypoth-mhd}
\mathbf{u}^{\,\prime}
=
-\,\xi\cdot\nabla\bar\mathbf{u}
\,,\quad
{D}^{\,\prime}
=
-\,\xi\cdot\nabla\bar{D}
\quad\hbox{and}\quad
\mathbf{B}^{\,\prime}
=
-\,\xi\cdot\nabla\bar\mathbf{B}
\,.
\nonumber
\end{equation}
Substituting these relations into the Lagrangian
(\ref{lag-bar-incomp-MHD}) for incompressible $g\ell{m}$ MHD 
leads again to equation (\ref{glm-incomp-mhd}) with the
re-definitions,
\begin{equation}\label{vee-def-close-mhd}
\bar\mathbf{v}
=
\bar\mathbf{u}
-\,\hat\Delta\,\bar\mathbf{u}
\,,\quad
\overline{ \mathbf{J}^{\,\prime\prime} }
=
-\,{\rm curl}\,\hat\Delta\,\bar\mathbf{B}
\,,\,,\quad
\hat\Delta
=
\partial_{\,l}\,\overline{\xi^k\xi^l}\,\partial_k
\,.
\nonumber
\end{equation}
The auxiliary equations are 
\begin{equation}
{\rm div}\,\bar\mathbf{u}
=
0
\,,\quad
\partial_t\, \bar\mathbf{B}
=
{\rm curl}\,
(\bar\mathbf{u}\times\bar\mathbf{B})
\quad\hbox{with}\quad
{\rm div}\,\bar\mathbf{B} = 0
\,.\nonumber
\end{equation}

Other incompressible MHD$-\alpha$ closure options are
available, corresponding to other Taylor hypotheses for
$\mathbf{u}^{\,\prime}$ and 
approximations for $\mathbf{B}^{\,\prime}$. The present 
closure option is a straight-forward generalization of the
incompressible Euler$-\alpha$ model. As one may verify, the
corresponding conserved energy in this case is
\begin{equation}\label{erg-bar-incomp-MHD}
\bar{\cal E}
=
\frac{1}{2}
\int \Big[\
|\bar\mathbf{u}|^2
+
\overline{\xi^k\xi^l}\
\bar\mathbf{u}_{,\,k}
\cdot
\bar\mathbf{u}_{,\,l}
+
|\bar\mathbf{B}|^2
+
\overline{\xi^k\xi^l}\
\bar\mathbf{B}_{,\,k}
\cdot
\bar\mathbf{B}_{,\,l}
\Big]\,d^3x
\,.
\end{equation}
This is the $H_1$ norm in both $\bar\mathbf{u}$ and 
$\bar\mathbf{B}$, when one chooses the subcase
$\overline{\xi^k\xi^l}=\alpha^2\delta^{kl}$ and
$\alpha^2=$constant. Thus, in this subcase, the presence of
$\alpha\ne0$ regularizes {\it both} the mean velocity and the mean
magnetic field.

The incompressible $\overline{g\ell{m}}$ MHD case may be modified to
include compressibility by following the steps for the compressible
case without magnetic fields discussed in section \ref{baro-comp-sec}.

\end{document}